\documentclass[12pt]{article}
\usepackage{latexsym}
\usepackage{amsmath,amsfonts}
\usepackage{times}

\catcode`\@=11

\newcount\hour
\newcount\minute
\newtoks\amorpm \hour=\time\divide\hour by 60\minute
=\time{\multiply\hour by 60 \global\advance\minute by-\hour}
\edef\standardtime{{\ifnum\hour<12 \global\amorpm={am}%
        \else\global\amorpm={pm}\advance\hour by-12 \fi
        \ifnum\hour=0 \hour=12 \fi
        \number\hour:\ifnum\minute<10
        0\fi\number\minute\the\amorpm}}
\edef\militarytime{\number\hour:\ifnum\minute<10
0\fi\number\minute}

\def\draftlabel#1{{\@bsphack\if@filesw {\let\thepage\relax
   \xdef\@gtempa{\write\@auxout{\string
      \newlabel{#1}{{\@currentlabel}{\thepage}}}}}\@gtempa
   \if@nobreak \ifvmode\nobreak\fi\fi\fi\@esphack}
        \gdef\@eqnlabel{#1}}
\def\@eqnlabel{}
\def\@vacuum{}
\def\marginnote#1{}
\def\draftmarginnote#1{\marginpar{\raggedright\scriptsize\tt#1}}
\def\draft{
        \pagestyle{plain}
        \overfullrule=2pt
        \oddsidemargin -.5truein
        \def\@oddhead{\sl \phantom{\today\quad\militarytime} \hfil
        \smash{\Large\sl DRAFT} \hfil \today\quad\militarytime}
        \let\@evenhead\@oddhead
        \let\label=\draftlabel
        \let\marginnote=\draftmarginnote
        \def\ps@empty{\let\@mkboth\@gobbletwo
        \def\@oddfoot{\hfil \smash{\Large\sl DRAFT} \hfil}
        \let\@evenfoot\@oddhead}
        \def\@eqnnum{(\theequation)\rlap{\kern\marginparsep\tt\@eqnlabel}%
        \global\let\@eqnlabel\@vacuum}  }

\newcommand{\rf}[1]{(\ref{#1})}
\renewcommand{\theequation}{\thesection.\arabic{equation}}
\renewcommand{\thefootnote}{\fnsymbol{footnote}}
\newcommand{\newsection}{    
\setcounter{equation}{0}\section}

\def\appendix#1{\addtocounter{section}{1}\setcounter{equation}{0}
\renewcommand{\thesection}{\Alph{section}}
\section*{Appendix \thesection\protect\indent \parbox[t]{11.15cm}{#1}}
\addcontentsline{toc}{section}{Appendix \thesection\ \ \ #1}}

\def\nline{\,\nabla\kern -0.7em\raise0.2ex\hbox{/}\,\,}
\def\yline{\,y\kern -0.47em /}
\def\aline{\,a\kern -0.49em /}
\def\parline{\,\partial\kern -0.55em /\,\,}

\newcommand{\Po}{\mathbb{P}}

\def\Pt{{\widetilde{P}}}

\def\be{\begin{equation}}
\def\ee{\end{equation}}
\def\beq{\begin{eqnarray}}
\def\eeq{\end{eqnarray}}

\def\sma{{\scriptscriptstyle (a)}}
\def\smaa{{\scriptscriptstyle (aa)}}
\def\smab{{\scriptscriptstyle (ab)}}
\def\smb{{\scriptscriptstyle (b)}}

\def\smaplusone{{\scriptscriptstyle (a+1)}}
\def\smaplustwo{{\scriptscriptstyle (a+2)}}

\def\smone{{\scriptscriptstyle (1)}}
\def\smtwo{{\scriptscriptstyle (2)}}
\def\smthree{{\scriptscriptstyle (3)}}

\def\smonetwo{{\scriptscriptstyle (12)}}
\def\smtwothree{{\scriptscriptstyle (23)}}
\def\smthreeone{{\scriptscriptstyle (31)}}

\def\smoneone{{\scriptscriptstyle (11)}}
\def\smtwotwo{{\scriptscriptstyle (22)}}
\def\smthreethree{{\scriptscriptstyle (33)}}

\def\smaaplusone{{\scriptscriptstyle (aa+1)}}
\def\smaplusoneaplustwo{{\scriptscriptstyle (a+1a+2)}}
\def\smaplustwoa{{\scriptscriptstyle (a+2a)}}

\def\smpt{{\scriptscriptstyle [2]}}
\def\smp3{{\scriptscriptstyle [3]}}
\def\smpf{{\scriptscriptstyle [4]}}
\def\smpn{{\scriptscriptstyle [n]}}

\def\jbf{{\bf j}}
\def\sbf{{\bf s}}
\def\Jbf{{\bf J}}
\def\Lbf{{\bf L}}
\def\Mbf{{\bf M}}
\def\Pbf{{\bf P}}
\def\Sbf{{\bf S}}

\def\betabf{{\boldsymbol{\beta}}}
\def\zetabf{{\boldsymbol{\zeta}}}

\def\BB{{\cal B}}
\def\II{{\cal I}}
\def\LL{{\cal L}}
\def\NN{{\cal N}}
\def\PP{{\cal P}}
\def\ZZ{{\cal Z}}

\def\Nsf{{\sf N}}

\def\nnu{\nu}

\def\kh{\widehat{k}}

\def\phiwt{\widetilde{\phi}}

\def\Gwt{\widetilde{G}}

\def\vac{|0\rangle}

\def\xb{{\bar{x}}}

\def\mas{{\rm m}}

\def\LL{{\cal L}}
\def\XX{{\cal X}}

\begin{document}


\begin{flushright}
FIAN/TD/19-05 \\
hep-th/0512342
\end{flushright}

\vspace{1cm}

\begin{center}

{\Large \bf Cubic interaction vertices for massive and massless

\medskip
higher spin fields}

\vspace{2.5cm}

R.R. Metsaev\footnote{ E-mail: metsaev@lpi.ru }

\vspace{1cm}

{\it Department of Theoretical Physics, P.N. Lebedev Physical
Institute, \\ Leninsky prospect 53,  Moscow 119991, Russia }

\vspace{3cm}

{\bf Abstract}

\end{center}

Using the light-cone formulation of relativistic dynamics, we develop
various methods for constructing cubic interaction vertices and apply
these methods to the study of higher spin fields propagating in flat
space of dimension greater than or equal to four. Generating
functions of parity invariant cubic interaction vertices for massive
and massless higher spin fields of arbitrary symmetry are obtained.
We derive restrictions on the allowed values of spins and the number
of derivatives, which provide a classification of cubic interaction
vertices for totally symmetric fields. As an example of application
of the light-cone formalism, we obtain simple expressions for the
minimal Yang-Mills and gravitational interactions of massive totally
symmetric arbitrary spin fields. We give the complete list of parity
invariant and parity violating cubic interaction vertices that can be
constructed for massless fields in five and six-dimensional spaces.

\vspace{3cm}

Keywords: Light-cone formalism, interaction vertices, higher spin
fields.

\bigskip
PACS-2006:  11.10.-z;  11.30-j; 11.30.Cp

\newpage
\renewcommand{\thefootnote}{\arabic{footnote}}
\setcounter{footnote}{0}

\section{Introduction}

The light-cone formalism \cite{Dirac:1949cp}-\cite{Goddard:1973qh}
offers conceptual and technical simplifications of approaches  to
various problems of modern quantum field and string theories. This
formalism hides some of the symmetries and makes the notation
somewhat cumbersome but eventually turns out to be rather effective.
A number of important problems have been solved in the framework of
this formalism. For example, we mention the solution to the
light-cone gauge string field theory
\cite{Kaku:1974zz}-\cite{Green:1984fu} and the construction of a
superfield formulation for some versions of supersymmetric theories
\cite{Brink:1982pd}-\cite{Brink:1983pf}. Theories formulated within
this formalism may sometimes be a good starting point for deriving a
Lorentz covariant formulation
\cite{Hata:1986jd}-\cite{Siegel:1988yz}. Another attractive
application of the light-cone formalism is the construction of
interaction vertices in the theory of massless higher spin fields
\cite{Bengtsson:1983pd}-\cite{Fradkin:1991iy}. Some interesting
applications of the light-cone formalism to field theory such as QCD
are reviewed in \cite{Brodsky:1997de}. Discussions of super
$p$-branes and string bit models in the light-cone gauge is given in
\cite{deWit:1988ig,Bergshoeff:1988hw} and \cite{Bergman:1995wh}
respectively.

In this paper, we apply the light-cone formalism to study interaction
vertices for higher spin fields. Considerable progress has been
achieved in the problem of constructing the theory describing the
interaction of massless higher spin fields with gravity. In
Ref.\cite{Fradkin:1987ks}, cubic interaction vertices for massless
higher spin fields propagating in $AdS_4$ space were constructed; in
Ref.\cite{Vasiliev:1990en}, nonlinear equations of motion to all
orders in the coupling constant for massless higher spin fields in
$AdS_4$ were found. Nonlinear equations of motion for massless
totally symmetric higher spin fields in $AdS_d$ space ($d\geq4$) were
found in Ref.\cite{Vasiliev:2003ev} (see
\cite{Vasiliev:2004cp},\cite{Bekaert:2005vh} for a recent review). It
now becomes apparent that constructing a self-consistent theory of
massless higher spin fields interacting with gravity requires
formulating the theory in $AdS$ space. Unfortunately, despite the
efforts, an action that leads to the above-mentioned nonlinear
equations of motion has not yet been obtained. To quantize these
theories and investigate their ultraviolet behavior, it would be
important to find an appropriate action. Since the massless higher
spin field theories correspond quantum mechanically to non-local
point particles in a space of certain auxiliary variables, it is
conjectured that such theories may be ultraviolet finite
\cite{Vasiliev:1991rj}. We believe that the light-cone formulation
may be helpful in understanding these theories better. The situation
here may be analogous to that in string theory; a covariant
formulation of the closed string field theories is non-polynomial and
is not useful for practical calculations, while the light-cone
formulation restricts the string action to the cubic order in string
fields.

In this paper, keeping these extremely important applications in
mind, we develop various methods for constructing cubic interaction
vertices and use these methods to find cubic vertices for massive and
massless arbitrary spin fields propagating in flat space. We believe
that most of our approach to massless higher spin fields can be
relatively straightforwardly generalized to the case of massless
higher spin fields in $AdS$ space. The light-cone gauge approach to
dynamics of free fields in $AdS$ space was developed in
\cite{Metsaev:1999ui} (see also
\cite{Metsaev:2002vr},\cite{Metsaev:2003cu}). Although the light-cone
approach in $AdS$ space is complicated compared to that in flat
space, it turns out that these approaches share many properties. We
therefore believe that methods developed in flat space might be
helpful in analyzing dynamics of interacting massless higher spin
fields in $AdS$ space. As regards our study of massive fields, we
note that our interest in light-cone gauge vertices for massive
higher spin fields in flat space is motivated, among other things, by
the potential of our approach for in-depth studies of the interaction
vertices of the light-cone gauge (super)string field theory.

At present, a wide class of cubic interaction vertices for fields
propagating in flat space is known. In particular, the
self-interaction cubic vertices for the massless spin 3 field were
found in \cite{Berends:1984wp}-\cite{Boulanger:2005br} and the
higher-derivative cubic vertex for massless spin 2 and spin 4 fields
was studied in \cite{Deser:1990bk}. More general examples of the
cubic interaction vertices for massless higher spin fields were
discovered in \cite{Bengtsson:1983pd,Bengtsson:1983pg,Berends:1985xx}
and the full list of cubic interaction vertices for massless higher
spin fields was given in \cite{Bengtsson:1986kh}. A wide list of
cubic interaction vertices for massive higher spin fields was
obtained in \cite{Weinberg:1969di} (see also
\cite{Weinberg:1964cn},\cite{Weinberg:1964ev}). With the exception of
Refs.\cite{Bekaert:2005jf,Boulanger:2005br} (devoted to spin 3 field
self-interactions) all the above-mentioned works were devoted to the
analysis of interaction vertices for higher spin fields in $4d$ flat
space. In view of possible applications of the higher spin field
theory to string theory, it is instructive to study cubic interaction
vertices for higher spin fields in space of dimension $d \geq 4$. We
do this in the present paper.

This paper is organized as follows. In Section \ref{freesec}, we
introduce the notation and describe the standard manifestly $so(d-2)$
covariant light-cone formulation of free massless and massive fields.

In Section \ref{GENVERsec},  we discuss arbitrary $n$-point
interaction vertices and find restrictions imposed by kinematical
symmetries of the  Poincar\'e algebra on these vertices.

In Section \ref{CUBversec}, we study restrictions imposed by
kinematical and dynamical symmetries of the Poincar\'e algebra on
cubic interaction vertices for massless and massive fields. We find
various forms of closed equations on cubic interaction vertices.

In Section \ref{Solcubintversec}, we present solution to equations
for parity invariant cubic interaction vertices of massless fields.
Section \ref{secMMO} is devoted to parity invariant cubic interaction
vertices for massless and massive fields. We apply our general
results to derive the minimal Yang-Mills and gravitational
interactions of massive arbitrary spin fields. Our approach allows us
to obtain simple expressions for vertices of these interactions. In
Section \ref{secMMM}, we discuss parity invariant cubic interaction
vertices for massive fields.  In Sections
\ref{Solcubintversec}-\ref{secMMM}, we also derive restrictions on
the allowed values of spins and the number of derivatives for cubic
interaction vertices of the totally symmetric fields.

In Section \ref{sod-4sec}, we develop the light-cone formalism with
manifestly realized $so(d-4)$ symmetries that allows us to study both
the parity invariant and parity violating interaction vertices on an
equal footing. To illustrate this formalism, we construct cubic
interaction vertices for massless totally symmetric fields in $5d$
space and for massless totally symmetric and mixed-symmetry fields in
$6d$ space. We present the complete list of cubic interaction
vertices that can be constructed for massless fields in $d=5,6$
dimensions.

Section \ref{CONsec} summarizes our conclusions and suggests
directions for future research. Appendices contain some mathematical
details and useful formulas.

\newsection{Free light-cone gauge massive and massless fields}\label{freesec}

The method suggested in Ref.\cite{Dirac:1949cp} reduces the problem
of finding a new (light-cone gauge) dynamical system  to the problem
of finding a new solution of defining symmetry algebra commutators%
\footnote{This method is the Hamiltonian version of the Noether
method for finding a new dynamical system. An interesting up-to date
discussion of the Noether method may be found in
\cite{Hurth:1998nq}.}.
Since in our case the defining symmetries are generated by the
Poincar\'e algebra, we begin with a discussion of the realization of
the Poincar\'e algebra on the space of massive and massless fields.
We focus on free fields in this section.

The Poincar\'e algebra of $d$-dimensional Minkowski space is spanned
by translation generators $P^A$ and rotation generators $J^{AB}$ (the
latter span the $so(d-1,1)$ Lorentz algebra). The Lorentz covariant
form of the non-trivial Poincar\'e algebra commutators is
\be \label{pj1} {} [P^A,\,J^{BC}]=\eta^{AB}P^C - \eta^{AC} P^B\,,
\qquad {} [J^{AB},\,J^{CD}] =\eta^{BC}J^{AD} + 3\hbox{ terms}\,, \ee
where $\eta^{AB}$ stands for the mostly positive flat metric tensor.
The generators $P^A$ are chosen to be hermitian, and the $J^{AB}$ to
be antihermitian. To develop the light-cone formulation, in place of
the Lorentz basis coordinates $x^A$ we introduce the light-cone basis
coordinates $x^\pm$, $x^I$ defined by\!
\footnote{ $A,B,C,D = 0,1,\ldots,d-1$ are $so(d-1,1)$ vector indices;
`transverse' indices $I,J,K=1,\ldots,d-2$ are $so(d-2)$ vector
indices; $i,j=1,\ldots,d-4$ are $so(d-4)$ vector indices.}
\be x^\pm \equiv \frac{1}{\sqrt{2}}(x^{d-1}  \pm x^0)\,,\qquad
x^I\,,\quad I=1,\ldots, d-2\,, \ee
and treat $x^{+}$ as an evolution parameter. In this notation, the
Lorentz basis vector $X^A$ is decomposed as $(X^+,X^-,X^I)$ and a
scalar product of two vectors is then decomposed as
\be \eta_{AB}X^A Y^B = X^+Y^- + X^-Y^+ +X^IY^I\,, \ee
where the covariant and contravariant components of vectors are
related as $X^+=X_-$, $X^-=X_+$, $X^I=X_I$. Here and henceforth, a
summation over repeated transverse indices is understood. In the
light-cone formalism, the Poincar\'e algebra  generators can be
separated into two groups:
\beq
&& \label{kingen} P^+,\quad P^I,\quad J^{+I},\quad J^{+-},\quad
J^{IJ}, \qquad \hbox{ kinematical generators}\,;
\\
&& \label{dyngen} P^-,\quad J^{-I}\,, \hspace{4.5cm} \hbox{ dynamical
generators}\,. \eeq
For $x^+=0$, the kinematical generators in the field realization are
quadratic in the physical fields%
\footnote{Namely, for $x^+=\!\!\!\!\!\!/\, 0 $ they have a structure
$G= G_1 + x^+ G_2$, where $G_1$ is quadratic in fields, while $G_2$
contains higher order terms in fields.},
while the dynamical generators receive higher-order
interaction-dependent corrections.

Commutators of the Poincar\'e algebra in light-cone basis can be
obtained from \rf{pj1} by using the light-cone metric having the
following non vanishing elements: $\eta^{+-}=\eta^{-+}=1$,
$\eta^{IJ}=\delta^{IJ}$. Hermitian conjugation rules of the
Poincar\'e algebra generators in light-cone basis take the form
\be P^{\pm \dagger}=P^\pm, \quad P^{I\dagger} = P^I, \quad
J^{IJ\dagger}=-J^{IJ}\,,\quad J^{+-\dagger}=-J^{+-}, \quad J^{\pm
I\dagger} = -J^{\pm I}\,. \ee
To find  a realization of the Poincar\'e algebra on the space of
massive and massless fields we use the light-cone gauge description
of those fields. We discuss massive and massless fields in turn.

{\it Mixed-symmetry massive fields}. In order to obtain the
light-cone gauge description of a massive mixed-symmetry field in an
easy--to--use form, let us introduce a finite set of the creation and
annihilation operators $\alpha_n^I$, $\alpha_n$ and
$\bar{\alpha}_n^I$, $\bar{\alpha}_n$ ($n=1,2,\ldots, \nnu $) defined
by the relations
\be\label{intver15}
[\bar\alpha_n^I,\,\alpha_m^J]=\delta_{nm}\delta^{IJ}\,,\qquad
[\bar{\alpha}_n,\,\alpha_m]=\delta_{nm}\,, \ee
\be \bar\alpha_n^I|0\rangle=0\,,\qquad\bar{\alpha}_n|0\rangle=0\,.\ee
The oscillators $\alpha_n^I$, $\bar\alpha_n^I$ and $\alpha_n$,
$\bar\alpha_n$ transform in the respective vector and scalar
representations of the $so(d-2)$ algebra. In $d$-dimensional
Minkowski space, the massive arbitrary spin field is labeled by the
mass parameter $\mas$ and spin labels $s_1,\ldots,s_\nnu $, $\nnu
 = [\frac{d-1}{2}]$\!
\footnote{ In $4d$ flat space, massive arbitrary spin field is
labeled by the mass parameter $\mas$ and by one spin label $s=s_1$.
Appearance of the spin labels $s_1,\ldots,s_\nnu $ is related to the
fact that physical spin D.o.F of massive field in $d$-dimensional
flat space are described by the $so(d-1)$ algebra irreps labeled by
$[\frac{d-1}{2}]$ Gelfand-Zetlin (or Dynkin) labels.}.
Physical D.o.F of the massive field labeled by spin labels
$s_1,\ldots,s_\nnu $ can be collected into a ket-vector defined by
\be
\label{intver16n1}
|\phi_{s_1 \ldots s_\nnu }(p,\alpha)\rangle \equiv
\prod_{n=1}^\nnu  \sum_{t_n=0}^{s_n}
\alpha_n^{I_1^n}\ldots\alpha_n^{I_{s_n-t_n}^n} \alpha_n^{t_n}\,
\phi_{s_1\ldots s_\nnu}^{I_1^1\ldots I_{s_1-t_1}^1\ldots I_1^\nnu
\ldots I_{s_\nnu - t_\nnu}^\nnu }(p)
|0\rangle\,. \ee
We note that the superscripts like $I_{s_n-t_n}^n$ in \rf{intver16n1}
denote the transverse indices, while $t_n$ is the degree of the
oscillator $\alpha_n$. In \rf{intver16n1} and the subsequent
expressions, $\alpha$ occurring in the argument of ket-vectors
$|\phi(p,\alpha)\rangle$ denotes a set of the oscillators
$\{\alpha_n^I\,,\alpha_n\}$, while $p$ occurring in the argument of
ket-vectors $|\phi(p,\alpha)\rangle$ and $\delta$- functions denotes
a set of the momenta $\{p^I\,,\beta\equiv p^+\}$. Also, we do not
explicitly show the dependence of the ket-vectors
$|\phi(p,\alpha)\rangle$ on the evolution parameter $x^+$. The
ket-vector \rf{intver16n1} is a degree $s_n$ homogeneous polynomial
in the oscillators $\alpha_n^I$, $\alpha_n$:
\be
\label{intver16nn1} \left(\alpha_n^I\bar\alpha_n^I +
\alpha_n\bar\alpha_n- s_n\right)|\phi_{s_1 \ldots s_\nnu
}(p,\alpha)\rangle=0\,,\qquad n=1,\ldots, \nu\,.\ee
As noted above, physical D.o.F of a massive field in $d$-dimensional
Minkowski space are described by irreps of the $so(d-1)$ algebra. For
the ket-vector \rf{intver16n1} to be a carrier of $so(d-1)$ algebra
irreps, some constraints must be imposed on the ket-vector
\rf{intver16n1}. But to avoid unnecessary complications, we do not
impose any constraints on the tensor fields \rf{intver16n1}, which
single out irreps of the $so(d-1)$ algebra from these fields%
\footnote{ For even $d$, these constraints are: a) $ (\alpha_m^I
\bar\alpha_n^I + \alpha_m\bar\alpha_n -s_n\delta_{mn})|\phi_{s_1
\ldots s_\nnu }\rangle=0$, $m\leq n$; b) $ ( \bar\alpha_m^I
\bar\alpha_n^I + \bar\alpha_m \bar\alpha_n )|\phi_{s_1 \ldots s_\nnu
}\rangle=0$; c) $s_1 \geq \ldots \geq s_{\nnu-1} \geq s_\nnu  \geq
0$, $\nnu = [(d-1)/2]$. For odd $d$ and $s_\nnu=0$, one can use the
constraints a),b),c), while for $s_\nnu\ne 0$ the label $s_\nnu$ in
a),c) should be replaced by $|s_\nu|$ and constraints a),b),c) should
be supplemented by appropriate self-duality constraints. After
imposing the constraints a),b),c) the labels $s_1,\ldots,s_\nnu $
become Gelfand-Zetlin labels. In sections 2-7 we assume $s_\nnu
\geq0$ and the constraints \rf{intver16nn1}.}.
This implies that the ket-vector \rf{intver16n1} actually describes a
finite set of massive fields. To develop the light-cone gauge
description of massive arbitrary spin fields on an equal footing we
use a ket-vector defined by
\be
\label{intver16} |\phi(p,\alpha)\rangle \equiv \sum_{s_1,\ldots,
s_\nnu = 0}^{\infty}\,\,|\phi_{s_1 \ldots s_\nnu
}(p,\alpha)\rangle\,. \ee

{\it Mixed-symmetry massless fields}. The light-cone gauge
description of a massless mixed-sym\-metry field can be realized by
using a finite set of the creation and annihilation operators
$\alpha_n^I$ and $\bar{\alpha}_n^I$ ($n=1,2,\ldots, \nnu$). In
$d$-dimensional Minkowski space, the massless arbitrary spin field is
labeled by spin labels $s_1,\ldots,s_\nnu $, $\nnu =
[\frac{d-2}{2}]$\!
\footnote{In $4d$ flat space, massless arbitrary spin field is
labeled by one spin label $s=s_1$. Appearance of the spin labels
$s_1,\ldots,s_\nnu $ is related to the fact that physical D.o.F of
massless field in $d$-dimensional flat space are described by
$so(d-2)$ algebra irreps labeled by $[\frac{d-2}{2}]$ Gelfand-Zetlin
(or Dynkin) labels.}.
Physical D.o.F of the massless field labeled by spin labels
$s_1,\ldots,s_\nnu $ can be collected into a ket-vector defined by
\be
\label{intver16n2} |\phi_{s_1 \ldots s_\nnu }^{\mas=0}
(p,\alpha)\rangle \equiv \prod_{n=1}^\nnu
\alpha_n^{I_1^n}\ldots\alpha_n^{I_{s_n}^n}\,
\phi_{s_1\ldots s_\nnu}^{I_1^1\ldots I_{s_1}^1\ldots I_1^\nnu \ldots
I_{s_\nnu }^\nnu }(p)|0\rangle\,, \ee
which is degree $s_n$ homogeneous polynomial in the oscillators
$\alpha_n^I$:
\be \label{intver16n3}  \left(\alpha_n^I\bar\alpha_n^I -
s_n\right)|\phi_{s_1 \ldots s_\nnu }^{\mas=0}
(p,\alpha)\rangle=0\,,\qquad n=1,\ldots, \nnu \,.\ee
In $d$-dimensional Minkowski space, physical D.o.F of massless field
are described by irreps of the $so(d-2)$ algebra. For the ket-vector
\rf{intver16n2} to be a carrier of $so(d-2)$ algebra irreps some
additional constraints must be imposed on this ket-vector\!
\footnote{ For odd $d$, these constraints are: a) $ (\alpha_m^I
\bar\alpha_n^I - s_n \delta_{mn} )|\phi_{s_1 \ldots s_\nnu
}\rangle=0$, $m\leq n$; b) $\bar\alpha_m^I \bar\alpha_n^I |\phi_{s_1
\ldots s_\nnu }\rangle=0$; c) $s_1 \geq \ldots \geq s_{\nnu-1} \geq
s_\nnu  \geq 0$, $\nnu = [(d-2)/2]$. For even $d$ and $s_\nnu=0$, one
can use the constraints a),b),c), while for $s_\nnu\ne 0$ the label
$s_\nnu$ in a),c) should be replaced by $|s_\nu|$ and constraints
a),b),c) should be supplemented by appropriate self-duality
constraints. In sections 2-7 we assume $s_\nnu \geq0$ and the
constraints \rf{intver16n3}.}.
But, as in the case of a massive field, to avoid unnecessary
complications we do not impose any constraints on the tensor fields
\rf{intver16n2}; therefore the ket-vector \rf{intver16n2} describes a
finite set of massless fields. By analogy with \rf{intver16}, the
ket-vectors of massless fields \rf{intver16n2} can be collected into
a ket-vector $|\phi^{\mas=0}(p,\alpha)\rangle$. We note that in
\rf{intver16n2} and the subsequent expressions, the letter $\alpha$
occurring in the argument of the ket-vectors of massless fields
$|\phi^{\mas=0}(p,\alpha)\rangle$ denotes a set of the oscillators
$\alpha_n^I$\,.

Below, unless otherwise specified, we keep the integer $\nnu$ to be
arbitrary for flexibility.

\medskip

{\it Totally symmetric massive and massless fields}. Totally
symmetric fields are popular in various studies because these fields,
being simpler than the mixed-symmetry fields, allow illustrating many
characteristic features of higher spin fields in a relatively
straightforward way. In order to obtain a description of massive and
massless totally symmetric fields it is sufficient to introduce one
sort of oscillators, i.e. we set $\nnu = 1$ in \rf{intver16n1} and
\rf{intver16n2} respectively. This is to say that physical D.o.F. of
massive and massless totally symmetric spin $s$ fields can be
collected into the respective ket-vectors
\beq
&& \label{intver16n4} |\phi_s(p,\alpha)\rangle = \sum_{t=0}^s
 \alpha^{I_1} \ldots \alpha^{I_{s-t}} \alpha^t \,  \phi^{I_1\ldots
I_{s-t}}(p)|0\rangle\,,
\\[5pt]
&& \label{intver16n5} |\phi_s^{\mas=0}(p,\alpha)\rangle =
\alpha^{I_1} \ldots \alpha^{I_s}\, \phi^{I_1 \ldots I_s}(p)
|0\rangle\,. \eeq
The ket-vector of massive field \rf{intver16n4} is degree $s$
homogeneous polynomial in oscillators $\alpha^I$, $\alpha$, while the
ket-vector of massless field \rf{intver16n5} is degree $s$
homogeneous polynomial in oscillator $\alpha^I$:
\beq
 \label{intver16n6} && \left(\alpha^I\bar\alpha^I + \alpha\bar\alpha - s
\right)|\phi_s(p,\alpha)\rangle =0\,,
\\[5pt]
\label{intver16n7} && \left(\alpha^I\bar\alpha^I - s
\right)|\phi_s^{\mas=0}(p,\alpha\rangle =0 \,.\eeq
As was said in $d$-dimensional Minkowski space physical D.o.F of
massive and massless fields are described by irreps of the $so(d-1)$
and $so(d-2)$ algebras respectively. In order for the fields
\rf{intver16n4} and \rf{intver16n5} to realize irreps of the
$so(d-1)$ and $so(d-2)$ algebras respectively we should impose the
respective tracelessness constraints
\be \label{intver16n8}
\left(\bar\alpha^I\bar\alpha^I + \bar\alpha\bar\alpha\right)
|\phi_s(p,\alpha)\rangle =0\,, \qquad
\bar\alpha^I\bar\alpha^I |\phi_s^{\mas=0}(p,\alpha)\rangle =0 \,.\ee
As in the case of mixed-symmetry fields in order to treat the totally
symmetric arbitrary spin fields on an equal footing it is convenient
to introduce ket-vectors for the respective towers of massive and
massless fields
\beq
\label{intver16n9} && |\phi(p,\alpha)\rangle \equiv
\sum_{s=0}^{\infty}\,\,|\phi_s(p,\alpha)\rangle\,,
\\
\label{intver16n10} && |\phi^{\mas=0}(p,\alpha)\rangle \equiv
\sum_{s=0}^{\infty}\,\,|\phi_s^{\mas=0}(p,\alpha)\rangle\,. \eeq

We proceed with the discussion of a realization of the Poincar\'e
algebra on the space of massive and massless fields. A representation
of the kinematical generators in terms of differential operators
acting on the ket-vector $|\phi\rangle$ is given by\!
\footnote{Throughout this paper, without loss of generality, we
analyze generators of the Poincar\'e algebra and their commutators
for $x^+=0$.}
\beq \label{intver6}&& P^I=p^I\,, \qquad  \qquad\quad P^+=\beta\,,
\\[3pt]
\label{intver9}&& J^{+I}=\partial_{p^I} \beta\,, \qquad \quad \
J^{+-}=\partial_\beta \beta\,,
\\[3pt]
\label{intver11}&&
J^{IJ}=p^I\partial_{p^J}-p^J\partial_{p^I}+M^{IJ}\,, \eeq
where a spin operator $M^{IJ}$ satisfies commutators of the $so(d-2)$
algebra
\be\label{intver13} [M^{IJ},M^{KL}] = \delta^{JK}M^{IL} + 3\hbox{
terms }\,,\ee
and we use the notation
\be \beta\equiv p^+\,,\qquad
\partial_\beta\equiv \partial/\partial \beta\,,
\quad
\partial_{p^I}\equiv \partial/\partial p^I\,. \ee
The representation of the dynamical generators in terms of
differential operators acting on the ket-vector $|\phi\rangle$ is
given by
\beq \label{intver8}&& P^-= p^-\,,\qquad p^- \equiv -\frac{p^Ip^I +
\mas^2}{2\beta}\,,
\\[3pt]
\label{intver12}&& J^{-I}=-\partial_{\beta}p^I + \partial_{p^I}P^-
+\frac{1}{\beta}(M^{IJ}p^J + \mas M^I)\,, \eeq
where $\mas$ is the mass parameter and $M^I$ is a spin operator
transforming in the vector representation of the $so(d-2)$ algebra.
This operator satisfies the commutators
\be\label{intver14} [M^I,M^{JK}] = \delta^{IJ}M^K -\delta^{IK}M^J \,,
\qquad  [M^I,M^J ] = -M^{IJ}\,. \ee
The spin operators $M^{IJ}$ and $M^I$ form commutators of the
$so(d-1)$ algebra (as it should be for the case of massive fields).
The particular form of $M^{IJ}$ and $M^I$ depends on the choice of
the realization of spin D.o.F of physical fields. For example, a
representation of the spin operators $M^{IJ}$ and $M^I$ for the
realization of the physical fields given in \rf{intver16} takes the
form
\beq && M^{IJ}=\sum_{n=1}^\nnu  (\alpha_n^I\bar{\alpha}_n^J-
\alpha_n^J\bar{\alpha}_n^I)\,,\qquad
%
M^I=\sum_{n=1}^\nnu
(\alpha_n^I\bar{\alpha}_n-\alpha_n\bar{\alpha}_n^I)\,.\eeq
As seen from \rf{intver12}, in the limit as $\mas \rightarrow 0$, the
Poincar\'e algebra generators are independent of the spin operator
$M^I$, i.e. the free light-cone gauge dynamics of massive fields have
a smooth limit, given by the dynamics of massless fields.

The above expressions provide a realization of the Poincar\'e algebra
in terms of  differential operators acting on the physical field
$|\phi\rangle$. We now write a field theoretical realization of this
algebra in terms of the physical field $|\phi\rangle$.  As mentioned
above the kinematical generators $G^{kin}$ are realized quadratically
in $|\phi\rangle$, while the dynamical generators $G^{dyn}$ are
realized non-linearly. At the quadratic level, both $G^{kin}$ and
$G^{dyn}$ admit the representation
\be \label{fierep} G_\smpt=\int \beta d^{d-1}p\, \langle\phi(-p)| G
|\phi(p)\rangle\,, \qquad d^{d-1}p \equiv d\beta d^{d-2}p\,,\ee
where $G$ are the differential operators given in
\rf{intver6}-\rf{intver11}, \rf{intver8}, \rf{intver12} and the
notation $G_\smpt$ is used for the field theoretical free generators.
The field $|\phi\rangle$ satisfies the Poisson-Dirac commutator
\be\label{bascomrel}
[\,|\phi(p,\alpha)\rangle\,,\,|\phi(p^\prime\,,\alpha^\prime)\rangle\,]
\bigl|_{equal\, x^+}=\bigr.\frac{\delta^{d-1}(p+p^\prime)}{2\beta}
|\rangle |\rangle'\,, \ee
\be |\rangle |\rangle' \equiv \exp\bigl(\sum_{n=1}^\nnu
(\alpha_n^I\alpha_n^{\prime I}+
\alpha_n\alpha_n^\prime)\bigr)\vac|0^\prime\rangle\,.\ee
With these definitions, we have the standard commutator
\be \label{phig} [ |\phi\rangle,G_\smpt\,] = G|\phi\rangle\,. \ee
In the framework of the Lagrangian approach the light-cone gauge
action takes the standard form
\be \label{lcact} S=\int dx^+  d^{d-1} p\,\, \langle \phi(-p)|{\rm
i}\, \beta
\partial^- |\phi(p)\rangle +\int dx^+ P^-\,, \ee
where $P^-$ is the Hamiltonian. This representation for the
light-cone action is valid for the free and for the interacting
theory. The free theory Hamiltonian can be obtained from relations
\rf{intver8}, \rf{fierep}.

Incorporation of the internal symmetry into the theory under
consideration resembles the Chan--Paton method in string theory
\cite{Paton:1969je}, and could be performed as in
\cite{Metsaev:1991nb}.

\newsection{General structure of $n$-point interaction
vertices and light-cone dynamical principle}\label{GENVERsec}

We begin with discussing the general structure of the Poincar\'e
algebra dynamical generators \rf{dyngen}. In theories of interacting
fields, the dynamical generators receive corrections involving higher
powers of physical fields, and we have the following expansion for
them:
\be\label{GDYN01} G^{dyn}=\sum_{n=2}^\infty G^{dyn}_\smpn\,, \ee
where $G_\smpn^{dyn}$ stands for the $n$ - point contribution (i.e.
the functional that has $n$ powers of physical fields) to the
dynamical generator $G^{dyn}$. The generators $G^{dyn}$ of classical
supersymmetric Yang-Mills theories do not receive corrections of the
order higher than four in fields
\cite{Brink:1982pd,Mandelstam:1982cb,Brink:1984ry}, while the
generators $G_\smpn^{dyn}$ of (super)gravity theories are nontrivial
for all $n\geq 2$ \cite{Goroff:1983hc,Hori:1985qy,Aragone:1989py}\!
\footnote{Generators of the closed string field theories, which
involve the graviton field, terminate at cubic correction
$G_\smp3^{dyn}$ \cite{Green:1983hw,Green:1984fu}. It is natural to
expect that generators of a general covariant theory should involve
all powers of the graviton field $h_{\mu\nu}$. The fact that the
closed string field theories do not involve vertices of the order
higher than tree in $h_{\mu\nu}$ implies that the general covariance
in these theories is realized in a nontrivial way. In string theory,
the general covariance manifests itself upon integrating over massive
string modes and going to the low energy expansion (see
\cite{Tseytlin:1986eq} for a discussion of this theme).}.

The `free' generators $G_\smpt^{dyn}$ \rf{GDYN01}, which are
quadratic in the fields, were discussed in Section 2. Here we discuss
the general structure of the `interacting' dynamical generators
$G_\smpn^{dyn}$, $n\geq 3$. Namely, we describe those properties of
the dynamical generators $G_\smpn^{dyn}$, $n\geq 3$, that can be
obtained from commutators between $G^{kin}$ and $G^{dyn}$. In other
words, we find restrictions imposed by kinematical symmetries on the
dynamical `interacting' generators. We proceed in the following way.

({\bf i}) {} We first consider restrictions imposed by kinematical
symmetries on the dynamical generator $P^-$. As seen from \rf{pj1},
the kinematical generators $P^I$, $P^+$, $J^{+I}$ have the following
commutators  with $P^-$: $[P^-,G_\smpt^{kin}]=G_\smpt^{kin}$. Since
$G_\smpt^{kin}$ are quadratic in the fields, these commutators imply
\be \label{gdynngkinr} [P_\smpn^-,G_\smpt^{kin}]=0\,, \qquad n\geq 3
\,. \ee
Commutators \rf{gdynngkinr} for $G_\smpt^{kin}=(P^I,P^+)$ lead to the
representation for $P_\smpn^-$ as
\beq \label{pm1} && P_\smpn^-  = \int d\Gamma_n  \langle \Phi_\smpn|
p_\smpn^-\rangle \,, \qquad n\geq 3 \,,\eeq
where we use the notation
\be  \label{pm1NN1} \langle \Phi_\smpn| \equiv \prod_{a=1}^n \langle
\phi(p_a,\alpha_a)|\,,\qquad\qquad
|p_\smpn^-\rangle \equiv p_\smpn^- \prod_{a=1}^n |0\rangle_a \,, \ee

\be
\label{delfun01}
d\Gamma_n \equiv (2\pi)^{d-1} \delta^{d-1}(\sum_{a=1}^np_a)
\prod_{a=1}^n \frac{d^{d-1} p_a}{(2\pi)^{(d-1)/2}} \,. \ee
Here and below, the indices $a,b=1,\ldots,n$ label $n$ interacting
fields and the $\delta$- functions in $d\Gamma_n$ \rf{delfun01}
respect conservation laws for the transverse momenta $p_a^I$ and
light-cone momenta $\beta_a$. Generic densities $p_\smpn^-$
\rf{pm1NN1} depend on the momenta $p_a^I$, $\beta_a$, and variables
related to the spin D.o.F, which we denote by $\alpha$:
\be \label{pmpmN1} p_\smpn^- = p_\smpn^- (p_a,\beta_a;\, \alpha)\,.
\ee

({\bf ii}) Commutators \rf{gdynngkinr} for $G_\smpt^{kin}=J^{+I}$
tell us that the generic densities $p_\smpn^-$ in \rf{pm1NN1} depend
on the momenta $p_a^I$ through the new momentum variables
$\Po_{ab}^I$ defined by
\be \label{pablab} {\Po }_{ab}^I\equiv p_a^I\beta_b-p_b^I\beta_a\,,
\ee
i.e. the densities $p_\smpn^-$ turn out to be functions of ${\Po
}_{ab}^I$ in place of $p_a^I$\!
\footnote{We note that due to momentum conservation laws not all
${\Po }_{ab}^I$ are independent. It easy to check that the $n$-point
vertex involves $n-2$ independent momenta $\Po_{ab}^I$.}:
\beq \label{pmpm} && p_\smpn^- =p_\smpn^-({\Po }_{ab},\beta_a;\,
\alpha)\,. \eeq

({\bf iii}) Commutators between $P^-$ and the remaining kinematical
generators $J^{IJ}$, $J^{+-}$ have the form $[P^-,J^{IJ}]=0$,
$[P^-,J^{+-}]=P^-$. Since $J^{IJ}$, $J^{+-}$  are quadratic in
physical fields, these commutators lead to
\be \label{gdynngkin}  [P_\smpn^-,J^{IJ}]= 0\,, \qquad
[P_\smpn^-,J^{+-}]= P_\smpn^-\,, \quad \qquad n\geq 3\,. \ee
It is straightforward to check that commutators \rf{gdynngkin} lead
to the respective equations for the generic densities $p_\smpn^- =
p_\smpn^- (p_a,\beta_a;\, \alpha)$ in \rf{pmpmN1}:
\beq
\label{JIJeq01} && \sum_{a=1}^n \left(p_a^I\partial_{p_a^J}
-p_a^J\partial_{p_a^I} + M^{\sma IJ}\right) |p_\smpn^-\rangle =0\,,
\\
\label{jmpgn2} && \sum_{a=1}^n \beta_a\partial_{\beta_a}
|p_\smpn^-\rangle = 0\,.
\eeq
Using \rf{pablab}, we rewrite Eqs.\rf{JIJeq01}, \rf{jmpgn2} in terms
of $p_\smpn^-= p_\smpn^-(\Po_{ab},\beta_a;\, \alpha)$ in \rf{pmpm} as
\beq
\label{JIJeq01nn} && \Bigl(\sum_{\{ a b \}}
\Po_{ab}^I\partial_{\Po_{ab}^J} - \Po_{ab}^J\partial_{\Po_{ab}^I} +
\sum_{a=1}^n M^{\sma IJ} \Bigr) |p_\smpn^-\rangle =0\,,
\\
\label{jmpgn2nn} && \Bigl(\sum_{\{ a b \}}
\Po_{ab}^I\partial_{\Po_{ab}^I} + \sum_{a=1}^n
\beta_a\partial_{\beta_a}\Bigr) |p_\smpn^-\rangle =0\,,
\eeq
where the notation $\{ab\}$ is used to label the $n-2$ independent
momenta $\Po_{ab}^I$.

({\bf iv}) To complete the description of the dynamical generators,
we consider the dynamical generator $J^{-I}$. Using commutators of
$J^{-I}$ with the kinematical generators, we obtain the
representation for $J_\smpn^{-I}$, $n\geq 3$ as

\be \label{npoi4} J_\smpn^{-I} =\int d\Gamma_n\,\Bigl(\langle
\Phi_\smpn | j_\smpn^{-I}\rangle +\frac{1}{n} \Bigl(\sum_{a=1}^n
\partial_{p_a^I} \langle\Phi_\smpn|\Bigr)|p_\smpn^-\rangle\Bigr)\,, \ee
where we introduce new densities $j_\smpn^{-I}$. From the commutators
of $J^{-I}$ with the kinematical generators, we learn that the
densities $j_\smpn^{-I}$ depend on the momenta $p_a^I$ through the
momenta ${\Po }_{ab}^I$ in \rf{pablab} and satisfy the equations
\beq
\label{neweqqqq01} && \Bigl(\sum_{\{ a b \}}
\Po_{ab}^I\partial_{\Po_{ab}^J} - \Po_{ab}^J\partial_{\Po_{ab}^I} +
\sum_{a=1}^n M^{\sma IJ} \Bigr) |j_\smpn^{-K}\rangle + \delta^{IK}
|j_\smpn^{-J}\rangle - \delta^{JK} |j_\smpn^{-I}\rangle =0\,,
\\
\label{neweqqqq02} && \Bigl(\sum_{\{ a b \}}
\Po_{ab}^I\partial_{\Po_{ab}^I} + \sum_{a=1}^n
\beta_a\partial_{\beta_a}\Bigr) |j_\smpn^{-K}\rangle =0\,.
\eeq

To summarize, the commutators between the kinematical and dynamical
generators yield the expressions for the dynamical generators
\rf{pm1}, \rf{npoi4}, where the densities $p_\smpn^-$, $j_\smpn^{-I}$
depend on ${\Po }_{ab}^I$, $\beta_a$, and spin variables $\alpha$ and
satisfy Eqs.\rf{JIJeq01nn}, \rf{jmpgn2nn},
\rf{neweqqqq01},\rf{neweqqqq02}.

To find the densities $p_\smpn^-$, $j_\smpn^{-I}$, we consider
commutators between the respective dynamical generators; the general
strategy of finding these densities consists basically of the
following three steps, to be referred to as the {\it light-cone
dynamical principle}:

\noindent {\bf a}) Find restrictions imposed by commutators of the
Poincar\'e algebra between the dynamical generators. Using these
commutators shows that the densities $j_\smpn^{-I}$ are expressible
in terms of the densities $p_\smpn^-$.

\noindent {\bf b}) Require the densities $p_\smpn^-$, $j_\smpn^{-I}$
to be polynomials in the momenta $\Po_{ab}^I$. We refer to this
requirement as the light-cone locality condition.

\noindent {\bf c}) Find those densities $p_\smpn^-$ that cannot be
removed by field redefinitions.

In what follows, we apply the light-cone dynamical principle to study
the density $p_\smp3^-$, which we refer to as the cubic interaction
vertex.

\newsection{ Equations for cubic interaction
vertices}\label{CUBversec}

Although many examples of cubic interaction vertices are known in the
literature, constructing cubic interaction vertices for concrete
field theoretical models is still a challenging procedure. General
methods essentially simplifying the procedure of obtaining cubic
interaction vertices were discovered in
\cite{Metsaev:1993gx,Metsaev:1993mj,Metsaev:1993ap}. In this section
we develop the approach in Ref.\cite{Metsaev:1993ap} and demonstrate
how our approach allows constructing cubic interaction vertices
systematically and relatively straightforwardly.

As was explained above (see \rf{pmpm}), the vertex $p_{\smp3}^-$
depends on the momenta $\Po_{ab}^I$, where $a,b=1,2,3$ label three
interacting fields in the cubic interaction vertex. But the momenta
${\Po }_{12}^I$, ${\Po }_{23}^I$, ${\Po }_{31}^I$ are not
independent. This is, using the momentum conservation laws for
$p_a^I$ and $\beta_a$,
\be   p_1^I + p_2^I + p_3^I = 0\,, \qquad \quad
\label{betaconlaw} \beta_1 +\beta_2 +\beta_3 =0 \,,\ee
it is easy to check that ${\Po }_{12}^I$, ${\Po }_{23}^I$, ${\Po
}_{31}^I$ can be expressed in terms of a new momentum ${\Po }^I$ as
\be\label{po122331} {\Po }_{12}^I ={\Po }_{23}^I ={\Po }_{31}^I={\Po
}^I \,, \ee
where the new momentum $\Po^I$ is defined by
\be \label{defpi} {\Po }^I \equiv
\frac{1}{3}\sum_{a=1}^3\check{\beta}_a p_a^I\,, \qquad
\check{\beta}_a\equiv \beta_{a+1}-\beta_{a+2}\,, \quad \beta_a\equiv
\beta_{a+3}\,. \ee
The use of ${\Po }^I$ is advantageous since ${\Po }^I$ is manifestly
invariant under cyclic permutations of the external line indices
$1,2,3$. Therefore the vertex $p_\smp3^-$ is eventually a function of
${\Po }^I$:

\be \label{p2v} p_\smp3^- = p_\smp3^-({\Po },\beta_a;\, \alpha)\,.
\ee
Before discussing the restrictions imposed by the light-cone
dynamical principle, we note that the kinematical symmetry equations
\rf{JIJeq01nn}, \rf{jmpgn2nn} take the following form  in terms of
vertex $p_\smp3^-$ \rf{p2v}:
\beq
\label{kinsod} &&  {\bf J}^{IJ} |p_\smp3^-\rangle =0\,,
\\
\label{honcon04} && (\Po^I\partial_{\Po^I} +
\sum_{a=1}^3\beta_a\partial_{\beta_a}) |p_\smp3^-\rangle =0\,,\eeq
where we use the notation
\beq
\label{JIJp3} && {\bf J}^{IJ} \equiv {\bf L}^{IJ}(\Po) + {\bf
M}^{IJ}\,,
\\[4pt]
\label{LIJ01} && {\bf L}^{IJ}(\Po) \equiv \Po^I \partial_{\Po^J}  -
\Po^J
\partial_{\Po^I} \,,\qquad  \ \ \ \
%
{\bf M}^{IJ}\equiv \sum_{a=1}^3 M^{\sma IJ}\,.
\eeq

We now proceed with discussing the restrictions imposed by the
light-cone dynamical principle on vertex $p_\smp3^-$ \rf{p2v}.
Following the procedure in the previous section, we first find the
restrictions imposed by the Poincar\'e algebra commutators between
the dynamical generators. All that is required is to consider the
commutators
\be  \label{cubeq01}
[\,P^-\,,\,J^{-I}\,]=0\,,\qquad\quad[\,J^{-I}\,,\,J^{-J}\,]=0\,, \ee
which in the cubic approximation take the form
\beq
\label{cubeq02} &&  {} [\, P_\smpt^- \,,\,J_\smp3^{-I}\,] +
[\,P_\smp3^-\,,\,J_\smpt^{-I}\,]=0\,,
\\[7pt]
\label{cubeq03} && [\,J_\smpt^{-I}\,,\,J_\smp3^{-J}\,] +
[\,J_\smp3^{-I}\,,\,J_\smpt^{-J}\,]=0\,. \eeq
Equation \rf{cubeq02} leads to the equation for the densities
$|p_\smp3^-(\Po,\beta_a;\alpha)\rangle$ and
$|j_\smp3^{-I}(\Po,\beta_a;\alpha)\rangle$,
\be\label{cubver3} \Pbf^- |j_\smp3^{-I}\rangle = \Jbf^{-I\dagger}
|p_\smp3^-\rangle\,, \ee
where we use the notation
\beq
&& \label{cubeq05}
{\bf P}^- \equiv \sum_{a=1}^3 p_a^-\,,\qquad \Jbf^{-I\dagger} \equiv
\sum_{a=1}^3 J_a^{-I\dagger}  \,,
\\[5pt]
&& p_a^- \equiv - \frac{p_a^Ip_a^I + \mas_a^2}{2\beta_a}\,,
\\
\label{cubver4}  && J_a^{-I\dagger} \equiv  p_a^I\partial_{\beta_a}
-p_a^-\partial_{p_a^I} -\frac{1}{\beta_a}(M^{\sma IJ}p_a^J + \mas_a
M^{\sma I})\,. \eeq
$\Pbf^-$ and the differential operator $\Jbf^{-I\dagger}$ in
\rf{cubeq05} can be expressed in terms of the momentum $\Po^I$ (see
Appendix A):
\beq
\label{cubver15}&&\Pbf^- \equiv \frac{\Po^I\Po^I}{2\hat{\beta}}
-\sum_{a=1}^3 \frac{\mas_a^2}{2\beta_a}\,,
\\
&& \label{cubeq04} \Jbf^{-I\dagger}|p_\smp3^-\rangle
=-\frac{1}{3\hat\beta}\, \XX^I |p_\smp3^-\rangle \,,\eeq
where we use the notation
\beq
&& \label{cubeq06}  \XX^I \equiv X^{IJ} \Po^J + X^I +
X\partial_{\Po^I}\,,
\\[3pt]
\label{harver02} && X^{IJ} \equiv \sum_{a=1}^3 \check\beta_a(
\beta_a\partial_{\beta_a} \delta^{IJ} - M^{\sma IJ})\,,
\\
\label{harver03} && X^I \equiv \sum_{a=1}^3 \frac{3\hat\beta
\mas_a}{\beta_a} M^{\sma I}\,,
\\
\label{harver04} && X \equiv -\sum_{a=1}^3 \frac{\hat{\beta}
\check{\beta}_a \mas_a^2}{2\beta_a}\,,
\\
\label{cubver16}&& \hat{\beta} \equiv \beta_1\beta_2\beta_3\,. \eeq
Taking \rf{cubeq04} into account we can rewrite Eq.\rf{cubver3} as
\be\label{cubver13} |j_\smp3^{-I}\rangle = - \frac{1}{3\hat\beta{\bf
P^-}} \XX^I |p_\smp3^-\rangle\,, \ee
which tells us that the density $j_\smp3^{-I}$ is not an independent
quantity but is expressible in terms of vertex $p_\smp3^-$ \rf{p2v}.
By substituting $j_\smp3^{-I}$ \rf{cubver13} into Eq.\rf{cubeq03}, we
can verify that Eq.\rf{cubeq03} is fulfilled. Thus, we exhaust all
commutators of the Poincar\'e algebra in the cubic approximation.
Equations \rf{kinsod}, \rf{honcon04} supplemented by relation
\rf{cubver13} provide the complete list of restrictions imposed by
commutators of the Poincar\'e algebra on the densities $p_\smp3^-$,
$j_\smp3^{-I}$. We see that the restrictions imposed by commutators
of the Poincar\'e algebra by themselves are not sufficient to fix the
vertex $p_\smp3^-$ uniquely. To choose the physically relevant
densities $p_\smp3^-$, $j_\smp3^{-I}$, i.e. to fix them uniquely, we
impose the light-cone locality condition: we require the densities
$p_\smp3^-$, $j_\smp3^{-I}$ to be polynomials in ${\Po }^I$. As
regards the vertex $p_\smp3^-$, we require this vertex to be local
(i.e. polynomial in ${\Po }^I$) from the very beginning. However it
is clear from relation \rf{cubver13} that a local $p_\smp3^-$ does
not lead automatically to a local density $j_\smp3^{-I}$. From
\rf{cubver13}, we see that the light-cone locality condition for
$j_\smp3^{-I}$ amounts to the equation
\be\label{loccon01} \XX{}^I |p_\smp3^-\rangle  = {\bf P}^-
|V^I\rangle\,, \ee
where a vertex $|V^I\rangle$ is restricted to be polynomial in
$\Po^I$. In fact, imposing the light-cone locality condition amounts
to requiring that the generators of the Poincar\'e algebra be local
functionals of the physical fields with respect to transverse
directions.

The last requirement we impose on the cubic interaction vertex is
related to field redefinitions. We note that by using local (i.e.
polynomial in the transverse momenta) field redefinitions, we can
remove the terms in the vertex $p_\smp3^-$ that are proportional to
${\bf P}^-$ (see Appendix B). Since we are interested in the vertex
that cannot be removed by field redefinitions, we impose the equation
\be\label{cubver17} |p_\smp3^-\rangle \ne {\bf P}^- |V\rangle\,,\ee
where a vertex $|V\rangle$ is restricted to be polynomial in $\Po^I$.
We note that Eqs.\rf{loccon01}, \rf{cubver17} amount to the
light-cone dynamical principle. If we restrict ourselves to low spin
$s=1,2$ field theories, i.e. Yang-Mills and Einstein theories, it can
then be verified that the light-cone dynamical principle and
Eqs.\rf{kinsod}, \rf{honcon04} allow fixing the cubic interaction
vertices unambiguously (up to several coupling constants). It then
seems reasonable to use the light-cone dynamical principle and
Eqs.\rf{kinsod}, \rf{honcon04} for studying the cubic interaction
vertices of higher spin fields.

To summarize the discussion in this section, we collect equations
imposed by the kinematical symmetries and the light-cone dynamical
principle on vertex $p_\smp3^-$ \rf{p2v}:
\beq
\label{basequ0001} &&  {\bf J}^{IJ} |p_\smp3^-\rangle = 0\,,
\hspace{4.5cm} so(d-2) \hbox{ invariance }
\\
\label{basequ0002} && (\Po^I\partial_{\Po^I} +
\sum_{a=1}^3\beta_a\partial_{\beta_a}) | p_\smp3^- \rangle
=0\,,\qquad \  \ \ \ \ \ \ \beta-\hbox{ homogeneity }
\\
&& \label{basequ0003}  \XX{}^I |p_\smp3^-\rangle  = {\bf P}^- |V^I
\rangle\,, \qquad \hspace{2.4cm} \hbox{ light-cone locality condition
}
\\[5pt]
&& \label{basequ0004} |p_\smp3^-\rangle \ne {\bf P}^-
|V\rangle\,,\eeq
where the vertices $|V\rangle$ and $|V^I\rangle$ are restricted to be
polynomials in $\Po^I$. Solving light-cone locality condition
\rf{basequ0003} leads to the representation for the density
$|j_\smp3^{-I}\rangle$ \rf{cubver13} as
\be\label{cubver13nn} |j_\smp3^{-I}\rangle = - \frac{1}{3\hat\beta}
|V^I\rangle\,. \ee
Equations \rf{basequ0001}-\rf{basequ0004} constitute a complete
system of equations on vertex $p_\smp3^-$ \rf{p2v}. Equation
\rf{basequ0001} reflects the invariance of the vertex
$|p_\smp3^-\rangle$ under $so(d-2)$ rotations, and Eq.\rf{basequ0002}
tells us that $|p_\smp3^-\rangle$ is a zero-degree homogeneity
function in $\Po^I$ and $\beta_a$. Equations \rf{basequ0003},
\rf{basequ0004} and the representation for the density
$|j_\smp3^{-I}\rangle$ \rf{cubver13nn} are obtainable from the
light-cone dynamical principle.

\subsection{Equations for cubic interaction vertices in the
harmonic scheme for an arbitrary realization of the spin degrees of
freedom }\label{equharmschem}

Kinematical symmetry equations \rf{basequ0001}, \rf{basequ0002}
present no difficulties.  For example, the solution of
Eq.\rf{basequ0001} can be written simply  as $p_\smp3^- =
p_\smp3^-(\II)$, where $\II$ is the complete set of the $so(d-2)$
algebra invariants, which can be constructed using $\Po^I$ and the
variables describing spin degrees of freedom. A real difficulty is
then to choose such a $p_\smp3^-(\II)$ that satisfies the light-cone
locality condition \rf{basequ0003} and Eq.\rf{basequ0004}. Since
Eqs.\rf{basequ0003}, \rf{basequ0004} are inconvenient to use it is
preferable to recast these equations into some explicit differential
equations. In turns out that this becomes possible by using the
special, so called harmonic scheme. Our purpose now is therefore to
formulate equations for the cubic interaction vertex in the harmonic
scheme.

We begin with defining the harmonic scheme. By definition, the vertex
$|p_\smp3^-\rangle$ is a polynomial in the momentum $\Po^I$. It is
well known that an {\it arbitrary polynomial} in $\Po^I$ can be made
a {\it harmonic polynomial} in $\Po^I$ by adding a suitable
polynomial proportional to $\Po^I\Po^I$. We also recall that a
polynomial proportional to $\Po^I\Po^I$ can be generated using field
redefinitions. In other words, via field redefinitions, the vertex
$|p_\smp3^-\rangle$ can be made a harmonic polynomial in $\Po^I$ (see
Appendix B). {\it The scheme in which the vertex $|p_\smp3^-\rangle$
satisfies the harmonic equation
\be \label{harmcon01}
\partial_{\Po^I}^{}\partial_{\Po^I}^{}|p_\smp3^-\rangle=0
\ee
is referred to as the harmonic scheme}.

We proceed with the discussion of Eqs.\rf{basequ0003},
\rf{basequ0004} in the harmonic scheme. In the harmonic scheme,
Eq.\rf{basequ0004} is satisfied automatically because a harmonic
polynomial in $\Po^I$ cannot be represented as $\Pbf^- |V\rangle$,
where $|V\rangle$ is a polynomial in $\Po^I$. It then remains to
analyze the light-cone locality condition \rf{basequ0003} in the
harmonic scheme. It turns out that this condition leads to the
differential equations for vertex $p_\smp3^-$ \rf{p2v} (see Appendix
C):
\beq \label{harver01} && \Bigl( X^{IJ} \PP^J + X^I +
\Bigl(X\delta^{IJ} + \frac{\hat\beta}{2\kh + N}\sum_{a=1}^3
\frac{\mas_a^2}{\beta_a}X^{IJ}\Bigr)
\partial_{\Po^J} \Bigr) |p_\smp3^-\rangle   = 0\,,
\eeq
where $X^{IJ}$, $X^I$, $X$ are defined in \rf{harver02}-\rf{harver04}
and we use the notation
\be \label{khdef}
\PP^I  \equiv \Po^I - \Po^J\Po^J \frac{1}{2\kh +
N}\partial_{\Po^I}\,,\qquad
\kh \equiv \Po^I\partial_{\Po^I}\,,\qquad
N\equiv d-2\,.\ \ee
In what follows, we refer to Eqs.\rf{harver01} as locality equations.
A remarkable property of the harmonic scheme is that it allows
writing closed expression for the density $|j_\smp3^{-I}\rangle$ in
terms of vertex $|p_\smp3^-\rangle$ \rf{p2v} without specifying the
spin operators $M^{IJ}$, $M^I$:
\be \label{clorepj3}|j_\smp3^{-I}\rangle = -\frac{2}{3(2 \kh + N)}
X^{IJ}
\partial_{\Po^J}|p_\smp3^-\rangle\,. \ee
To summarize, the complete set of equations to be solved in the
harmonic scheme is given by \rf{basequ0001}, \rf{basequ0002},
\rf{harmcon01}, \rf{harver01}. These equations are used in Section
\ref{sod-4sec} to develop the so called $so(d-4)$ light-cone
approach.

\subsection{Equations for parity invariant cubic interaction vertices
in the minimal scheme and for the oscillator realization of spin
degrees of freedom}

In this section, we develop an alternative approach to the analysis
of the equations for the cubic interaction vertex
\rf{basequ0001}-\rf{basequ0004} based on a scheme we refer to as the
minimal scheme. This scheme, to be defined below, turns out to be
convenient for the classification of cubic interaction vertices.

To proceed, we use the oscillator realization of physical fields in
terms of the ket-vectors in \rf{intver16n1}, \rf{intver16}. In this
case, the spin arguments of vertex $p_\smp3^-$ \rf{p2v} denoted by
$\alpha$ become oscillators $\alpha_n^{\sma I}$, $\alpha_n^\sma$,
$a=1,2,3$, and vertex $p_\smp3^-$ \rf{p2v} takes the form%
\footnote{Throughout this paper, unless otherwise specified, the
subscripts $m,n,q$ takes the values $1,\ldots,\nu$. The short
notation like $p_\smp3^-(x^\sma)$ is used to indicate the dependence
of $p_\smp3^-$ on $x^\smone$, $x^\smtwo$, $x^\smthree$.}:
\be \label{varrep5} |p_\smp3^-\rangle = p_\smp3^- (\Po^I,\beta_a ;\,
\alpha_n^{\sma I}, \, \alpha_n^\sma
)|0\rangle_1|0\rangle_2|0\rangle_3\,. \ee
We now analyze Eqs.\rf{basequ0001}-\rf{basequ0004} in turn.

{\bf i}) We first analyze the restrictions imposed by the $so(d-2)$
invariance equations \rf{basequ0001}. These equations tell us that
vertex $p_\smp3^-$ \rf{varrep5} depends on the $so(d-2)$ algebra
invariants that can be constructed using $\Po^I$ and the oscillators
$\alpha_n^{\sma I}$. It is clear that the following $so(d-2)$
invariants can be constructed:
\be \Po^I\Po^I\,,\qquad \alpha_n^{\sma I}\Po^I\,,\qquad
\alpha_n^{\sma I}\alpha_m^{\smb I}\,.\ee
We note that there are additional invariants constructed using the
antisymmetric Levi-Civita symbol $\epsilon^{I_1\ldots I_{d-2}}$. {\it
The vertices not involving the antisymmetric Levi-Civita symbol
are said to be parity invariant vertices}%
\footnote{ The methods of manifest covariantization of light-cone
vertices \cite{Hata:1986jd}-\cite{Siegel:1988yz} are most suitable
for studying the parity invariant vertices. Thus, we expect that all
our parity invariant vertices could be covariantized in a relatively
straightforward way.},
and {\it vertices involving one antisymmetric Levi-Civita symbol are
said to be parity violating vertices}. We focus on the parity
invariant vertices (below we shall discuss the cases where the parity
invariant vertices provide a complete list of vertices). Thus, we
restrict our attention to the vertex
\be\label{varrep6} p_\smp3^- = p_\smp3^-(\Po^I\Po^I,\, \beta_a\,;\,
\alpha_n^{\sma I}\Po^I,\, \alpha_n^\sma\,;\,
\alpha_{mn}^\smaaplusone,\,\, \alpha_{mn}^{\smaa} )\,, \ee
where%
\be\label{amnabdef} \alpha_{mn}^{\smab}\equiv \alpha_m^{\sma I}
\alpha_n^{\smb I}\,.\ee

\bigskip
{\bf ii}) The second step is to analyze the restrictions imposed by
Eq.\rf{basequ0004}. Using field redefinitions, we can remove the
terms in \rf{varrep6} that  are proportional to $\Po^I\Po^I$. Thus,
we can drop down the dependence on $\Po^I\Po^I$ in $p_\smp3^-$
\rf{varrep6}. {\it The scheme in which the vertex $p_\smp3^-$ is
independent of $\Po^I\Po^I$ is said to be the minimal scheme}.
Obviously, in the minimal scheme, vertex $p_\smp3^-$ \rf{varrep6},
being polynomial in $\Po^I$, satisfies Eq.\rf{basequ0004}
automatically.

Before analyzing the light-cone locality condition, in place of the
variables used in \rf{varrep6},
\be\label{var01}  \beta_a\,, \quad \alpha_n^{\sma I}\Po^I\,,\quad
\alpha_n^\sma\,,\quad \alpha_{mn}^\smaaplusone, \quad
\alpha_{mn}^\smaa\,,  \ee
we introduce the new variables
\be \label{var02} {}~ \ \ \  \beta_a\,,\quad \ \ B_n^\sma, \qquad
\alpha_n^\sma\,,\quad \alpha_{mn}^\smaaplusone\,, \quad
Q_{mn}^\smaa\,, \ \ \ee
where the new `improved' $so(d-2)$ invariants $B_m^\sma$,
$Q_{mn}^\smaa$ are defined by
\beq
\label{var02N1} && B_n^\sma \equiv \frac{\alpha_n^{\sma
I}\Po^I}{\beta_a}+ \frac{\check{\beta}_a}{2\beta_a}\mas_a
\alpha_n^\sma\,,
\\[5pt]
\label{Qmnaadef} && Q_{mn}^\smaa \equiv \alpha_{mn}^\smaa +
\alpha_m^\sma \alpha_n^\sma\,.
\eeq
The use of the variables $Q_{mn}^\smaa$ \rf{var02} instead of
$\alpha_{mn}^\smaa$ \rf{var01} is preferable because the variables
$Q_{mn}^\smaa$ commute with the spin operators of the $so(d-1)$
algebra
\beq \label{MaIJdef} && M^{\sma IJ} = \sum_{n=1}^\nu (\alpha_n^{\sma
I} \bar{\alpha}_n^{\sma J} - \alpha_n^{\sma J}\bar{\alpha}_n^{\sma
I})\,,\qquad
M^{\sma I} = \sum_{n=1}^\nu ( \alpha_n^{\sma I}\bar{\alpha}_n^\sma  -
\alpha_n^\sma\bar{\alpha}_n^{\sma I})\,, \ \ \ \eeq
i.e. $Q_{mn}^\smaa$ are invariants of the $so(d-1)$ algebra. The
advantages of the variables $B_n^\sma$ \rf{var02} over the variables
$\alpha_n^{\sma I} \Po^I$ \rf{var01} are to be explained shortly in
the course of the analysis of the light-cone locality condition.
Thus, we use the `improved' representation for the vertex
\be\label{varrep7} p_\smp3^- = p_\smp3^-(  \beta_a\,,\, B_n^\sma,\,
\alpha_n^\sma ;\, \alpha_{mn}^\smaaplusone,\, Q_{mn}^\smaa)\,. \ee

{\bf iii}) We next analyze the restrictions imposed by the light-cone
locality condition \rf{basequ0003}. For this, we derive the following
formula for action of the operator $\XX^I$ \rf{cubeq06} on vertex
$p_\smp3^-$ \rf{varrep7}:
\be\label{basrel1}
\frac{1}{3\hat{\beta}}\XX^I | p_\smp3^-\rangle =  -{\bf P}^-
\sum_{a=1,2,3\atop n=1,\dots ,\nu} \frac{2\check{\beta}_a}{3\beta_a}
\alpha_n^{\sma I}\partial_{B_n^\sma} |p_\smp3^-\rangle  +
\sum_{a=1,2,3\atop n=1,\dots ,\nu} \frac{1}{\beta_a}\alpha_n^{\sma I}
G_{a n} +  \Po^I E\,,\ee
where we use the notation
\beq
\label{Gandef} G_{a n} & \equiv & \Bigl\{ -\frac{1}{2}(\mas_{a+1}^2-
\mas_{a+2}^2)\partial_{B_n^\sma} + \mas_a
\partial_{\alpha_n^\sma}\Bigr.
\nonumber\\
\Bigl. & + &\sum_{m=1}^\nnu  (B_m^\smaplusone + \frac{1}{2}
\mas_{a+1} \alpha_m^\smaplusone )
\partial_{\alpha_{nm}^\smaaplusone }-(B_m^\smaplustwo -
\frac{1}{2} \mas_{a+2} \alpha_m^\smaplustwo )
\partial_{ \alpha_{mn}^\smaplustwoa} \Bigr\}
|p_\smp3^-\rangle\,,\ \ \ \ \ \
\\
E & \equiv & \frac{1}{3\hat{\beta}} \sum_{a=1}^3 \check{\beta}_a
\beta_a \partial_{\beta_a} |p_\smp3^-\rangle\,.
\eeq
It follows from relation \rf{basrel1} that the light-cone locality
condition \rf{basequ0003} amounts to the equations
\beq
\label{loc1}&& G_{a n} =0\,,\qquad  a=1,2,3; \qquad n=1,\ldots,\nnu;
\\[6pt]
\label{loc2}&& E=0\,. \eeq
We note that in deriving relation \rf{basrel1} we use the fact that
the operator $\XX^I$ does not act on the variables $Q_{mn}^\smaa$
because these variables commute with the spin operators \rf{MaIJdef}.
The use of the variables $B_n^\sma$ is advantageous because
$B_n^\sma$ satisfy the equations
\beq
\label{Bhelequ1} && (\Po^I\partial_{\Po^I} +
\sum_{a=1}^3\beta_a\partial_{\beta_a}) B_n^\smb=0\,,
\\[6pt]
\label{Bhelequ2} && \sum_{a=1}^3 \Bigl\{ \check{\beta}_a \Bigr( \Po^I
\beta_a
\partial_{\beta_a} + \sum_{n=1}^\nnu
\Po^J \alpha_n^{\sma J} \bar{\alpha}_n^{\sma I}\Bigl)-
\frac{3\hat{\beta} \mas_a}{\beta_a} \sum_{n=1}^\nnu  \alpha_n^\sma
\bar{\alpha}_n^{\sma I}\Bigr\} B_m^\smb|0\rangle =0\,.
\eeq
Equations \rf{Bhelequ1} and \rf{Bhelequ2} are very helpful for
solving the $\beta$-homogeneity equation \rf{basequ0002} and deriving
formula \rf{basrel1} respectively.

\bigskip
{\bf iv}) We finally analyze the restrictions imposed by the
$\beta$--homogeneity equation \rf{basequ0002} and Eq.\rf{loc2}. In
terms of vertex $p_\smp3^-$ \rf{varrep7}, Eqs.\rf{basequ0002},
\rf{loc2} become

\be
\label{betdep1} \sum_{a=1}^3 \beta_a
\partial_{\beta_a} p_\smp3^-=0\,,\qquad\quad
\sum_{a=1}^3 \check{\beta}_a\beta_a\partial_{\beta_a} p_\smp3^-=0\,.
\ee
Equations \rf{betdep1} imply that vertex $p_\smp3^-$ \rf{varrep7} is
independent of $\beta_a$, $a=1,2,3$:

\be\label{varrep8} p_\smp3^- = p_\smp3^-(B_n^\sma, \alpha_n^\sma\,,
\alpha_{mn}^\smaaplusone, Q_{mn}^\smaa )\,, \ee
while Eqs.\rf{cubver13}, \rf{basrel1}, \rf{loc1}, \rf{loc2} lead to
the following expression for $|j_\smp3^{-I}\rangle$:

\be
\label{varrep8N1} |j_\smp3^{-I}\rangle = \sum_{a=1,2,3\atop n=1,\dots
,\nu} \frac{2\check{\beta}_a}{3\beta_a} \alpha_n^{\sma
I}\partial_{B_n^\sma} |p_\smp3^-\rangle\,. \ee

{\bf Summary} of analysis in Eqs.\rf{basequ0001}-\rf{basequ0004} in
the minimal scheme is given in formula \rf{varrep8}, with the
equations still to be solved given in \rf{loc1}. Up to this point our
treatment has been applied to vertices for massive as well as
massless fields. From now on, we separately consider vertices for the
massless fields, vertices involving both massless and massive fields,
and vertices for the massive fields. The vertices in Sections
\ref{Solcubintversec}-\ref{secMMM} constitute the complete lists of
parity invariant cubic interaction vertices for massless and massive
fields in $d\geq 4$ dimensions.

\newsection{Parity invariant cubic interaction vertices for massless fields }
\label{Solcubintversec}

We begin with discussing the parity invariant cubic interaction
vertex for the massless mixed-symmetry fields. We consider vertex
\rf{varrep8} for three massless fields:

\be\label{0001} \mas_1 = \mas_2 = \mas_3=0\,.\ee
Equations for the vertex involving three massless fields are
obtainable from Eqs.\rf{loc1} by letting $\mas_a \rightarrow 0 $,
$a=1,2,3$, in Eqs.\rf{loc1} . The general solution for vertex
\rf{varrep8} then takes the form (see Appendix D)

\be\label{0002} p_\smp3^- = p_\smp3^-(B_n^\sma;
\alpha_{mn}^{\smaa};\, Z_{mnq})\,,\ee
where we use the notation

\be \label{0003}
B_n^\sma \equiv \frac{\alpha_n^{\sma I}\Po^I}{\beta_a}\,,\qquad\quad
Z_{mnq}\equiv  B_m^\smone \alpha_{nq}^\smtwothree +  B_n^\smtwo
\alpha_{qm}^\smthreeone + B_q^\smthree \alpha_{mn}^\smonetwo \,,
\ee
and $\alpha_{mn}^{\smab}$ is defined in \rf{amnabdef}. This solution
provides the complete list of parity invariant cubic interaction
vertices for both totally symmetric and mixed-symmetry fields.

We now comment on the solution obtained. Vertex $p_\smp3^-$ \rf{0002}
depends on $B_n^\sma$, $\alpha_{mn}^\smaa$ and $Z_{mnq}$, which are
the respective degree 1, 2, and 3 homogeneous polynomials in
oscillators. Henceforth, degree 1, 2, and 3 homogeneous polynomials
in oscillators are referred to as linear, quadratic, and cubic forms
respectively. We emphasize, however, that the contribution of the
$\alpha_{mn}^\smaa$-terms to the Hamiltonian $P_\smp3^-$ vanishes
when the ket-vector \rf{intver16n2} is subjected to the tracelessness
constraint. This is, the physical massless fields, being irreps of
the $so(d-2)$ algebra, satisfy the tracelessness constraints

\be\label{traceless01} \bar\alpha_m^{\sma I}\bar\alpha_n^{\sma
I}|\phi_a^{\mas_a=0}\rangle = 0\,,\qquad a=1,2,3\,.\ee
It is then clear that the $\alpha_{mn}^\smaa$-terms in \rf{0002} do
not contribute to the Hamiltonian $P_\smp3^-$ \rf{pm1}%
\footnote{ We keep $\alpha_{mn}^\smaa$-terms in the general solution
\rf{0002} because in certain applications it is convenient to deal
with ket-vectors $|\phi^{\mas=0}\rangle$, which are not subjected to
the tracelessness constraint \rf{traceless01}. It is clear that such
ket-vectors describe a collection of massless fields.}.
Thus, in case of massless fields belonging to irreps of the $so(d-2)$
algebra, vertex  \rf{0002} is governed by the linear forms $B_n^\sma$
and by the cubic forms $Z_{mnq}$. To understand the remaining
important properties of solution \rf{0002} we consider cubic vertices
for massless totally symmetric fields.

\subsection{ Cubic interaction vertices for massless totally symmetric
fields}\label{SolcubintversecN1}

In this section we restrict attention to the parity invariant cubic
interaction vertices for massless totally symmetric fields. To
consider the totally symmetric fields it suffices to use one sort of
oscillators, i.e. to set $\nnu = 1$ in \rf{0002}, \rf{0003}. To
simplify formulas we drop oscillator's subscript $n=1$ and use the
simplified notation $\alpha^I\equiv \alpha_1^I$. The cubic
interaction vertex can then be obtained from the general solution
\rf{0002} by using the identifications

\be\label{simnot01} \alpha^{\sma I} \equiv \alpha_1^{\sma I}\,,\qquad
a=1,2,3\,,\ee
in \rf{0002} and ignoring contribution of oscillators carrying a
subscript $n>1$. Adopting the simplified notation \rf{simnot01} for
linear forms $B^\sma \equiv B_1^\sma$ \rf{0003}, quadratic forms
$\alpha^\smab \equiv \alpha_{11}^\smab$ \rf{amnabdef},  and cubic
form $Z\equiv Z_{111}$ \rf{0003}, we obtain

\be\label{0008} B^\sma \equiv \frac{\alpha^{\sma I}
\Po^I}{\beta_a}\,,\qquad \alpha^\smab \equiv  \alpha^{\sma
I}\alpha^{\smb I}\,,\qquad Z\equiv \sum_{a=1}^3 B^\sma
\alpha^\smaplusoneaplustwo \,, \ee
and vertex \rf{0002} becomes

\be \label{sec05nn1} p_\smp3^- = p_\smp3^-(B^\sma;\, \alpha^\smaa;\,
Z )\,.\ee
Vertex \rf{sec05nn1} describes interaction of towers of massless
fields \rf{intver16n10}. We now obtain vertex for massless totally
symmetric  spin $s^\smone$, $s^\smtwo$, $s^\smthree$ fields. The
massless totally symmetric spin $s^\sma$ fields are described by the
respective ket-vectors $|\phi_{s^\sma }^{\mas_a=0}\rangle$. The
ket-vectors of massless fields $|\phi_{s^\sma }^{\mas_a=0}\rangle$,
$a=1,2,3$, are obtainable from \rf{intver16n5} by replacement
$s\rightarrow s^\sma $, $\alpha^I\rightarrow \alpha^{\sma I}$ in
\rf{intver16n5}. Taking into account that the ket-vectors
$|\phi_{s^\sma}^{\mas_a=0}\rangle$ are the respective degree $s^\sma
$ homogeneous polynomials in the oscillators $\alpha^{\sma I}$ (see
\rf{intver16n7}) it is easy to see that the vertex we are interested
in must satisfy the equations

\be \label{sec05nn2} (\alpha^{\sma I}\bar\alpha^{\sma I}  - s^\sma
)|p_\smp3^-\rangle  = 0\,,\qquad a=1,2,3, \ee
which tell us that the vertex $p_\smp3^-$ should be degree $s^\sma $
homogeneous polynomial in the oscillators $\alpha^{\sma I}$. Taking
into account that the forms $B^\sma$ and $Z$ \rf{0008} are the
respective degree 1 and 3 homogeneous polynomials in oscillators, we
find the general solution of Eq.\rf{sec05nn2}

\be\label{0006} p_\smp3^-(s^\smone,s^\smtwo,s^\smthree;k) =
Z^{\frac{1}{2}(\sbf - k)} \prod_{a=1}^3 (B^\sma)^{s^\sma +
\frac{1}{2}(k - \sbf) }\,, \ee
where we use the notation%
\footnote{ We ignore the contribution of $\alpha^\smaa$-terms
\rf{sec05nn1} to vertex \rf{0006}. Because of the tracelessness
constraint (see the second relation in \rf{intver16n8}) the
contribution of these terms to the Hamiltonian $P_\smp3^-$ \rf{pm1}
vanishes.}

\be\label{0007}  \sbf \equiv \sum_{a=1}^3 s^\sma \,, \ee
and integer $k$ is a freedom in our solution. The integer $k$ labels
all possible cubic vertices that can be built for massless spin
$s^\smone$, $s^\smtwo$, $s^\smthree$ fields and has a clear physical
interpretation. Taking into account that the forms $B^\sma$ and $Z$
\rf{0008} are degree 1 homogeneous polynomials in the momentum
$\Po^I$,%
\footnote{ It is this property of the forms $B^\sma$ and $Z$ that
allows us to introduce the vertex that is the homogeneous polynomial
in $\Po^I$. A completely different type of a situation occurs in the
case of massive fields, whose cubic interaction vertices depend on
forms that are non-homogeneous polynomials in the $\Po^I$.}
it is easy to see that vertex \rf{0006} is a degree $k$ homogeneous
polynomial in $\Po^I$. To summarize, the vertex
$p_\smp3^-(s^\smone,s^\smtwo,s^\smthree;k)$ describes interaction of
three massless spin $s^\smone$, $s^\smtwo$, $s^\smthree$ fields
having $k$ powers of the momentum $\Po^I$. In the Lorentz covariant
approach, the integer $k$ is equal to the number of the derivatives
with respect to space-time coordinates.

\medskip
We now discuss the restrictions to be imposed on the spin values
$s^\smone$, $s^\smtwo$, $s^\smthree$ and the integer $k$. The powers
of the forms $B^\sma$ and $Z$ in \rf{0006} must be non--negative
integers. For this to be the case, it is necessary to impose the
following restrictions on the allowed spin values $s^\smone$,
$s^\smtwo$, $s^\smthree$ and the number of powers of the momentum
$\Po^I$ (the number of the derivatives):

\beq
\label{0009} &&  k\leq \sbf \,,\qquad\quad \sbf - k \leq 2 s^\sma
\,, \qquad a=1,2,3\,,
\\[9pt]
\label{00011} && \sbf - k \qquad\qquad \qquad   \hbox{ even
integer}\,. \eeq
Restrictions \rf{0009} can be rewritten equivalently as

\be\label{00012} \sbf - 2s_{min} \leq k \leq \sbf \,, \quad \qquad
s_{min}\equiv \min_{a=1,2,3} s^\sma\,. \ee

\bigskip

A few remarks are in order.

\bigskip

{\bf i}) The restriction $k\leq \sbf$ in \rf{00012} expresses the
fact that in the minimal scheme, which does not admit
$\Po^I\Po^I$-terms, the total number of the transverse indices of
fields that enter the cubic Hamiltonian $P_\smp3^-$ cannot be less
than the number of powers of the momentum $\Po^I$ in the vertex.

\bigskip

{\bf ii}) If $k=2$, then the restriction $\sbf - 2s_{min} \leq 2$ in
\rf{00012} is precisely the restriction that leaves no place for the
gravitational interaction of massless higher spin fields ($s>2$).
Indeed, the gravitational interaction of a massless spin $s$ field
could be described by the vertex
$p_\smp3^-(s^\smone,s^\smtwo,s^\smthree;k)$ with
$s^\smone=s^\smtwo=s>2$, $s^\smthree=2$, $k=2$. For these values
$s^\sma$ we obtain $s_{min} = 2$, $\sbf = 2s+2$ and therefore
restrictions \rf{00012} take the form

\be 2s  - 2 \leq k \leq 2s+2 \,. \ee
On the one hand, these restrictions tell us that for $s>2$, the
gravitational interaction, i.e. the case $k=2$, is not allowed. On
the other hand, we see that all allowed cubic interactions vertices
for graviton and higher spin $s>2$ fields involve higher derivatives,
$k>2$.

\bigskip

{\bf iii}) Restrictions \rf{00011}, \rf{00012} lead to a surprisingly
simple result for the number of allowed parity invariant cubic
interaction vertices $p_\smp3^-(s^\smone,s^\smtwo,s^\smthree;k)$.
Indeed, we see from \rf{00011} and \rf{00012} that for spin values
$s^\smone$, $s^\smtwo$, $s^\smthree$, the integer $k$ takes the
values

\be\label{kval01} k = \sbf,\, \sbf - 2,\, \sbf - 4,\, \ldots\, , \sbf
- 2s_{min}\,,\qquad \hbox{ for } \ \ d>4\,. \ee
This implies that given spin values $s^\smone$, $s^\smtwo$,
$s^\smthree$, the number of parity invariant cubic interaction
vertices $p_\smp3^-(s^\smone,s^\smtwo,s^\smthree;k)$ that can be
constructed is given by

\be\label{number01} \Nsf (s^\smone,s^\smtwo,s^\smthree) = s_{min}  +
1\,, \qquad \hbox{ for } \ \ d>4 \,. \ee

\bigskip

{\bf iv}) Vertices \rf{0006}, with $k$ in \rf{kval01}, constitute the
complete list of vertices for $d>4$. For $d=4$, the number of allowed
vertices is decreased. This is, if $d=4$ then for spin values
$s^\smone$, $s^\smtwo$, $s^\smthree$, the integer $k$ takes the values%
\footnote{ For $d=4$, the vertices \rf{0006} with $\sbf > k > \sbf -2
s_{min}$ (see \rf{kval01}) are proportional to $\Po^2$ (and can be
removed by field redefinitions) or to $\alpha_a^I\alpha_a^I$ (and do
not contribute to the Hamiltonian in view of tracelessness constraint
\rf{intver16n8}). This can be demonstrated straightforwardly using
helicity formalism in Ref.\cite{Bengtsson:1983pd}.}

\be \label{kval02} k = \sbf,\,\, \sbf - 2s_{min}\,,\qquad \hbox{ for
} \ \ d=4. \ee
This implies that for spin values $s^\smone$, $s^\smtwo$,
$s^\smthree$, the number of parity invariant cubic vertices that can
be built for massless fields in four dimensions is equal one if
$s_{min}=0$ and two if $s_{min}\ne 0$%
\footnote{ Values of $k$ \rf{kval02} explain the vanishing of the
vertices $p_\smp3^-(2,2,2;4)$ (see Table I) and $p_\smp3^-(3,3,3;5)$
(see table II) in $d=4$. The vanishing of the covariant counterpart
of our light-cone vertex $p_\smp3^-(3,3,3;5)$ in $d=4$ was discussed
in Ref.\cite{Bekaert:2005jf}.}.

\medskip

{\bf v}) We comment on the relation of our vertices to the vertices
for higher spin fields in $AdS$ space
\cite{Fradkin:1987ks}-\cite{Vasiliev:2003ev}. Clearly, direct
comparison of our vertices with AdS vertices is not possible because
of two reasons: 1) Our vertices are defined for fields in {\it flat
space}; 2) Cubic vertices in $AdS$ are given in terms of some
explicit {\it generating function}, but vertices for three fields
with arbitrary (but fixed) spin values are still not available in the
literature. Nevertheless, it seems likely that: 1) All our vertices
\rf{0006},\rf{kval01} (and \rf{kval02} for $d=4$) allow a smooth
extension to $AdS$ space and these vertices, being supplemented by
appropriate cosmological constant dependent corrections, coincide
with some $AdS$ vertices; 2) The remaining AdS vertices are singular
in the flat space limit and do not have flat space counterparts.

Formula \rf{0006} not only provides a surprisingly simple form for
the vertices of massless higher spin fields but also gives a simple
form for the vertices of the well-studied massless low spin $s=0,1,2$
fields. By way of example, we consider cubic vertices that describe
the self-interaction of spin $s$ field having $k=s$ powers of
$\Po^I$. In the literature, such cubic vertices are referred to as
Yang-Mills like interaction vertices%
\footnote{ Such vertices for spin $s>2$ fields in $4d$ flat space
were built by using light-cone approach in \cite{Bengtsson:1983pd}.
Our formula \rf{YMHSver} gives alternative simple expression for
these vertices in $d=4$ and provides an extension to arbitarry $d>4$
dimensions on an equal footing.}.
We consider vertices with $s^\smone = s^\smtwo = s^\smthree = s$ and
$k=s$ and formula \rf{0006} leads to
\be \label{YMHSver} p_\smp3^-(s,s,s;s)=Z^s\,.\ee
For spin $s=1$ and $s=2$ fields, these vertices describe the
respective cubic interaction vertices of Yang-Mills and Einstein
theories,
\beq
\label{YMcovL} && p_\smp3^-(1,1,1;1)=Z \sim
(F_{\mu\nu}F^{\mu\nu})_\smp3,
\\
\label{EcovL} && p_\smp3^-(2,2,2;2)=Z^2 \sim (\sqrt{g}R)_\smp3\,,\eeq
where the subscript $[3]$ of Yang-Mills and Einstein covariant
Lagrangians is used to indicate the cubic vertices of these theories.
Taking into account the complicated structure of the cubic vertices
of Yang-Mills and Einstein theories in covariant approaches, we see
that the light-cone gauge approach gives a simpler representation for
such vertices. Another attractive property of the light-cone approach
is that it allows treating the interaction vertices of Yang-Mills and
Einstein theories on an equal footing.

Formula \rf{0006} provides a convenient representation for other
well-known parity invariant cubic interaction vertices of massless
low spin fields. These vertices and their Lorentz covariant
counterparts are collected in Table I. In Table II, we present
light-cone vertices \rf{0006} involving higher spin fields whose
Lorentz covariant descriptions are available in the literature.

\noindent{\sf Table I. Parity invariant cubic vertices for massless
low spin $s=0,1,2$ fields. \small In the 3rd column, $\phi$ stands
for the scalar field, $F_{\mu\nu}$ and $D_\mu$ stand for the
respective Yang-Mills field strength and covariant derivative $D_\mu
=
\partial_\mu +A_\mu$, and $R_{\mu\nu\rho\sigma}$ stands
for the Riemann tensor}%
\footnote{ $\omega_\mu{}^{\nu\rho}$ and $R_{\mu\nu\rho\sigma}$ in
covariant vertices (1,2,2;3), (1,2,2;5) stand for the linearized
Lorentz connection, $\omega_\mu{}^{\nu\rho}= -
\omega_\mu{}^{\rho\nu}$, and the Riemann tensor of the {\it charged}
(w.r.t. the Yang-Mills gauge group) spin 2 field. These covariant
vertices and the vertex (0,1,2;3) are invariant under linearized
on-shell gauge transformations.}
{\small
\begin{center}
\begin{tabular}{|l|c|c|}
\hline        & &
\\ [-3mm] \  Spin values and   & \ \ \ Light-cone  \ \ \  & Covariant
\\
  number of derivatives  & vertex   & Lagrangian
\\
 \ \ \ $ s^\smone , s^\smtwo , s^\smthree ;\, k $
 &  $p_\smp3^-(s^\smone,s^\smtwo,s^\smthree;k)$ &
\\ [2mm]\hline
&&
\\[-3mm]
\ \ \ \ \ \ $ 0,0,0;\, 0$   & $ 1 $  & $ \phi^3  $
\\[2mm]\hline
&&
\\[-3mm]
\ \ \ \ \ \ $ 0,0,1;\,1 $   & $ B^\smthree $  & $( D_\mu\phi D^\mu
\phi)_\smp3$
\\[2mm]\hline
&&
\\[-3mm]
\ \ \ \ \ \ $ 0,0,2;\,2$   & $  (B^\smthree)^2 $  &
$(\sqrt{g}g^{\mu\nu}
\partial_\mu\phi\partial_\nu\phi)_\smp3$
\\[2mm]\hline
&&
\\[-3mm]
\ \ \ \ \ \ $0,1,1;\, 2$   & $ B^\smtwo B^\smthree  $  & $(\phi
F_{\mu\nu} F^{\mu\nu})_\smp3$
\\[2mm]\hline
&&
\\[-3mm]
\ \ \ \ \ \ $ 0,1,2;\,3$   & $ B^\smtwo (B^\smthree)^2 $  &
$(\partial^\mu \phi F_{\nu\rho}\omega_\mu{}^{\nu\rho})_\smp3$
\\[2mm]\hline
&&
\\[-3mm]
\ \ \ \ \ \ $ 0,2,2;\,4$   & $  (B^\smtwo B^\smthree)^2 $  &
$(\sqrt{g} \phi
 R_{\mu\nu\rho\sigma} R^{\mu\nu\rho\sigma})_\smp3$
\\[2mm]\hline
&&
\\[-3mm]
\ \ \ \ \ \ $ 1,1,1;\,1$   & $ Z $  & $(F_{\mu\nu} F^{\mu\nu})_\smp3$
\\[2mm]\hline
&&
\\[-3mm]
\ \ \ \ \ \ $ 1,1,1;\,3$   & $B^\smone B^\smtwo B^\smthree $  &
$(F_{\mu\nu} F^{\nu\rho} F_\rho{}^\mu)_\smp3$
\\[2mm]\hline
&&
\\[-3mm]
\ \ \ \ \ \ $ 1,1,2;\,2 $   & $B^\smthree Z$   &
$(\sqrt{g}g^{\mu\rho}g^{\nu\sigma}F_{\mu\nu}F_{\rho\sigma})_\smp3$
\\[2mm]\hline
&&
\\[-3mm]
\ \ \ \ \ \ $ 1,1,2;\,4 $   & $ B^\smone B^\smtwo (B^\smthree)^2$  &
$(\sqrt{g}F_{\mu\nu}F_{\rho\sigma}R^{\mu\nu\rho\sigma})_\smp3$
\\[2mm]\hline
&&
\\[-3mm]
\ \ \ \ \ \ $ 1,2,2;\,3 $   & $ B^\smtwo B^\smthree Z $  &
$F_{\mu\nu} (\omega^{\mu,\rho\sigma} \omega^\nu{}_{\rho\sigma} -
\omega^{\rho,\sigma\mu} \omega_{\rho,\sigma}{}^\nu)_\smp3$
\\[2mm]\hline
&&
\\[-3mm]
\ \ \ \ \ \ $ 1,2,2;\,5$   & $ B^\smone (B^\smtwo)^2 (B^\smthree)^2$
& $(F^{\mu\nu}
R_\mu{}^{\rho\sigma\lambda}R_{\nu\rho\sigma\lambda})_\smp3$
\\[2mm]\hline
&&
\\[-3mm]
\ \ \ \ \ \ $ 2,2,2;\,2$   & $ Z^2 $  & $(\sqrt{g}R)_\smp3$
\\[2mm]\hline
&&
\\[-3mm]
\ \ \ \ \ \ $ 2,2,2;\,4$   & $ B^\smone B^\smtwo B^\smthree Z $  &
$(\sqrt{g}R_{\mu\nu\rho\sigma} R^{\mu\nu\rho\sigma})_\smp3$
\\[2mm]\hline
&&
\\[-3mm]
\ \ \ \ \ \ $ 2,2,2;\,6$   & $(B^\smone B^\smtwo B^\smthree)^2$  &
$(\sqrt{g}R^{\mu\nu}_{\rho\sigma} R^{\rho\sigma}_{\lambda\tau}
R^{\lambda\tau}_{\mu\nu})_\smp3$
\\[2mm]\hline
\end{tabular}
\end{center}
}

\medskip

We note that vertices with $k=\sbf$ correspond to gauge theory cubic
interaction vertices built entirely in terms of gauge field
strengths%
\footnote{Our result for the vertex $p_\smp3^-(2,2,2;6)$ implies that
there is only one Lorentz covariant $R_{....}^3$ vertex ($R_{....}$
is the Riemann tensor) that gives a non-trivial contribution to the
3-point scattering amplitude. In Ref.\cite{Metsaev:1986yb}, it was
demonstrated that this is indeed the case.}.
The vertices with $k<\sbf$ cannot be built entirely in terms of gauge
field strengths. It is the vertices with $k<\sbf$ that are difficult
to construct in Lorentz covariant approaches. The light-cone approach
treats all vertices on an equal footing.

We finish with a discussion of the completeness of solution
\rf{0006}. Our solution \rf{0006} provides the complete list of
parity invariant cubic vertices for the massless totally symmetric
fields in $d \geq 4$ dimensions. In $d>6$ dimensions,
$so(d-2)$-invariants constructed out of the antisymmetric Levi-Civita
symbol $\epsilon^{I_1\ldots I_{d-2}}$, the oscillators $\alpha^{\sma
I}$, and the momentum $\Po^I$ are equal to zero, and therefore there
are no parity violating cubic vertices for the massless totally
symmetric fields. For $d=4,5,6$ there are nontrivial $so(d-2)$
invariants constructed out of the antisymmetric Levi-Civita symbol.
It turns out that for $d=5$, these invariants allow building parity
violating cubic vertices for the massless totally symmetric fields%
\footnote{ Complete list of cubic vertices for massless fields in
$4d$ was obtained in \cite{Bengtsson:1986kh} and we do not discuss
this case.},
while for $d=6$, solution \rf{0006} provides the complete list of
cubic vertices for the massless totally symmetric fields (i.e. there
are no parity violating cubic vertices for the massless totally
symmetric fields in $d=6$). The complete list of cubic vertices for
massless fields in $d=5,6$ is to be obtained in the respective
Sections
\ref{5dtheor} and \ref{6dtheor}%
\footnote{ Complete list of cubic vertices for massless fields in
$d=5,6$ was announced in Refs.\cite{Metsaev:1993gx,Metsaev:1993mj}.
In Refs.\cite{Metsaev:1993gx,Metsaev:1993mj} we mistakenly thought
that our solution \rf{0006} provides the complete list of cubic
vertices for massless fields in $d=5$.}.

\medskip
\noindent{\sf Table II. Parity invariant cubic interaction vertices
for massless higher spin fields.}

{\small
\begin{center}
\begin{tabular}{|l|c|c|}
\hline        & &
\\ [-3mm]\ Spin values and   & \ \ \ Light-cone  \ \ \  & Covariant
\\
number of derivatives  & vertex   & Lagrangian
\\
\ \ \ $ s^\smone,s^\smtwo,s^\smthree;\,k$ &
$p_\smp3^-(s^\smone,s^\smtwo,s^\smthree;k)$    &
\\[1mm] \hline
&&
\\[-3mm]
\ \ \ \ \ \ $ 2,2,4;\,4 $   & $ (B^\smthree)^2 Z^2 $  &
$\LL(\hbox{see Ref.\cite{Deser:1990bk}})$
\\[2mm]\hline
&&
\\[-3mm]
\ \ \ \ \ \ $ 3,3,3;\,3 $   & $Z^3 $  & $\LL (\hbox{see
Ref}.\cite{Berends:1984wp}$
\\[2mm]\hline
&&
\\[-3mm]
\ \ \ \ \ \ $  3,3,3;\,5 $   & $B^\smone B^\smtwo B^\smthree Z^2 $ &
$\LL(\hbox{see Ref}.\cite{Bekaert:2005jf})$
\\[2mm]\hline
&&
\\[-3mm]
\ \ \ \ \ \ $ s,s,s';\,k' $   & $(B^\smone B^\smtwo)^{s-s_{min}}
(B^\smthree)^{s'-s_{min}} Z^{s_{min}} $ & $\LL(\hbox{see
Refs.\cite{Berends:1984rq,Berends:1985xx}})$
\\[1mm]
&&
\\[-3mm]
$ k'= 2s+s' -2s_{min} $   &   &
\\
&&
\\[-3mm]
$ s_{min} \equiv \min (s,s') $   &   &
\\[2mm]\hline
\end{tabular}
\end{center}
}

\newsection{Parity invariant cubic interaction vertices for massless and massive
fields}\label{secMMO}

We now study cubic interaction vertices for massless and massive
field. We consider cubic vertices for one massive field and two
massless fields and cubic vertices for one massless field and two
massive fields. In other words, we consider vertices for fields with
the following mass values:
\beq
&&  \mas_1 = \mas_2 = 0,\qquad \mas_3 \ne  0\,;
\\
&& \mas_1 = \mas_2 \equiv \mas   \ne 0 ,\qquad \mas_3 = 0\,;
\\
&& \mas_1 \ne 0 ,\qquad \mas_2\ne  0,\qquad  \mas_1 \ne  \mas_2,
\qquad \mas_3= 0\,.
\eeq
We study these cases in turn.

\subsection{ Cubic interaction vertices for two massless and one massive
fields}

We start with the cubic interaction vertex \rf{varrep8} for three
fields with the mass values
\be\label{00m1} \mas_1 = \mas_2 = 0,\qquad \mas_3 \ne  0\,, \ee
i.e. the {\it massless} fields carry external line indices $a=1,2$,
while the {\it massive} field corresponds to $a=3$. Equations for the
vertex involving two massless fields can be obtained from
Eqs.\rf{loc1} in the limit as $\mas_1 \rightarrow 0 $, $\mas_2
\rightarrow 0 $.
The general solution for vertex \rf{varrep8} is then found to be (see
Appendix D)
\be\label{00m2} p_\smp3^- = p_\smp3^-( B_n^\smthree;\,
Q_{mn}^\smaaplusone,\, \alpha_{mn}^\smoneone,\,
\alpha_{mn}^\smtwotwo,\, Q_{mn}^\smthreethree)\,, \ee
where we use the notation
\footnote{ We recall that the short notation like
$p_\smp3^-(Q^\smaaplusone)$ is used to indicate a dependence of
$p_\smp3^-$ on $Q^\smonetwo$, $Q^\smtwothree$, $Q^\smthreeone$.}
\beq
&& \label{00m6} B_n^\sma \equiv \frac{\alpha_n^{\sma I}
\Po^I}{\beta_a}\,,\quad a=1,2;\qquad \ \ \
B_n^\smthree\equiv \frac{\alpha_n^{\smthree I}\Po^I}{\beta_3}+
\frac{\check{\beta}_3}{2\beta_3} \mas_3 \alpha_n^\smthree\,,\eeq
\beq
\label{00m3} && Q_{mn}^\smonetwo \equiv \alpha_{mn}^\smonetwo -
\frac{2}{\mas_3^2} B_m^\smone  B_n^\smtwo \,,
\\
\label{00m4}&& Q_{mn}^\smtwothree \equiv \alpha_{mn}^\smtwothree
+\frac{\alpha_n^\smthree}{\mas_3} B_m^\smtwo + \frac{2}{\mas_3^2}
B_m^\smtwo B_n^\smthree\,,
\\
\label{00m5}&& Q_{mn}^\smthreeone \equiv \alpha_{mn}^\smthreeone -
\frac{\alpha_m^\smthree}{\mas_3} B_n^\smone + \frac{2}{\mas_3^2}
B_m^\smthree B_n^\smone\,, \eeq
and $\alpha_{mn}^\smab$, $Q_{mn}^\smaa$ are defined in \rf{amnabdef},
\rf{Qmnaadef}. This solution describes cubic interaction vertices for
both totally symmetric and mixed-symmetry fields.

We note that all forms in \rf{00m2} that depend on $\Po^I$ (the
linear forms $B_m^\smthree$ and the quadratic forms
$Q_{mn}^\smaaplusone$) are non-homogeneous polynomials in $\Po^I$.
Therefore, as seen from \rf{00m2}-\rf{00m5}, there is no possibility
to construct a cubic vertex that would be a homogeneous polynomial in
$\Po^I$. In other words, the dependence on the linear forms
$B_m^\smthree$ and the quadratic forms $Q_{mn}^\smaaplusone$ leads to
the cubic vertices that are non-homogeneous polynomials in $\Po^I$.
The appearance of massive field interaction vertices involving
different powers of derivatives is a well-known fact (see e.g.
\cite{GR,Ferrara:1992yc}). Thus, we see that the light-cone formalism
gives a very simple explanation to this phenomenon by means of the
linear forms $B_m^\smthree$ and the quadratic forms
$Q_{mn}^\smaaplusone$. To understand the remaining characteristic
properties of solution \rf{00m2}, we consider vertices for the
totally symmetric fields.

\subsubsection{ Cubic interaction vertices for totally symmetric fields}

In this section, we restrict ourselves to cubic interaction vertices
for two massless totally symmetric  fields and one massive totally
symmetric field. To consider the totally symmetric fields, it is
sufficient to use one sort of oscillators and we set $\nnu = 1$ in
\rf{00m2}-\rf{00m5}. To simplify the formulas we drop the
oscillator's subscript $n=1$ and use the simplified notation for
oscillators: $\alpha^I \equiv \alpha_1^I$, $\alpha \equiv \alpha_1$.
The cubic interaction vertex for totally symmetric fields under
consideration can then be obtained from the general solution
\rf{00m2}  by making the identifications
\beq
\label{00m7} && \alpha^{\sma I} \equiv \alpha_1^{\sma I}, \quad  \ a
=1,2\,;  \qquad \alpha^{\smthree I} \equiv  \alpha_1^{\smthree I}
\qquad \alpha^\smthree \equiv \alpha_1^\smthree\,, \eeq
in \rf{00m2}-\rf{00m5} and ignoring the contribution of oscillators
carrying a subscript $n>1$. Adopting simplified notation \rf{00m7}
for forms \rf{00m6}-\rf{00m5}:
\be \label{00m9}  B^\sma \equiv B_1^\sma\,, \qquad Q^\smab \equiv
Q_{11}^\smab\,, \qquad \alpha^\smab \equiv \alpha_{11}^\smab\,, \ee
we see that vertex \rf{00m2} takes the form
\be \label{00m10} p_\smp3^- = p_\smp3^-(B^\smthree;\,
Q^\smaaplusone,\, \alpha^\smoneone,\, \alpha^\smtwotwo,\,
Q^\smthreethree)\,.\ee
Vertex \rf{00m10} describes the interaction of the towers of massive
and massless fields \rf{intver16n9}, \rf{intver16n10}. We now obtain
vertex for two massless totally symmetric spin $s^\smone$, $s^\smtwo$
fields and one massive totally symmetric spin $s^\smthree$ field. The
massless totally symmetric spin $s^\smone$ and $s^\smtwo$ fields are
described by respective ket-vectors $|\phi_{s^\smone
}^{\mas_1=0}\rangle$ and $|\phi_{s^\smtwo }^{\mas_2=0}\rangle$, while
the massive totally symmetric spin $s^\smthree$ field is described by
a ket-vector $|\phi_{s^\smthree }\rangle$. The ket-vectors of
massless fields $|\phi_{s^\sma }^{\mas_a=0}\rangle$, $a=1,2$, can be
obtained from \rf{intver16n5} by the replacement $s\rightarrow s^\sma
$, $\alpha^I\rightarrow \alpha^{\sma I}$, $a=1,2$, in
\rf{intver16n5}, while the ket-vector of the massive field
$|\phi_{s^\smthree }\rangle$ can be obtained from \rf{intver16n4} by
the replacement $s\rightarrow s^\smthree $, $\alpha^I\rightarrow
\alpha^{\smthree I}$, $\alpha\rightarrow \alpha^\smthree$ in
\rf{intver16n4}. Taking into account that the ket-vectors
$|\phi_{s^\sma }^{\mas_a=0}\rangle$, $a=1,2$, are the respective
degree $s^\sma $ homogeneous polynomials in the oscillators
$\alpha^{\sma I}$ (see \rf{intver16n7}), while the ket-vector
$|\phi_{s^\smthree }\rangle$ is a degree $s^\smthree$ homogeneous
polynomial in the oscillators $\alpha^{\smthree I}$,
$\alpha^\smthree$ (see \rf{intver16n6}) it is easy to understand that
the vertex we are interested in must satisfy the equations
\beq
&& \label{00m11}  (\alpha^{\sma I}\bar\alpha^{\sma I}  - s^\sma )
|p_\smp3^-\rangle = 0\,,\qquad a=1,2,
\\[3pt]
&& \label{00m12}  (\alpha^{\smthree I} \bar\alpha^{\smthree I}  +
\alpha^\smthree\bar\alpha^\smthree - s^\smthree ) |p_\smp3^-\rangle =
0\,. \eeq
These equations tell us that the vertex must be a degree $s^\sma $
homogeneous polynomial in the respective oscillators. Taking into
account that the forms $B^\smthree$ and $Q^\smaaplusone$ are the
respective degree 1 and 2 homogeneous polynomials in the oscillators
we find the general solution of Eqs.\rf{00m11}, \rf{00m12} as%
\footnote{ We ignore the contribution of $\alpha^\smoneone$-,
$\alpha^\smtwotwo$-, $Q^\smthreethree$-terms of \rf{00m10} to vertex
\rf{intver30}. Because of the tracelessness constraints
\rf{intver16n8} the contribution of these terms to the Hamiltonian
$P_\smp3^-$ \rf{pm1} vanishes.}
%
%
\be \label{intver30} p_\smp3^-(s^\smone,s^\smtwo,s^\smthree;x)  =
(B^\smthree)^x \prod_{a=1}^3 (Q^\smaaplusone)^{y^\smaplustwo } \,,
\ee
where integers $y^\sma $ are expressible in terms of the $s^\sma $
and an integer $x$ by the relations
\beq
\label{yexp01}&& y^\sma  = \frac{\sbf - x}{2} -s^\sma \,,\qquad
a=1,2\,,
\\
\label{yexp03} && y^\smthree  = \frac{\sbf + x}{2} -s^\smthree \,.
\eeq
The integer $x$ expresses the freedom of the solution and labels all
possible cubic interaction vertices that can be constructed for the
fields under consideration. For vertex \rf{intver30} to be sensible,
we impose the restrictions
\beq
\label{restr01}&&  x\geq 0\,,\qquad y^\sma  \geq 0\,, \quad a=1,2,3;
\\
\label{restr03} && \sbf - x \qquad \hbox{ even integer}\,,\eeq
which amount to the requirement that the powers of all forms in
\rf{intver30} be non--negative integers. We note that using relations
\rf{yexp01}, \rf{yexp03} allows rewriting the restrictions
\rf{restr01} as%
\footnote{ If $x=0$, then restrictions \rf{restr03NNN1} become the
restrictions well known in the angular momentum theory: $ |s^\smone -
s^\smtwo| \leq s^\smthree \leq s^\smone + s^\smtwo$.}
\be \label{restr03NNN1} \max(0, s^\smthree - s^\smone - s^\smtwo)
\leq x \leq s^\smthree - |s^\smone - s^\smtwo|\,. \ee
Compared to the vertex for three massless fields \rf{0006}, vertex
\rf{intver30} is a non-homogeneous polynomial in $\Po^I$. An
interesting property of vertex \rf{intver30} is that the maximal
number of powers of the momentum $\Po^I$, denoted by $k_{max}$, is
independent of $x$ and is determined only by $\sbf$,%
\footnote{ Expressions for $B^\smthree$ and $Q^\smaaplusone$
\rf{00m6}-\rf{00m5} imply that $k_{max} = x + 2\sum_{a=1}^3 y^\sma$.
Taking expressions for $y^\sma $ \rf{yexp01}, \rf{yexp03} into
account we find \rf{kmaxN1}.}
\be \label{kmaxN1} k_{max}= \sbf\,.\ee

\subsection{ Cubic interaction vertices for one massless and two massive fields
with the same mass values}\label{equalmasses}

The case under consideration is most interesting because it involves
the minimal Yang-Mills and gravitational interactions of massive
arbitrary spin fields as particular cases. We now consider the cubic
interaction vertex \rf{varrep8} for one massless field and two
massive fields with the same mass values,
\be\label{eqmas00007NN1} \mas_1 = \mas_2 \equiv \mas  \ne 0 ,\qquad
\mas_3 = 0\,, \ee
i.e. the {\it massive} fields carry external line indices $a=1,2$,
while the {\it massless} field corresponds to $a=3$. The analysis of
equations for the vertex is straightforward and the general solution
is found to be (see Appendix D)
\be \label{intvereqmas01} p_\smp3^- = p_\smp3^-(L_n^\smone,
L_n^\smtwo, B_n^\smthree;\, Q_{mn}^\smonetwo, Q_{mn}^\smoneone,
Q_{mn}^\smtwotwo, \alpha_{mn}^\smthreethree\,;\, Z_{mnq})\,, \ee
where we use the notation
\beq
\label{eqmas00007} &&  L_n^\smone \equiv B_n^\smone + \frac{1}{2}\mas
\alpha_n^\smone\,,\qquad\quad
%
%
L_n^\smtwo \equiv B_n^\smtwo - \frac{1}{2}\mas \alpha_n^\smtwo\,,
\\
\label{eqmas00009} && B_n^\sma \equiv \frac{\alpha_n^{\sma
I}\Po^I}{\beta_a}+ \frac{\check{\beta}_a}{2\beta_a} \mas
\alpha_n^\sma\,,\qquad a=1,2;
\\
\label{eqmas00010} && B_n^\smthree\equiv \frac{\alpha_n^{\smthree
I}\Po^I}{\beta_3}\,,
\\
\label{eqmas00006} &&  Q_{mn}^\smonetwo \equiv \alpha_{mn}^\smonetwo
+ \frac{\alpha_n^\smtwo}{\mas} B_m^\smone -
\frac{\alpha_m^\smone}{\mas} B_n^\smtwo\,,
\\[5pt]
\label{eqmas00005} && Z_{mnq} \equiv L_m^\smone
\alpha_{nq}^\smtwothree + L_n^\smtwo \alpha_{qm}^\smthreeone +
B_q^\smthree ( \alpha_{mn}^\smonetwo - \alpha_m^\smone\alpha_n^\smtwo
)\,,
\eeq
and $\alpha_{mn}^\smab$, $Q_{mn}^\smaa$ are defined in \rf{amnabdef},
\rf{Qmnaadef}. Thus, we see that vertex \rf{intvereqmas01} depends,
among other things, on linear forms $B_n^\smthree$ \rf{eqmas00010},
which are degree 1 {\it homogeneous} polynomials in the momentum
$\Po^I$. This implies that cubic interaction vertices that are
homogeneous polynomials in $\Po^I$ can be constructed for certain
fields. This also implies that the minimal number of powers of
$\Po^I$ in \rf{intvereqmas01} is not equal to zero in general (for
example, the dependence on $B_m^\smthree$ leads to an increasing
number of powers of the momentum $\Po^I$). All the remaining forms
that depend on the momentum $\Po^I$ and enter the cubic vertex (the
linear forms $L_n^\smone$, $L_n^\smtwo$ and the quadratic forms
$Q_{mn}^\smonetwo$) are non-homogeneous polynomials in $\Po^I$. To
discuss the remaining important properties of solution
\rf{intvereqmas01} we restrict attention to cubic vertices for the
totally symmetric fields.

\subsubsection{Cubic interaction vertices for totally symmetric
fields}\label{tsymMM0}

In this section, we restrict ourselves to cubic interaction vertices
for the totally symmetric fields with mass values given in
\rf{eqmas00007NN1}. As usual, we use one sort of oscillators, i.e. we
set $\nnu = 1$ in \rf{intvereqmas01}-\rf{eqmas00005} and simplify
formulas by dropping the oscillator's subscript $n=1$: $\alpha^I
\equiv \alpha_1^I$, $\alpha \equiv \alpha_1$. The cubic interaction
vertex for the totally symmetric fields under consideration can be
obtained from the general solution \rf{intvereqmas01} by making the
identifications
\be
\label{mm014nn} \alpha^{\sma I} \equiv \alpha_1^{\sma I}, \qquad
\alpha^\sma \equiv \alpha_1^\sma\,,  \quad a =1,2; \qquad
\alpha^{\smthree I} \equiv  \alpha_1^{\smthree I} \,, \ee
in \rf{intvereqmas01}-\rf{eqmas00005} and ignoring the contribution
of oscillators carrying a subscript $n>1$. Adopting the simplified
notation \rf{mm014nn} for forms \rf{eqmas00007}-\rf{eqmas00005},
\be \label{mm016nn}  L^\sma \equiv L_1^\sma\,, \qquad B^\sma \equiv
B_1^\sma\,, \qquad \alpha^\smab \equiv \alpha_{11}^\smab\,,\qquad
Q^\smab \equiv Q_{11}^\smab\,, \qquad Z \equiv Z_{111}\,, \ee
we see that vertex \rf{intvereqmas01} takes the form
\be \label{intvereqmas01N1} p_\smp3^- = p_\smp3^-(L^\smone, L^\smtwo,
B^\smthree;\, Q^\smonetwo, Q^\smoneone, Q^\smtwotwo,
\alpha^\smthreethree\,;\, Z)\,. \ee
Vertex \rf{intvereqmas01N1} describes the interaction of the towers
of massive and massless fields \rf{intver16n9}, \rf{intver16n10}. We
next obtain the vertex for two massive totally symmetric spin
$s^\smone$, $s^\smtwo$ fields and one massless totally symmetric spin
$s^\smthree$ field. Two massive totally symmetric spin $s^\smone$ and
$s^\smtwo$ fields are described by the respective ket-vectors
$|\phi_{s^\smone }\rangle$ and $|\phi_{s^\smtwo }\rangle$, while one
massless totally symmetric spin $s^\smthree$ field is described by a
ket-vector $|\phi_{s^\smthree }^{\mas_3=0}\rangle$. The ket-vectors
of massive fields $|\phi_{s^\sma }\rangle$, $a=1,2$, can be obtained
from \rf{intver16n4} by the replacement $s\rightarrow s^\sma $,
$\alpha^I\rightarrow \alpha^{\sma I}$, $\alpha\rightarrow
\alpha^\sma$, $a=1,2$, in \rf{intver16n4}, while the ket-vector of
massless field $|\phi_{s^\smthree }^{\mas_3=0}\rangle$ can be
obtained from \rf{intver16n5} by the replacement $s\rightarrow
s^\smthree $, $\alpha^I\rightarrow \alpha^{\smthree I}$ in
\rf{intver16n5}. Taking into account that the ket-vectors
$|\phi_{s^\sma }\rangle$, $a=1,2$, are the respective degree $s^\sma
$ homogeneous polynomials in the oscillators $\alpha^{\sma I}$,
$\alpha^\sma $ (see \rf{intver16n6}), while the ket-vector
$|\phi_{s^\smthree }^{\mas_3=0}\rangle$ is a degree $s^\smthree$
homogeneous polynomial in the oscillator $\alpha^{\smthree I}$ (see
\rf{intver16n7}) it is easy to understand that the vertex we are
interested in must satisfy the equations
\beq
&& \label{mm018}  (\alpha^{\sma I}\bar\alpha^{\sma I} +\alpha^\sma
 \bar\alpha^\sma  - s^\sma ) |p_\smp3^-\rangle = 0\,,\qquad a=1,2\,,
\\[3pt]
&& \label{mm019}  (\alpha^{\smthree I} \bar\alpha^{\smthree I}  -
s^\smthree ) |p_\smp3^-\rangle = 0\,. \eeq
These equations tell us that the vertex must be a degree $s^\sma $
homogeneous polynomial in the respective oscillators. Taking into
account that the forms $L^\smone$, $L^\smtwo$, $B^\smthree$
\rf{mm016nn}  are degree 1 homogeneous polynomials in the
oscillators, while the forms $Q^\smonetwo$ and $Z$ \rf{mm016nn} are
respective degree 2 and 3 homogeneous polynomials in the oscillators
we find the general solution of Eqs.\rf{mm018}, \rf{mm019} as%
\footnote{ We ignore the contribution of $Q^\smoneone$-,
$Q^\smtwotwo$-, $\alpha^\smthreethree$-terms of \rf{intvereqmas01N1}
to vertex \rf{intvereqmass01}. Because of the tracelessness
constraints \rf{intver16n8} the contribution of these terms to the
Hamiltonian $P_\smp3^-$ \rf{pm1} vanishes.}
\be \label{intvereqmass01}
p_\smp3^-(s^\smone,s^\smtwo,s^\smthree\,;\,k_{min},k_{max}) =
(L^\smone)^{x^\smone  } (L^\smtwo)^{x^\smtwo }
(B^\smthree)^{x^\smthree } (Q^\smonetwo)^{y^\smthree } Z^y\,, \ee
where the parameters $x^\smone  $, $x^\smtwo $, $x^\smthree $,
$y^\smthree $, $y$ are given by
\beq
\label{xadefmm0}
&& x^\smone   = k_{max} - k_{min} - s^\smtwo \,,
\\
&& x^\smtwo  = k_{max} - k_{min} - s^\smone \,,
\\
&& x^\smthree  = k_{min}\,,
\\
&& y^\smthree =  \sbf - 2s^\smthree  - k_{max} + 2 k_{min}\,,
\\
\label{y3def}
&& y= s^\smthree  -k_{min}\,,
\eeq
and $\sbf$ is defined in \rf{0007}. New integers $k_{min}$ and
$k_{max}$ in \rf{intvereqmass01}-\rf{y3def} are the freedom in our
solution. In general, vertex \rf{intvereqmass01} is a non-homogeneous
polynomial in the momentum $\Po^I$ and the integers $k_{min}$ and
$k_{max}$ are the respective minimal and maximal numbers of powers of
the momentum $\Po^I$ in \rf{intvereqmass01}\!
\footnote{ This can be checked by taking into account that the forms
$L^\smone$, $L^\smtwo$, $Q^\smonetwo$ and $Z$ are degree 1
polynomials in $\Po^I$, while the form $B^\smthree$ is degree 1
homogeneous polynomial in $\Po^I$ (see
\rf{eqmas00007}-\rf{eqmas00005}).}.
As noted above, the minimal number of powers of the momentum $\Po^I$
is not equal to zero in general.
For vertex \rf{intvereqmass01} to be sensible, we should impose the
restrictions
\beq
\label{restrict01}&& x^\sma  \geq 0\,, \qquad a=1,2,3;
\\[5pt]
\label{restrict03} &&  y^\smthree  \geq 0\,,\qquad y \geq 0\,, \eeq
which amount to requiring the powers of all forms in
\rf{intvereqmass01} to be non--negative integers. With
\rf{xadefmm0}-\rf{y3def}, restrictions \rf{restrict01},
\rf{restrict03} can be rewritten in a more convenient form as
\be \label{restrict04} k_{min}  + \max_{a=1,2} s^\sma \leq k_{max}
\leq \sbf- 2s^\smthree  + 2 k_{min}\,, \ee
\be \label{restrict05} 0\leq k_{min} \leq s^\smthree \,.\ee

\subsubsection{ Minimal Yang-Mills interaction of massive totally symmetric
arbitrary spin field}

We now apply our results in Section \ref{tsymMM0} to the discussion
of the minimal Yang-Mills interaction of the massive totally
symmetric arbitrary spin field. We first present the list of {\it
all} cubic vertices for the massive totally symmetric spin $s$ field
interacting with the massless spin 1 field (Yang-Mills field). This
is, we consider the vertices \rf{intvereqmass01} with the spin values
\be \label{intvereqmass02N1}  s^\smone  = s^\smtwo  = s\,,\qquad
s^\smthree =1\,.\ee
Restrictions \rf{restrict05} lead to two allowed values of $k_{min}$:
$k_{min} =0,1$. Substituting these values of $k_{min}$ in
\rf{restrict04}, we obtain two families of vertices
\beq
\label{xxxN1N1} && k_{min} = 1 \,, \qquad  s + 1 \leq \ k_{max} \leq
\ 2s+1 \,,\qquad \ \ \ \quad s\geq 0\,;
\\[3pt]
\label{yyyN1N1} && k_{min} = 0 \,, \qquad  \ \ \ s \ \ \leq \ \
k_{max} \leq \ \ \ 2s-1 \,,\qquad\quad \ \ s\geq 1\,.
\eeq
We now discuss those vertices from the list in \rf{xxxN1N1},
\rf{yyyN1N1} that correspond to the minimal Yang-Mills interaction of
the massive arbitrary spin field. We consider various spin fields in
turn.

\noindent {\bf a}) Spin $s=0$ field. The vertices for the spin $s=0$
field fall in the family of vertices given in \rf{xxxN1N1}. Plugging
$s=0$ in \rf{xxxN1N1} we obtain $k_{min}=k_{max}=1$ and therefore the
cubic vertex of the minimal Yang-Mills interaction of the massive
scalar field is a degree 1 homogeneous polynomial in derivatives.
Relations \rf{intvereqmass01}-\rf{y3def} lead to the minimal
Yang-Mills interaction of the massive scalar field
\be\label{minint001} p_\smp3^-(0,0,1; 1, 1 ) = B^\smthree\,.\ee

\noindent {\bf b}) Spin $s\geq 1$ field. All vertices given in
\rf{xxxN1N1}, \rf{yyyN1N1} are candidates for the minimal Yang-Mills
interaction of the spin $s\geq 1$ field. We therefore impose an
additional requirement, which allows us to choose one suitable
vertex: given spin $s$, we look for the vertex with the minimal value
of $k_{max}$. It can be seen that such a vertex is given by
\rf{yyyN1N1} with $k_{max} = s$. The choice of the vertex from
\rf{yyyN1N1} implies $k_{min}=0$ and we obtain from
\rf{intvereqmass01}-\rf{y3def} the minimal Yang-Mills
interaction of the massive spin $s\geq 1$ field%
\footnote{ A gauge invariant description of the electromagnetic
interaction of the massive spin $s=2$ field was obtained in
\cite{Klishevich:1997pd}. The application of the approach in
\cite{Klishevich:1997pd} to the massive arbitrary spin $s$ field can
be found in \cite{Klishevich:1998ng}. The derivation of the
electromagnetic interaction of massive spin $s=2,3$ fields from
string theory is given in \cite{Argyres:1989cu,Klishevich:1998sr}. In
these references, the electromagnetic field is treated as an external
(non-dynamical) field.},
\be \label{minintss1} p_\smp3^-(s,s,1; 0, s ) = (Q^\smonetwo)^{s-1}
Z\,,\qquad s\geq 1\,.\ee

A few remarks are in order.

i) The forms $B^\smthree$ \rf{eqmas00007} and $Z$ \rf{eqmas00005}
have smooth massless limit ($\mas \rightarrow 0$). Therefore, the
minimal Yang-Mills interactions of the massive low spin $s=0,1$
fields given in \rf{minint001}, \rf{minintss1} have a smooth massless
limit, as they should. These interactions in the massless limit
coincide with  the respective interactions of the massless spin
$s=0,1$ fields in Table I.

ii) The form $Q^\smonetwo$  \rf{eqmas00006} does not have a smooth
massless limit ($\mas \rightarrow 0$). This implies that the minimal
Yang-Mills interaction of the massive spin $s > 1$ field
\rf{minintss1} does not admit a sensible massless limit; in
light-cone approach, it is contribution of $Q^\smonetwo$ that
explains why the minimal Yang-Mills interaction of the massive spin
$s>1$ field does not admit the massless limit. As was expected, the
minimal Yang-Mills interaction of the massive spin $s>1$ field
\rf{minintss1} involves higher derivatives. The appearance of the
higher derivatives in \rf{minintss1} can be seen from the expression
for $Q^\smonetwo$ \rf{eqmas00006}.

\subsubsection{ Gravitational interaction of massive totally symmetric
arbitrary spin field}

We proceed with the discussion of the gravitational interaction of
the massive totally symmetric arbitrary spin field. We first present
the list of {\it all} cubic vertices for the massive totally
symmetric spin $s$ field interacting with the massless spin 2 field.
This is, we consider vertices \rf{intvereqmass01} with the spin
values
\be \label{intvereqmass02}  s^\smone  = s^\smtwo  = s\,,\qquad
s^\smthree =2\,.\ee
Restrictions \rf{restrict05} lead to three allowed values of
$k_{min}$: $k_{min} =0,1,2$. Plugging these values of $k_{min}$ in
restrictions \rf{restrict04}, we obtain three families of vertices
\beq
\label{intvereqmass03} && k_{min} = 2 \,, \qquad  s + 2 \leq \
k_{max} \leq \ 2s+2 \,,\qquad \ \quad s\geq 0\,;
\\[3pt]
\label{intvereqmass04} && k_{min} = 1 \,, \qquad  s+1 \leq \ k_{max}
\leq \ \ \ 2s \,,\qquad\qquad \ \ s\geq 1\,;
\\[3pt]
\label{intvereqmass05} && k_{min} = 0 \,, \qquad \ \ \ s \ \ \ \ \leq
\ k_{max} \ \leq \ 2s- 2 \,,\qquad \ \ \ \ s\geq 2\,.
\eeq
We now discuss those vertices from the list given in
\rf{intvereqmass03}-\rf{intvereqmass05} that correspond to the
gravitational interaction of the massive arbitrary spin field. We
consider various spin fields in turn.

\noindent {\bf a}) Spin $s=0$ field. The gravitational interaction of
the massive scalar field is given by \rf{intvereqmass03}. Plugging
$s=0$ in \rf{intvereqmass03}, we obtain the well-known relation
$k_{min}=k_{max}=2$, which tells us that the cubic vertex of the
gravitational interaction of the massive scalar field is a degree 2
homogeneous polynomial in the derivatives. Formulas
\rf{intvereqmass01}-\rf{y3def} lead to the gravitational interaction
of the massive scalar field,
\be \label{intvereqmass06} p_\smp3^-(0,0,2; 2, 2 ) =
(B^\smthree)^2\,.\ee

\noindent {\bf b}) Spin $s=1$ field. The obvious candidates for the
gravitational interaction vertices of the massive vector field are
given in \rf{intvereqmass03}, \rf{intvereqmass04}. If $s=1$, then
restrictions \rf{intvereqmass03} lead to $3\leq k_{max}\leq 4$, and
therefore vertices \rf{intvereqmass03} involve higher derivatives.
But from the covariant approach, it is well known that the
gravitational interaction of the massive vector field does not
involve higher derivatives. We therefore restrict attention to the
vertices given in \rf{intvereqmass04}. Plugging $s=1$ in
\rf{intvereqmass04} we obtain $k_{max}=2$. Formulas
\rf{intvereqmass01}-\rf{y3def} then lead to the gravitational
interaction of the massive vector field
\be \label{intvereqmass07} p_\smp3^-(1,1,2; 1, 2 ) = B^\smthree
Z\,.\ee

\noindent {\bf c}) Spin $s\geq 2$ field. All vertices given in
\rf{intvereqmass03}-\rf{intvereqmass05} are candidates for the
gravitational interaction of spin $s\geq 2$ field. We should impose
some additional requirement that would allow us to choose one
suitable vertex. Our additional requirement is that given a spin $s$,
we look for vertex with the minimal value of $k_{max}$. It can be
seen that such a vertex is given by \rf{intvereqmass05} with $k_{max}
= s$. We note that $k_{min}=0$ and relations
\rf{intvereqmass01}-\rf{y3def} lead to the gravitational interaction
of the massive spin $s\geq 2$ field,
\be \label{intvereqmass08} p_\smp3^-(s,s,2; 0, s ) =
(Q^\smonetwo)^{s-2} Z^2\,,\qquad s\geq 2\,.\ee

A few remarks are in order.

i) Since the forms $B^\smthree$ \rf{eqmas00007} and $Z$
\rf{eqmas00005} have a smooth massless limit ($\mas \rightarrow 0$),
the gravitational interactions of the massive low spin $s=0,1,2$
fields \rf{intvereqmass06}-\rf{intvereqmass08} have smooth massless
limit, as they should. These gravitational interactions in the
massless limit reduce to the corresponding interactions of the
massless spin $s=0,1,2$ fields given in Table I.

ii) Since the form $Q^\smonetwo$ \rf{eqmas00006} does not have a
smooth massless limit ($\mas \rightarrow 0$), the gravitational
interaction of the massive higher spin $s > 2$ field
\rf{intvereqmass08} does not admit a sensible massless limit; it is
the form $Q^\smonetwo$ that explains why the gravitational
interaction of the massive higher spin field does not admit the
massless limit. Higher derivatives in the gravitational interaction
of the massive higher spin field are related to the contribution of
$Q^\smonetwo$ \rf{eqmas00006}%
\footnote{ Gauge invariant formulations of the gravitational
interaction of massive fields are studied e.g. in
\cite{Cucchieri:1994tx,Klishevich:1998wr}. Interesting discussion of
various aspects of the massive spin 2 field in gravitational
background may be found in
\cite{Buchbinder:1999ar,Buchbinder:2000fy}.}.

\subsection{ Cubic interaction vertices for one massless and two massive fields
with\\ different mass values}

We now consider the cubic interaction vertex \rf{varrep8} for fields
with the following mass values:
\be \label{mm08} \mas_1 \ne 0 ,\qquad \mas_2\ne  0,\qquad  \mas_1 \ne
\mas_2, \qquad \mas_3= 0, \ee
i.e. the {\it massive} fields carry external line indices $a=1,2$,
while the {\it massless} field corresponds to $a=3$. Equations for
the vertex involving one massless field can be obtained from
Eqs.\rf{loc1} in the limit as $\mas_3 \rightarrow 0 $. The general
solution for vertex \rf{varrep8} then takes the form (see Appendix D)
\be \label{mm09ex01} p_\smp3^- = p_\smp3^-(L_n^\smone, L_n^\smtwo;\,
Q_{mn}^\smaaplusone,\,
Q_{mn}^\smoneone,\,Q_{mn}^\smtwotwo,\,\alpha_{mn}^\smthreethree )\,,
\ee
where we use the notation
\be
\label{mm012} L_n^\smone \equiv B_n^\smone +
\frac{\mas_2^2}{2\mas_1}\alpha_n^\smone\,,\qquad
L_n^\smtwo \equiv B_n^\smtwo -
\frac{\mas_1^2}{2\mas_2}\alpha_n^\smtwo\,, \ee
\beq
&& B_n^\sma \equiv \frac{\alpha_n^{\sma I}\Po^I}{\beta_a}+
\frac{\check{\beta}_a}{2\beta_a} \mas_a \alpha_n^\sma\,,\qquad a=1,2;
\\
\label{mm013ex01}&& B_n^\smthree\equiv \frac{\alpha_n^{\smthree
I}\Po^I}{\beta_3}\,,\eeq
\vspace{-0.5cm}
\beq
\label{mm09}&& Q_{mn}^\smonetwo \equiv \alpha_{mn}^\smonetwo +
\frac{\alpha_n^\smtwo}{\mas_2} B_m^\smone
 -\frac{\alpha_m^\smone}{\mas_1} B_n^\smtwo\,,
\\
\label{mm010}&& Q_{mn}^\smtwothree \equiv \alpha_{mn}^\smtwothree
+\frac{\mas_2\alpha_m^\smtwo}{\mas_1^2 - \mas_2^2} B_n^\smthree -
\frac{2}{\mas_1^2 - \mas_2^2} B_m^\smtwo B_n^\smthree\,,
\\
\label{mm011}&& Q_{mn}^\smthreeone \equiv \alpha_{mn}^\smthreeone
+\frac{\mas_1\alpha_n^\smone}{\mas_1^2 - \mas_2^2} B_m^\smthree +
\frac{2}{\mas_1^2 - \mas_2^2} B_m^\smthree B_n^\smone\,, \eeq
and $\alpha_{mn}^\smab$, $Q_{mn}^\smaa$ are defined in \rf{amnabdef},
\rf{Qmnaadef}. An interesting property of the solution obtained is
the appearance of expressions like $\mas_1^2 - \mas_2^2$ in the
denominators of the quadratic forms $Q^\smtwothree$ \rf{mm010} and
$Q^\smthreeone$ \rf{mm011}; the forms $Q^\smtwothree$,
$Q^\smthreeone$ are therefore singular as $\mas_1\rightarrow \mas_2$.
For this reason, we considered the case of $\mas_1 = \mas_2$
separately in Section \ref{equalmasses}.

As can be seen from \rf{mm09ex01}-\rf{mm011}, it is impossible to
construct a cubic vertex that would be a homogeneous polynomial in
the momentum $\Po^I$. All forms that depend on $\Po^I$ and enter the
vertex (i.e. $L_n^\smone$, $L_n^\smtwo$, and $Q_{mn}^\smaaplusone$)
are non-homogeneous polynomials in $\Po^I$. This implies that the
cubic vertex is a non-homogeneous polynomial in $\Po^I$ in general.
To understand the remaining characteristic properties of solution
\rf{mm09ex01}, we consider the vertices for the totally symmetric
fields.

\subsubsection{ Cubic interaction vertices for totally symmetric fields}
\label{tsymMM0noneqmas}

The discussion of cubic interaction vertices for two massive totally
symmetric fields with different mass values and one massless totally
symmetric field largely follows that in Section \ref{tsymMM0}. The
cubic vertex we are interested in  can be obtained from the general
solution \rf{mm09ex01} by making identifications \rf{mm014nn} in
\rf{mm09ex01}-\rf{mm011} and ignoring the contribution of oscillators
carrying a subscript $n>1$. From \rf{mm09ex01}, adopting the
simplified notation for forms \rf{mm012}-\rf{mm011}:
\be  L^\sma \equiv L_1^\sma\,, \qquad B^\sma \equiv B_1^\sma\,,
\qquad Q^\smab \equiv Q_{11}^\smab\,, \qquad \alpha^\smab \equiv
\alpha_{11}^\smab\,, \ee
we obtain the vertex that describes the interaction of towers of
massive and massless totally symmetric fields
\be \label{xxxnnn1} p_\smp3^- = p_\smp3^-(L^\smone, L^\smtwo;\,
Q^\smaaplusone,\, Q^\smoneone,\,Q^\smtwotwo,\,\alpha^\smthreethree
)\,. \ee
The vertices for two massive totally symmetric spin $s^\smone$,
$s^\smtwo$ fields $|\phi_{s^\smone }\rangle$, $|\phi_{s^\smtwo
}\rangle$ with different mass values and one massless totally
symmetric spin $s^\smthree$ field $|\phi_{s^\smthree
}^{\mas_3=0}\rangle$ can be obtained by solving Eqs.\rf{mm018},
\rf{mm019} with $p_\smp3^-$ given in \rf{xxxnnn1}. We then obtain the
cubic vertex%
\footnote{ We ignore the contribution of $Q^\smoneone$-,
$Q^\smtwotwo$-, $\alpha^\smthreethree$-terms of \rf{xxxnnn1} to
vertex \rf{pint40}. Because of the tracelessness constraints
\rf{intver16n8}, the contribution of these terms to the Hamiltonian
$P_\smp3^-$ \rf{pm1} vanishes.}
\be\label{pint40} p_\smp3^-(s^\smone,s^\smtwo,s^\smthree\,;\,x^\smone
,x^\smtwo ) = (L^\smone)^{x^\smone  } (L^\smtwo)^{x^\smtwo }
(Q^\smonetwo)^{y^\smthree }(Q^\smtwothree)^{y^\smone
}(Q^\smthreeone)^{y^\smtwo }\,, \ee
where the parameters $y^\sma $ are given by
\beq
\label{mm014}&& y^\smone = \frac{1}{2}(s^\smtwo  + s^\smthree
-s^\smone  + x^\smone   -x^\smtwo )\,,
\\
\label{mm015}&& y^\smtwo = \frac{1}{2}(s^\smone  + s^\smthree
-s^\smtwo  - x^\smone   + x^\smtwo )\,,
\\
\label{mm016}&& y^\smthree = \frac{1}{2}(s^\smone  + s^\smtwo
-s^\smthree  - x^\smone   -x^\smtwo )\,. \eeq
Two integers $x^\smone  $, $x^\smtwo $ are the freedom of our
solution. For fixed spin values $s^\smone$, $s^\smtwo$, $s^\smthree$,
these integers label all possible cubic interaction vertices that can
be built for the fields under consideration. For vertex \rf{pint40}
to be sensible we impose the restrictions
\beq
\label{restr0101}&& y^\sma  \geq 0\,,\qquad a=1,2,3;
\\[4pt]
\label{restr0202} && x^\smone  \geq 0 \,,\qquad x^\smtwo \geq 0\,,
\\[4pt]
\label{restr0303} &&\sbf - x^\smone   - x^\smtwo   \qquad \hbox{ even
integer}\,,\eeq
which amount to the requirement that the powers of all forms in
\rf{pint40} be non--negative integers. The maximal number of powers
of $\Po^I$ in \rf{pint40}, which is denoted by $k_{max}$, is given by%
\footnote{ Expressions for $L^\sma$ and $Q^\smaaplusone$
\rf{mm012}-\rf{mm011} imply that $k_{max} = x^\smone + x^\smtwo
+2y^\smone + 2 y^\smtwo + y^\smthree$. Relations for $y^\sma$
\rf{mm014}-\rf{mm016} then lead to \rf{kmaxmm0N1}.}
\be\label{kmaxmm0N1}  k_{max} = \frac{1}{2}(s^\smone  + s^\smtwo   +
3s^\smthree + x^\smone  + x^\smtwo )\,. \ee
We note that using \rf{mm014}-\rf{mm016} allows rewriting
restrictions
\rf{restr0101} in the equivalent form%
\footnote{ If $x^\smone = x^\smtwo=0$, then restrictions \rf{mm017}
become the  restrictions well known in the angular momentum theory: $
|s^\smone - s^\smtwo| \leq s^\smthree \leq s^\smone + s^\smtwo$.}
\be
\label{mm017} |s^\smone  -s^\smtwo  -x^\smone   +x^\smtwo | \leq
s^\smthree \leq s^\smone + s^\smtwo -x^\smone   -x^\smtwo  \,.\ee

\newsection{ Parity invariant cubic
interaction vertices for massive fields }\label{secMMM}

We finally consider the cubic interaction vertex \rf{varrep8} for
three massive fields:
\be   \mas_1 \ne 0 ,\qquad  \mas_2 \ne 0,\qquad \mas_3\ne 0. \ee
The general solution for vertex  \rf{varrep8} is found to be (see
Appendix D)
\be\label{mmm1} p_\smp3^- = p_\smp3^-(L_n^\sma ;\,
Q_{mn}^\smaaplusone, Q_{mn}^\smaa\,)\,, \ee
where we use the notation
\beq
\label{mmm3} && L_n^\sma\equiv B_n^\sma + \frac{\mas_{a+1}^2 -
\mas_{a+2}^2}{2 \mas_a}\alpha_n^\sma\,,\qquad
B_n^\sma\equiv \frac{\alpha_n^{\sma I}\Po^I}{\beta_a}+
\frac{\check{\beta}_a}{2\beta_a}\mas_a \alpha_n^\sma\,,
\\[6pt]
\label{mmm2} && Q_{mn}^\smaaplusone \equiv \alpha_{mn}^\smaaplusone +
\frac{\alpha_n^\smaplusone }{\mas_{a+1}} B_m^\sma -
\frac{\alpha_m^\sma}{\mas_a} B_n^\smaplusone -
\frac{\mas_{a+2}^2}{2\mas_a \mas_{a+1}}\alpha_m^\sma
\alpha_n^\smaplusone\,, \eeq
and $\alpha_{mn}^\smab$, $Q_{mn}^\smaa$ are defined in \rf{amnabdef},
\rf{Qmnaadef}. From the expressions for the quadratic forms
$Q_{mn}^\smaaplusone$ \rf{mmm2}, it follows that the cubic vertex for
massive fields is singular as $\mas_a\rightarrow 0$, $a=1,2,3$. The
remaining quadratic forms $Q_{mn}^\smaa$ do not contribute to the
Hamiltonian when the ket-vectors $|\phi_a\rangle$ are restricted to
be traceless. We note, however, that it is sometimes convenient to
formulate interacting fields theories in terms of ket-vectors that
are not subjected to the tracelessness constraint. For example, the
ket-vectors of the light-cone gauge string field theories are not
subjected to the tracelessness constraint. We now restrict attention
to vertices for the totally symmetric fields.

\subsection{ Cubic interaction vertices for totally symmetric fields}

To obtain the cubic interaction vertex for the massive totally
symmetric fields we simply set $\nnu = 1$ in relations
\rf{mmm1}-\rf{mmm2}. To simplify the formulas we drop the
oscillator's subscript $n=1$ and use the simplified notation
$\alpha^I =\alpha_1^I$, $\alpha =\alpha_1$. The expression for the
cubic vertex can then be obtained from the general solution \rf{mmm1}
by using the identifications
\be\label{simnot01N1} \alpha^{\sma I} \equiv \alpha_1^{\sma I}\,,
\qquad \alpha^\sma \equiv \alpha_1^\sma\,,\qquad a=1,2,3\,,\ee
in \rf{mmm1} and ignoring contribution of oscillators carrying a
subscript $n>1$. Adopting the simplified notation \rf{simnot01N1} for
linear forms $L^\sma \equiv L_1^\sma$, $B^\sma \equiv B_1^\sma$
\rf{mmm3}, and quadratic forms $Q^\smab \equiv Q_{11}^\smab $
\rf{mmm2}, we see that vertex \rf{mmm1} takes the form
\be\label{mmmN2} p_\smp3^- = p_\smp3^-(L^\sma ;\, Q^\smaaplusone,
Q^\smaa\,)\,. \ee
Vertex \rf{mmmN2} describes the interaction of the towers of massive
totally symmetric fields \rf{intver16n9}. We next obtain the vertex
for massive totally symmetric spin $s^\smone$, $s^\smtwo$,
$s^\smthree$ fields. The massive totally symmetric spin $s^\sma$
fields are described by the respective ket-vectors $|\phi_{s^\sma
}\rangle$. The ket-vectors of massive fields $|\phi_{s^\sma
}\rangle$, $a=1,2,3$, can be obtained from \rf{intver16n4} by
replacement $s\rightarrow s^\sma $, $\alpha^I\rightarrow \alpha^{\sma
I}$, $\alpha\rightarrow \alpha^\sma$ in \rf{intver16n4}. Because
$|\phi_{s^\sma }\rangle$ are respective degree $s^\sma $ homogeneous
polynomials in $\alpha^{\sma I}$, $\alpha^\sma$ (see
\rf{intver16n6}), it is obvious that the vertex we are interested in
must satisfy the equations
\be \label{mmmN3} (\alpha^{\sma I}\bar\alpha^{\sma I} +
\alpha^\sma\bar\alpha^\sma  - s^\sma )|p_\smp3^-\rangle  = 0\,,\qquad
a=1,2,3, \ee
which tell us that the vertex $p_\smp3^-$ must be a degree $s^\sma $
homogeneous polynomial in the oscillators $\alpha^{\sma I}$,
$\alpha^\sma$. Taking into account that the forms $L^\sma$ and
$Q^\smaaplusone$ are respective degree 1 and 2 homogeneous
polynomials in oscillators we obtain the general solution of
Eqs.\rf{mmmN3} as%
\footnote{ We ignore the contribution of $Q^\smaa$-terms of
\rf{mmmN2} to vertex \rf{intvermmm01}. Because of the tracelessness
constraint (see the first relation in \rf{intver16n8}) the
contribution of these terms to the Hamiltonian $P_\smp3^-$ \rf{pm1}
vanishes.}
\be\label{intvermmm01}
p_\smp3^-(s^\smone,s^\smtwo,s^\smthree;x^\smone,x^\smtwo,x^\smthree)
= \prod_{a=1}^3 (L^\sma)^{x^\sma }(Q^\smaaplusone)^{y^\smaplustwo}\,,
\ee
where integers $y^\sma $ are expressible in terms of $s^\sma $ and
three integers $x^\sma$ labeling the freedom of our solution,
\be\label{mmm12} y^\sma  = \frac{1}{2}(\sbf + x^\sma - x^\smaplusone
- x^\smaplustwo)  -s^\sma \,, \qquad a=1,2,3\,,\ee
and $\sbf$ is given in \rf{0007}. The maximal number of powers of
$\Po^I$ in \rf{intvermmm01}, denoted by $k_{max}$, is given by%
\footnote{ Expressions for $L^\sma$ and $Q^\smaaplusone$ \rf{mmm3},
\rf{mmm2} imply that $k_{max} =\sum_{a=1}^3 (x^\sma  + y^\sma )$.
Taking $y^\sma$ \rf{mmm12} into account we then find \rf{mmmkmax}.}
\be \label{mmmkmax} k_{max} = \frac{1}{2}\bigl(\sbf + \sum_{a=1}^3
x^\sma\bigr)\,.\ee
Requiring the powers of the forms $L^\sma$ and $Q^\smaaplusone$ in
\rf{intvermmm01} to be non--negative integers gives the restrictions
\beq
\label{mmm13} && x^\sma  \geq 0 \,,\qquad   y^\sma  \geq 0\,, \qquad
a =1,2,3\,;
\\
\label{mmm14} && \sbf + \sum_{a=1}^3 x^\sma  \quad \hbox{ even
integer}\,. \eeq
Using relations \rf{mmm12} allows rewriting restrictions \rf{mmm13}
as%
\footnote{ If $x^\sma=0$, $a=1,2,3$, then restrictions \rf{mmm15}
become the restrictions well known in the angular momentum theory: $
|s^\smone - s^\smtwo| \leq s^\smthree \leq s^\smone + s^\smtwo$.}
\be \label{mmm15} s^\smthree  - s^\smone - s^\smtwo  + x^\smone +
x^\smtwo \leq x^\smthree \leq s^\smthree - | s^\smone - s^\smtwo -
x^\smone + x^\smtwo  |\,. \ee

\newsection{ ${\bf so(d-4)}$ light-cone formalism}\label{sod-4sec}

In the preceding sections we constructed parity invariant cubic
interaction vertices for massive and massless higher spin fields. We
studied cubic vertices for both the mixed-symmetry and totally
symmetric fields. For totally symmetric fields in Minkowski space
with dimension $d>6$, the antisymmetric Levi-Civita symbol does not
give a contribution to cubic vertices and therefore the parity
invariant vertices we obtained constitute the complete list of cubic
vertices. For totally symmetric fields in $d=4,5,6$ dimensions and
mixed-symmetry fields in $d\geq 6$ dimensions, the antisymmetric
Levi-Civita symbol admits new invariants and we should therefore
develop a method for deriving the complete lists of cubic vertices in
a systematic way. In the theories of higher spin fields, it is
important to know the complete lists of cubic vertices. This is
related to the fact that one needs to use all interaction vertices
for constructing full to all orders in coupling constant theories of
higher spin fields. We also note that vertices involving the
antisymmetric Levi-Civita symbol are unavoidable in supersymmetric
theories. The $\NN=4$, $4d$ supersymmetric Yang-Mills theory in
light-cone superspace and most supergravity theories are important
examples of such theories. Another very important example of a
dynamical system whose cubic vertices involve the antisymmetric
Levi-Civita symbol are superstring field theories. Cubic vertices of
the superstring field theories take the form $A\exp B$, where the
factor $A$ involves the antisymmetric Levi-Civita symbol. To
summarize, with the prospects of potentially interesting applications
to supersymmetric Yang-Mills theories, supergravity, superstring
theory, and supersymmetric higher spin field theories, it is
desirable to develop a method for constructing cubic vertices that
allows analyzing all possible cubic vertices on an equal footing. In
this section we develop such a method. Because one of the
characteristic features of our method is reducing the manifest
transverse $so(d-2)$ symmetry to the $so(d-4)$ symmetry we call it
the
$so(d-4)$ light-cone approach%
\footnote{ In the preceding studies \cite{Green:1984fu}, reducing the
manifest $so(d-2)$ symmetry to the $so(d-4)$ symmetry was used to
formulate superfield theory of $IIA$ superstrings. In
\cite{Green:1984fu}, the reduction was motivated by the desire to
obtain an unconstrained superfield formulation. In our study, the
main motivation for the reduction is the desire to obtain the most
general solution to cubic vertices for arbitrary spin fields in a
Poincar\'e invariant theory. It is worth noticing that our method is
especially convenient for studying the interaction vertices of
supersymmetric theories whose unconstrained superfield formulation is
based on reducing the manifest $so(d-2)$ symmetry to the $so(d-4)$
symmetry. The application of our method to the study of $11d$
supergravity can be found in \cite{Metsaev:2004wv}.}.

To develop the $so(d-4)$ light-cone approach we use equations for
cubic interaction vertices in the harmonic scheme (see Section
\ref{equharmschem}). To keep the discussion from becoming unwieldy,
we restrict our attention to the case of massless fields. All that is
then required is to solve the equations given in \rf{basequ0001},
\rf{basequ0002}, \rf{harmcon01}, \rf{harver01}:
\beq
\label{d43}&& {\bf J}^{IJ}|p_\smp3^-\rangle=0\,,
\\
\label{d44}&& (\Po^I\partial_{\Po^I} + \sum_{a=1}^3
\beta_a\partial_{\beta_a} )|p_\smp3^-\rangle=0\,,
\\
\label{d44N1} && \partial_{\Po^I} \partial_{\Po^I} |p_\smp3^-\rangle
= 0\,,
\\[4pt]
\label{m0basequ01} && X^{IJ} \PP^J  |p_\smp3^-\rangle = 0\,,
\\[3pt]
&& \label{p2vN1} |p_\smp3^-\rangle \equiv  p_\smp3^-({\Po
},\beta_a;\, \alpha)|0\rangle_1|0\rangle_2|0\rangle_3\,,
\eeq
where the angular momentum $\Jbf^{IJ}$ is defined in \rf{JIJp3}, and
Eqs.\rf{m0basequ01} are obtainable from Eqs.\rf{harver01} by setting
$\mas_a=0$, $a=1,2,3$. To proceed, we decompose the momentum $\Po^I$,
which is an $so(d-2)$ vector, as
\be\label{d46} \Po^I \  \ \rightarrow  \ \ \Po^i\,,\qquad
\Po^R\,,\qquad \Po^L \,, \qquad i=1,\ldots,d-4\,,\ee
where the momentum $\Po^i$ is an $so(d-4)$ vector and complex-valued
momenta $\Po^R$, $\Po^L $ are defined by
\be\label{d47} \Po^R =\frac{1}{\sqrt{2}}(\Po^{d-2}+{\rm i}\Po^{d-3}),
\qquad \Po^L =\frac{1}{\sqrt{2}}(\Po^{d-2} - {\rm i}\Po^{d-3})\,. \ee
In what follows, in place of the momenta $\Po^i$, $\Po^R $, $\Po^L $,
we prefer to use a dimensionfull momentum ${\Po }^L$ and
dimensionless momentum variables $q^i$, $\rho$ defined by
\be \label{newvar} q^i\equiv \frac{{\Po }^i}{{\Po }^L}\,, \qquad
\rho\equiv \frac{{\Po }^i{\Po }^i+2{\Po }^R{\Po }^L} {2({\Po
}^L)^2}\,, \qquad \frac{{\Po }^R}{{\Po }^L}= \rho - \frac{q^2}{2}\,,
\ee
where $q^2\equiv q^iq^i$. In terms of the new momenta, vertex
\rf{p2vN1} takes form
\be \label{p3v} p_\smp3^-=({\Po }^L)^k
V(q\,,\rho\,,\beta_a\,;\,\alpha)\,, \ee
which implies that the vertex $p_\smp3^-$ is a degree $k$ monomial in
${\Po }^L$. In terms of momenta \rf{newvar}, various components of
the orbital momentum \rf{LIJ01} take the form
\beq \label{Lrl} && \Lbf^{RL} =q^i\partial_{q^i} + 2\rho\partial_\rho
-{\Po }^{L}
\partial_{{\Po }^L}\,,\qquad\quad
\\
\label{Lij} &&
\Lbf^{ij} =q^i\partial_{q^j}-q^j\partial_{q^i}\,,
\\
\label{Lli} && \Lbf^{Li} =\partial_{q^i}\,,
\\
\label{Lri} && \Lbf^{Ri} =(\rho-\frac{q^2}{2})\partial_{q^i} +q^i
(q^j \partial_{q^j} + 2\rho\partial_\rho -{\Po }^L\partial_{{\Po
}^L})\,. \eeq
To demonstrate the main idea of introducing the variable $q^i$ we
focus on $Li$-part of Eqs.\rf{d43}. Plugging vertex $p_\smp3^-$
\rf{p3v} and $\Lbf^{Li}$ \rf{Lli} in the $Li$-part of Eqs.\rf{d43},
we obtain the equation
\be \label{Li2} (\partial_{q^i} + \Mbf^{Li})V(q\,,\rho\,,\beta_a\,;\,
\alpha)=0\,. \ee
The solution of Eq.\rf{Li2} is easily found to be
\be \label{d416} V(q\,,\rho\,,\beta_a\,;\, \alpha) = \widehat{E}_q\,
\widetilde{V}(\rho\,, \beta_a\,;\, \alpha)\,, \qquad\quad
\widehat{E}_q\equiv \exp(-q^i {\bf M}^{L i})\,. \ee
Collecting the above expressions, we obtain the following
representation for the vertex $p_\smp3^-$:
\be \label{p3int} p_\smp3^- = ({\Po }^L)^k \widehat{E}_q
\widetilde{V}(\rho\,, \beta_a\,;\, \alpha)\,, \ee
and note that in terms of the vertex $\widetilde{V}$ the
$\beta$-homogeneity equation \rf{d44} becomes
\be\label{d420} (\sum_{a=1}^3
\beta_a\partial_{\beta_a}+k)\widetilde{V}(\rho,\beta_a\,;\,
\alpha)=0\,. \ee
Next step is to find the dependence on the momentum $\rho$.  For
this, we use the $RL$, $Ri$ and $ij$ parts of Eqs.\rf{d43}:
\be \label{d417}
\Jbf^{RL}p_\smp3^- = 0\,,\qquad
\Jbf^{Ri}p_\smp3^-   = 0\,,\qquad
\Jbf^{ij}p_\smp3^- =0\,. \ee
It turns out that kinematical equations \rf{d417} allow finding
dependence on $\rho$. We thus obtain the following representation for
the vertex $\widetilde{V}$ (see Appendix E):
\beq
&& \label{d422} \widetilde{V}(\rho,\beta_a\,;\, \alpha)
=\widehat{E}_\rho\widetilde{V}_0(\beta_a\,;\, \alpha)\,,
\\[6pt]
&& \label{d423} \widehat{E}_\rho\equiv\sum_{n=0}^{k}
(-\rho)^n\frac{\Gamma(\frac{d-4}{2}+k-n)}{2^n
n!\Gamma(\frac{d-4}{2}+k)} ({\bf M}^{L i}{\bf M}^{L i})^n\,. \eeq
In addition, the kinematical equations \rf{d417} lead to equations
for the new vertex $\widetilde{V}_0(\beta_a\,;\, \alpha)$ \rf{d422},
\beq \label{d424} && ({\bf M}^{RL}-k)\widetilde{V}_0(\beta_a\,;\,
\alpha)=0\,,
\\[3pt]
\label{d425}&& {\bf M}^{Ri}\widetilde{V}_0(\beta_a\,;\, \alpha)=0\,,
\\[3pt]
\label{d426}&& {\bf M}^{ij}\widetilde{V}_0(\beta_a\,;\, \alpha)=0\,,
\eeq
while the $\beta$-homogeneity equation \rf{d420} takes the form
\be\label{d426ex} (\sum_{a=1}^3
\beta_a\partial_{\beta_a}+k)\widetilde{V}_0(\beta_a\,;\, \alpha)=0\,.
\ee
The dependence on the transverse space momentum ${\Po }^I$ is thus
found explicitly and we obtain the following representation for the
cubic interaction vertex:
\be \label{p3v0} p_\smp3^-({\Po }, \beta_a\,;\, \alpha) =({\Po }^L)^k
\widehat{E}_q \widehat{E}_\rho \widetilde{V}_0(\beta_a\,;\,
\alpha)\,, \ee
where $\widetilde{V}_0$ satisfies Eqs.\rf{d424}-\rf{d426ex}. One can
make sure that vertex \rf{p3v0} satisfies the harmonic equation
\rf{d44N1} (see Appendix E). We now proceed to the last step of our
method.

The last step is to find the dependence of the vertex
$\widetilde{V}_0(\beta_a\,;\, \alpha)$ on the three light-cone
momenta, $\beta_1$, $\beta_2$, $\beta_3$. Finding the dependence of
$\widetilde{V}_0(\beta_a\,;\, \alpha)$ on the momenta $\beta_a$ is
the most difficult point in the framework of the light-cone approach
because the vertices  $p_\smp3^-$ and $\widetilde{V}_0(\beta_a\,;\,
\alpha)$ are not polynomials in the light-cone momenta $\beta_a$ in
general, i.e. there is no locality condition with respect to the
light-cone coordinate $x^-$. But  our approach, which is algebraic in
nature, allows finding simple representation of the dependence on
$\beta_a$. We proceed as follows. Because of the second relation in
\rf{betaconlaw}, the vertex $\widetilde{V}_0$ depends on two
light-cone momenta. Therefore, we need two equations to find
$\widetilde{V}_0$. One of equations is given in \rf{d426ex}. Our
basic observation is that the second equation for $\widetilde{V}_0$
can be obtained from locality equations \rf{m0basequ01}. It is easy
to understand if the $so(d-2)$ invariance equations \rf{d43} are
satisfied then in order to respect all locality equations
\rf{m0basequ01} it is sufficient to solve the $L$-part of locality
equations \rf{m0basequ01}, which in the $so(d-4)$ notation takes the
form
\be\label{Llocequ01}  (X^{LR} \PP^L + X^{Li}\PP^i) |p_\smp3^-\rangle
= 0\,.\ee
Using the representation for $p_\smp3^-$ given in \rf{p3v0}, we can
prove that locality equation \rf{Llocequ01} amounts to requirement
that the vertex $\widetilde{V}_0$ satisfies the equation (see
Appendix E)
\be \label{d427} \sum_{a=1}^3 \check{\beta}_a\left( \beta_a
\partial_{\beta_a}^{\vphantom{5pt}}
 + M^{\sma RL} \right)\widetilde{V}_0 (\beta_a\,;\, \alpha) = 0\,.\ee
We note that the consistence requirement for Eqs.\rf{d425}, \rf{d427}
leads to the equation
\be \label{d428} \sum_{a=1}^3 \check{\beta}_a M^{\sma Ri}
\widetilde{V}_0(\beta_a\,;\, \alpha) = 0\,. \ee
It is easy to see that Eqs.\rf{d424}, \rf{d426ex}, \rf{d427} allow
finding the dependence of $\widetilde{V}_0$ on the light-cone momenta
$\beta_a$ completely:
\be
\label{d429} \widetilde{V}_0(\beta_a\,;\, \alpha)=
\widehat{E}_\beta\bar{V}_0(\alpha)\,,\qquad\quad
\widehat{E}_\beta\equiv \prod_{a=1}^3 \beta_a^{- M^{\sma RL}}\,. \ee
Using \rf{d429}, it can be shown that Eqs.\rf{d425}, \rf{d428} amount
to the following equations for $\bar{V}_0(\alpha)$:
\be\label{d431} M^{\smone Ri} \bar{V}_0(\alpha) = M^{\smtwo Ri}
\bar{V}_0(\alpha) = M^{\smthree Ri} \bar{V}_0(\alpha)\,,\ee
while  Eqs.\rf{d424}, \rf{d426} lead to the equations
\beq
&& \label{d430} ( {\bf M}^{RL} - k )\,\bar{V}_0(\alpha)=0\,,
\\[3pt]
&& \label{d432} {\bf M}^{ij}\,\bar{V}_0(\alpha)=0\,.\eeq
Collecting all the steps above, we obtain the following
representation for the cubic vertex:
\be \label{p3v001} p_\smp3^-({\Po }, \beta_a\,;\, \alpha) =({\Po
}^L)^k \widehat{E}_q \widehat{E}_\rho\widehat{E}_\beta
\bar{V}_0(\alpha)\,, \ee
where the vertex $\bar{V}_0(\alpha)$ depends only on the spin degrees
of freedom, denoted by $\alpha$, and satisfies
Eqs.\rf{d431}-\rf{d432}. The dependence on the transverse space
momentum ${\Po }^I$ and the light-cone momenta $\beta_a$ is thus
fixed explicitly. An attractive feature of the representation
\rf{p3v001} for the vertex is that it is valid for an arbitrary
realization of spin degrees of freedom. Because we used the general
form of the angular momentum $\Jbf^{IJ}$ in deriving \rf{p3v001}, the
solution for $p_\smp3^-$ \rf{p3v001} is universal and is valid for an
arbitrary Poincar\'e invariant theory. Various theories differ by:
(i) the spin operators $M^{IJ}$; (ii) the vertex $\bar{V}_0(\alpha)$
that depends only on spin variables $\alpha$.

We now demonstrate that the remaining Eqs.\rf{d431}-\rf{d432} can be
recast into a form that admits a purely group theoretical
interpretation. For this, we use the well-known fact that each irrep
of the $so(d-2)$ algebra can be realized as an induced representation
by inducing from the $so(d-4) \otimes so(2)$ subalgebra. Let $M^{IJ}$
be the $so(d-2)$ algebra generators realized in the $so(d-2)$ algebra
irreps labeled by Gelfand-Zetlin labels $s_1,s_2,\ldots,s_\nu$. Then
the generators $M^{IJ}$ obtained via the method of induced
representations take the form
\beq
&& \label{d433} M^{Ri} = \bar{\zeta}^i\,,
\\[3pt]
&& \label{d434}  M^{RL} =  -\zeta\bar{\zeta} + s_1\,,
\\[3pt]
&& \label{d435} M^{ij} = \zeta^i \bar{\zeta}^j -\zeta^j
\bar{\zeta}^i+S^{ij}\,,
\\[3pt]
&& \label{d436}  M^{L i} =  -\frac{1}{2}\zeta^2 \bar{\zeta}^i +
\zeta^i \zeta\bar{\zeta} +S^{ij}\zeta^j - s_1 \zeta^i\,, \eeq
where $\zeta\bar\zeta\equiv \zeta^i\bar\zeta^i$, $\zeta^2 \equiv
\zeta^i\zeta^i$ and $S^{ij}$ stands for the $so(d-4)$ algebra
generators. The generators $S^{ij}$ are realized in the $so(d-4)$
algebra irreps labeled by the Gelfand-Zetlin labels $s_2,\ldots,
s_\nu$. The oscillators $\zeta^i$, $\bar{\zeta}^i$, being vectors of
the $so(d-4)$ algebra, satisfy the commutator
$[\bar{\zeta}^i,\zeta^j]=\delta^{ij}$\!
\footnote{ Alternative convenient realization of $\zeta^i$ and
$\bar{\zeta}^i$ is to treat $\zeta^i$ as complex-valued vector and
$\bar{\zeta}^i$ as derivative in $\zeta^i$: $\bar{\zeta}^i \equiv
\partial_{\zeta^i}$.}.

To apply the spin operators $M^{IJ}$ \rf{d433}-\rf{d436} to the
analysis of cubic vertices, we attach external line index to all
quantities given in \rf{d433}-\rf{d436}, i.e. we introduce $M^{\sma
IJ}$, $\zeta^{\sma i} $, $S^{\sma ij}$, $s_1^\sma$, $a=1,2,3$. Using
the representation for the spin operators $M^{\sma Ri}$ \rf{d433} we
then obtain the solution of Eqs.\rf{d431}
\be \label{d438} \bar{V}_0(\alpha) = G(\zetabf, \alpha_S)\,, \qquad \
\ \ \
\zetabf^i\equiv \sum_{a=1}^3 \zeta^{\sma i}\,, \ee
where $\alpha_S$ stands for spin variables related to the $so(d-4)$
symmetries. Plugging $\bar{V}_0(\alpha)$ \rf{d438} into
Eqs.\rf{d430}, \rf{d432} we find the equations for the vertex $G$,
\beq \label{d440} &&(\Lbf^{ij}(\zetabf)+ \Sbf^{ij})G=0\,,\qquad
\Sbf^{ij} \equiv \sum_{a=1}^3 S^{\sma ij}\,,
\\
\label{d441}&&(\zetabf^i \partial_{\zetabf^i} + k- \sum_{a=1}^3
 s_1^\sma )G = 0\,, \eeq
where $\Lbf^{ij}(\zetabf)$ is defined similarly to \rf{LIJ01}.
Equations \rf{d440} for the vertex $G$ are purely group theoretical
equations. In the group theoretical language, $G$ is the generating
function of Clebsch -Gordan coefficients connecting one vector
representation $\zetabf^i$ and three representations whose spin
matrices are given by $S^{\sma ij}$, $a=1,2,3$. Thus, the final
expression for the vertex $p_\smp3^-$ is given by
\be\label{d443} p_\smp3^-= (\Po^L)^k\widehat{E}_\rho V_0 \,,\qquad
V_0 \equiv \widehat{E}_q \widehat{E}_\beta G(\zetabf, \alpha_S)\,.
\ee
In this formula, the operators $\widehat{E}_q$, $\widehat{E}_\rho$,
$\widehat{E}_\beta$ are given in \rf{d416}, \rf{d423}, \rf{d429},
while the spin operators $M^{IJ}$ are given in \rf{d433}-\rf{d436}.
The vertex $G$ is fixed by Eqs.\rf{d440}, \rf{d441}. We next
demonstrate how the general solution \rf{d443} can be used in
concrete applications.

\subsection{ Cubic interaction  vertices for massless totally symmetric
fields in $5d$ Minkowski space}\label{5dtheor}

$5d$ flat space is the simplest case where the advantages of the
$so(d-4)$ light-cone approach can be demonstrated. In $5d$ flat space
all physical massless fields are classified by irreps of the $so(3)$
algebra. Since irreps of the $so(3)$ algebra are labeled by one
label, all massless fields in $5d$ flat space can be described by
totally symmetric tensors fields of the $so(3)$ algebra; we therefore
restrict our attention to the study of cubic interaction vertices for
the totally symmetric fields.

For the $5d$ space, the indices $i,j$ that label $d-4$ directions
take one value: $i,j=1$. To simplify our expressions, we use the
short notation for the dimensionless momentum $q^i$ \rf{newvar} and
the spin variable $\zeta^i$:
\be q\equiv q^1\,,\qquad  \zeta \equiv \zeta^1 \,. \ee
The spin operators $M^{IJ}$ \rf{d433}-\rf{d436} then take the form
\be \label{d444}  M^{RL}=-\zeta\bar{\zeta} + s \,,\qquad M^{R 1}=
\bar\zeta\,,\qquad M^{L 1} =\frac{1}{2}\zeta^2\bar{\zeta}- s \zeta\,,
\ee
where $s\equiv s_1$. In considering cubic vertices, the quantities
$M^{IJ}$, $\zeta$, and $s$ should be equipped with an external line
index to become $M^{\sma IJ}$, $\zeta^\sma$, and $s^\sma$, $a=1,2,3$.
With this convention, the solution of Eq.\rf{d441} takes the form
\be\label{d445} G= \zetabf^{\sbf - k}\,, \qquad \zetabf \equiv
\sum_{a=1}^3 \zeta^\sma\,,\qquad  \sbf \equiv \sum_{a=1}^3
s^\sma\,.\ee
Using relations for action of the spin operators $M^{\sma RL}$,
$M^{\sma L1}$:

\beq \label{d446} && \prod_{a=1}^3 \beta_a^{-M^{\sma RL}}f(\zetabf)
=\prod_{a=1}^3 \beta_a^{-s^\sma}f( \betabf\cdot\zetabf)\,, \qquad
\betabf \cdot \zetabf \equiv \sum_{a=1}^3 \beta_a\zeta^\sma \,,
\\
\label{d447} && e^{-q(\zeta^2\bar{\zeta}+b\zeta)}f(\zeta)
=(1+q\zeta)^{-b}f\Bigl(\frac{\zeta}{1+q\zeta}\Bigr)\,, \eeq
we obtain the following representation for vertex $V_0$ \rf{d443}:
\be\label{d448} V_0(s^\smone,s^\smtwo,s^\smthree;\,k) = \ZZ^{\,\sbf -
k} \prod_{a=1}^3 (\BB^\sma)^{s^\sma} \,, \ee
where we use the notation
\be \label{d448N1} \BB^\sma \equiv  \frac{1}{\beta_a}(1 + \frac{1}{2}
q \zeta^\sma )^2\,, \quad \qquad \ZZ \equiv
\sum_{a=1}^3\frac{\beta_a\zeta^\sma }{1 + \frac{1}{2}q \zeta^\sma
}\,. \ee
For vertex $V_0$ \rf{d448} to be sensible it should be polynomial
with respect to the spin variables $\zeta^\sma$. We therefore impose
the restrictions on spin values $s^\sma$ and the number of
derivatives $k$:
\be \label{0009N1}   \sbf - k \geq 0\,, \qquad  2s^\sma \geq \sbf
-k\,, \qquad a=1,2,3\,,\ee
which coincide with those in \rf{0009}. We rewrite restrictions
\rf{0009N1} as
\be\label{d449} \sbf - 2 s_{min} \leq k \leq \sbf\,,\qquad
s_{min}\equiv \min_{a=1,2,3} s^\sma\,. \ee
Comparing with restrictions \rf{0009}, \rf{00011}, we see that {\it
the number $\sbf - k$ is not restricted to be even integer} in the
case under consideration. This is, for fixed spin values $s^\smone$,
$s^\smtwo$, $s^\smthree$, the integer $k$ takes the values (see
\rf{d449})
\be k = \sbf,\, \sbf-1,\, \sbf - 2,\, \ldots\, , \sbf - 2s_{min}\,.
\ee
This implies that for fixed spin values $s^\smone$, $s^\smtwo$,
$s^\smthree$, the number of allowed vertices
$p_\smp3^-(s^\smone,s^\smtwo,s^\smthree;k)$ that can be constructed
is given by
\be\label{Nallow1} \Nsf(s^\smone,s^\smtwo,s^\smthree) = 2 s_{min} +
1\,. \ee

Comparing \rf{Nallow1} with \rf{number01}, we conclude that the
$so(d-4)$ formalism gives additional $s_{min}$ vertices compared with
those obtained in the $so(d-2)$ formalism of Section
\ref{SolcubintversecN1} without using the antisymmetric Levi-Civita
symbol. It is these additional $s_{min}$ vertices that could be built
using the antisymmetric Levi-Civita symbol $\epsilon^{IJK}$ in the
$so(d-2)$ ($so(3)$ for $d=5$) formalism%
\footnote{ We recall that vertices not involving the antisymmetric
Levi-Civita symbol are referred to as parity invariant vertices,
while vertices involving the antisymmetric Levi-Civita symbol are
referred to as parity violating vertices.}.

To summarize, {\it the complete list of cubic vertices for the
massless spin $s^\smone,s^\smtwo,s^\smthree$ fields in $5d$ space
involves $s_{min}+1$ parity invariant vertices
$p_\smp3^-(s^\smone,s^\smtwo,s^\smthree;k)$} with the number of
derivatives given by
\be k = \sbf,\, \, \sbf-2,\, \ldots\, , \sbf - 2s_{min} \,,
\hspace{1cm}\qquad \hbox{ for parity invariant vertices}\,, \ \ \ \ee
{\it and $s_{min}$ parity violating vertices
$p_\smp3^-(s^\smone,s^\smtwo,s^\smthree;k)$} with the number of
derivatives given by
\be k = \sbf - 1,\, \sbf - 3,\, \ldots\, , \sbf - 2s_{min} +
1\,,\qquad \hbox{ for parity violating vertices}\,. \ee
A remarkable property of the $so(d-4)$ formalism is that it allows
constructing both the parity invariant and parity violating vertices
on an equal footing. This is especially important in applications to
supersymmetric theories that involve vertices of both these types. In
Table III we present those {\it parity violating} cubic vertices
whose Lorentz covariant counterparts are available in the literature%
\footnote{ In the literature, we have not found the parity violating
covariant Lagrangian for low spin $s=1,2$ fields that corresponds to
our light-cone vertex $V_0(1,1,2;3)$. The covariant Lagrangian in the
4th line of Table III is invariant only under linearized gauge
transformations.}.
The {parity invariant} cubic vertices for arbitrary $d\geq 4$ were
given in Tables I, II.

\bigskip
\noindent {\sf Table III. Parity violating cubic interaction vertices
for massless totally symmetric fields in $5d$ space. \small In the
3rd column, $A\mu$ stands for the Abelian spin 1 field, the matrices
$R_{\mu\nu}$ and $\omega_\mu$ stand for the Riemann tensor
$R_{\mu\nu}^{AB}$ and Lorentz connection $\omega_\mu^{AB}$, and Tr
denotes trace over Lorentz indices $A,B$}.
{\small
\begin{center}
\begin{tabular}{|l|c|c|}
\hline        & &
\\ [-3mm]\ Spin values and  \  & \ \ \ Light-cone  \ \ \  & Covariant
\\
number of derivatives  & vertex    & Lagrangian
\\
\ \ $ s^\smone,s^\smtwo,s^\smthree;\,k $ &
$V_0(s^\smone,s^\smtwo,s^\smthree;k)$ &
\\ \hline
&&
\\[-3mm]
\ \ \ \ \ \ \  $ 1,1,1;\,2$   & $\BB^\smone\BB^\smtwo\BB^\smthree \ZZ
$ & $
\epsilon^{\mu\nu\rho\sigma\lambda}F_{\mu\nu}F_{\rho\sigma}A_\lambda$
\\[2mm]\hline
&&
\\[-3mm]
\ \ \ \ \ \ \  $ 1,2,2;\,4$   & $
\BB^\smone(\BB^\smtwo\BB^\smthree)^2 \ZZ $ &
$\epsilon^{\mu\nu\rho\sigma\lambda} \hbox{ Tr } (R_{\mu\nu}
R_{\rho\sigma}) A_\lambda $
\\[2mm]\hline
&&
\\[-3mm]
\ \ \ \ \ \ \  $ 2,2,2;\,3 $   & $
(\BB^\smone\BB^\smtwo\BB^\smthree)^2 \ZZ^3 $ & $\LL(\hbox{see
Ref}.\cite{Boulanger:2000ni})$
\\[2mm]\hline
&&
\\[-3mm]
\ \ \ \ \ \ \  $ 2,2,2;\,5 $   & $
(\BB^\smone\BB^\smtwo\BB^\smthree)^2 \ZZ $ &
$\epsilon^{\mu\nu\rho\sigma\lambda} \hbox{Tr } ( R_{\mu\nu}
R_{\rho\sigma}\omega_\lambda) $
\\[2mm]\hline
&&
\\[-3mm]
\ \ \ \ \ \ \  $ 3,3,3;\,4$   & $
(\BB^\smone\BB^\smtwo\BB^\smthree)^3 \ZZ^5 $ & $\LL(\hbox{see
Ref}.\cite{Boulanger:2005br}) $
\\[2mm]\hline
\end{tabular}
\end{center}
}

\subsection{Cubic interaction  vertices for massless
totally symmetric and mixed-symmetry fields in $6d$ Minkowski
space}\label{6dtheor}

In $6d$ flat space all physical massless fields are classified by
irreps of the $so(4)$ algebra. Since irreps of the $so(4)$ algebra
are labeled by two Gelfand-Zetlin labels $s_1$, $s_2$, $s_1\geq
|s_2|$, massless fields in $6d$ flat space are described by $so(4)$
totally symmetric tensor fields ($s_1 \geq 0$, $s_2=0$) and $so(4)$
mixed-symmetry tensor fields ($s_2\ne 0$). For the massless {\it
mixed-symmetry} fields, $6d$ flat space is the simplest case where
advantages of the $so(d-4)$ light-cone approach can be demonstrated.
Our approach allows us to build cubic vertices for all massless
fields on an equal footing.

Since the indices $i,j$ labeling $d-4$ directions take two values for
$d=6$, $i,j=1,2$, we prefer to use the complex coordinates
\be x = \frac{1}{\sqrt{2}}(x^1+{\rm i}x^2)\,, \qquad \xb =
\frac{1}{\sqrt{2}}(x^1 - {\rm i}x^2)\,,\ee
in place of $d-4$ coordinates $x^1$, $x^2$. In the complex
coordinates, the indices $i,j$ range over $x$ and $\bar{x}$ and the
dimensionless momentum $q^i$ \rf{newvar} and the spin variable
$\zeta^i$ are decomposed as
\be q^i =  q^x,\,\, q^\xb \,; \qquad
\zeta^i = \, \zeta^x,\,\, \zeta^\xb \,;\ee
while the generator of the $so(2)$ algebra $S^{ij}$ is expressed in
terms of Gelfand-Zetlin label $s_2$,
\be S^{x\xb} = s_2\,. \ee
This implies the following representation for the spin operators
$M^{Li}$ \rf{d436}:
\be
\label{MLxb}  M^{L \xb } =(\zeta^{\bar{x}})^2\bar{\zeta}^x - (s_1 +
s_2) \zeta^{\bar{x}}\,,  \qquad M^{L x}=(\zeta^x)^2
\bar{\zeta}^{\bar{x}} - ( s_1 - s_2 ) \zeta^x\,. \ee
We note that the Gelfand-Zetlin labels $s_1$ and $s_2$ of the $so(4)$
algebra irreps  can be related to labels $j_1,j_2$ of irreps of
$so(3)_1$ and $so(3)_2$ algebras that enter the decomposition
$so(4)=so(3)_1 \oplus so(3)_2$:
\be\label{j1j2def} j_1 = \frac{1}{2}(s_1 + s_2)\,,\qquad
j_{2}=\frac{1}{2}(s_1 - s_2)\,.  \ee
To study cubic vertices the quantities $j_1$, $j_2$, $\zeta^i$ should
be equipped with external line index $a=1,2,3$, i.e. we should
introduce $j_1^\sma$, $j_2^\sma$, $\zeta^{\sma i}$. With this
convention, the solution of Eqs.\rf{d440}, \rf{d441} takes the form
\be G=(\zetabf^\xb)^{\jbf_1 - \frac{k}{2}} (\zetabf^x)^{\jbf_2 -
\frac{k}{2} }\,, \qquad \zetabf^i \equiv \sum_{a=1}^3 \zeta^{\sma
i}\,,\qquad  \quad \jbf_\sigma\equiv\sum_{a=1}^3 j{}_\sigma^\sma
\,.\ee
Using spin operators \rf{MLxb} and relations \rf{d446}, \rf{d447}, we
obtain the cubic vertex
\be \label{Vod6}
V_0(j_\sigma^\sma;\,k)= \prod_{\sigma=1,2} \ZZ_\sigma^{\jbf_\sigma
-\frac{k}{2}} \prod_{a=1,2,3\atop \sigma=1,2}
(\BB_\sigma^\sma)^{j_\sigma^\sma}\,,\ee
where we use the notation
\be
\BB_\sigma^\sma \equiv \frac{1}{\beta_a} (1+q_\sigma
\zeta{}_\sigma^\sma )^2 \,,\qquad
\ZZ_\sigma\equiv \sum_{a=1}^3 \frac{\beta_a\zeta{}_\sigma^\sma }{1 +
q_\sigma\zeta{}_\sigma^\sma }\,,
\ee
\be q_1\equiv q^x\,,\qquad q_2\equiv q^{\bar{x}}\,,\qquad
\zeta_1^\sma\equiv\zeta^{\sma \bar{x}}\,,\qquad \zeta_2^\sma
\equiv\zeta^{\sma x}\,. \ee
For vertex $V_0$ \rf{Vod6} to be polynomial in the spin variables
$\zeta_\sigma^\sma$, we impose the restrictions
\beq
\label{mixsymineq2} && 2(\jbf_\sigma - 2j{}_\sigma^\sma )\leq k \leq
2 \jbf_\sigma\,,\qquad a=1,2,3; \quad \sigma=1,2\,;
\\
\label{mixsymineq3} && 2\jbf_\sigma - k \quad \hbox{ even
integers}\,, \qquad \sigma = 1,2\,. \eeq
Restrictions \rf{mixsymineq2} amount to the restrictions
\be \label{mixsymineq4} 2\max_{\sigma=1,2}(\jbf_\sigma - 2
\min_{a=1,2,3} j{}_\sigma^\sma) \leq  k  \leq 2\min_{\sigma = 1,2}
\jbf_\sigma\,. \ee
From \rf{mixsymineq3}, \rf{mixsymineq4}, we see that for fixed spin
values $j_\sigma^\sma$, the number of cubic interaction vertices
$p_\smp3^-(j_\sigma^\sma;k)$ that can be constructed is given by
\be\label{Nd6} \Nsf(j_\sigma^\sma) = \min_{\sigma = 1,2} \jbf_\sigma
-\max_{\sigma=1,2}(\jbf_\sigma - 2 \min_{a=1,2,3} j{}_\sigma^\sma) +1
\,. \ee
To summarize, {\it for fixed  spin values $j_\sigma^\sma$,
restrictions \rf{mixsymineq3}, \rf{mixsymineq4} define all possible
(parity invariant and parity violating) cubic vertices} that can be
built for the massless totally symmetric and mixed-symmetry fields in
$6d$ flat space. The number of these vertices is given by \rf{Nd6}.

We now restrict our attention to the totally symmetric fields and
compare the complete list of vertices obtained  using the $so(d-4)$
method in this section and the list of parity invariant vertices
obtained by the $so(d-2)$ method in Section \ref{SolcubintversecN1}.
All that is required is to compare the restrictions \rf{mixsymineq3},
\rf{mixsymineq4} for the totally symmetric fields and restrictions
\rf{00011}, \rf{00012}. To adapt restrictions \rf{mixsymineq3},
\rf{mixsymineq4} to the totally symmetric fields we set $s_2^\sma=0$,
$a=1,2,3$. Relations \rf{j1j2def} then give $j_1^\sma=j_2^\sma
=s_1^\sma/2$. Using the identification $s_1^\sma \equiv s^\sma$, we
see that restrictions \rf{mixsymineq3}, \rf{mixsymineq4} for the
totally symmetric fields coincide with those of the $so(d-2)$
approach, \rf{00011}, \rf{00012}. This implies that for massless
totally symmetric fields in $6d$, the complete list of cubic vertices
obtained using the $so(d-4)$ method in this section coincides with
the list of parity invariant cubic vertices obtained using the
$so(d-2)$ method in Section \ref{SolcubintversecN1}, i.e. {\it all
cubic vertices for the massless totally symmetric fields in $6d$ are
parity invariant}. Thus, in contrast to the $5d$ case the
antisymmetric Levi-Civita symbol does not lead to new cubic vertices
for the massless totally symmetric fields in $6d$.

In a manifestly Lorentz covariant formulation, the massless
mixed-symmetry fields in $6d$ flat space are described by a set of
the tensor fields whose $SO(5,1)$ space-time tensor indices have the
structure of a Young tableaux with two rows. The study of interaction
vertices for massless mixed-symmetry fields for arbitrary $d$ in the
framework of covariant approach can be found in
\cite{Bekaert:2004dz}.

\newsection{Conclusions}\label{CONsec}

Using the light-cone formalism  we have developed various methods for
constructing cubic interaction vertices for higher spin fields
propagating in flat space. We applied these methods to construct a
wide class of cubic interaction vertices for massless and massive
arbitrary spin fields. For mixed-symmetry fields in space of
arbitrary dimension, we obtained the generating function of the
parity invariant cubic vertices. We believe that this generating
function involves all possible parity invariant vertices. To classify
these vertices (i.e. to find restrictions on powers of derivatives in
a vertex for three fields carrying various values of spins) it is
necessary to single out irreducible components of the reducible sets
of fields as this was done in the case of totally symmetric fields.
It seems likely that the easiest way to do that is to apply the
$so(d-4)$ method to the generating solution for vertices. We hope to
study this issue elsewhere. We emphasize that the generating form of
the cubic vertices is very convenient for studying higher order
interaction corrections in the theories of higher spin fields. We
believe that the study these corrections will allow us to find the
generating function explicitly.

Our results should have a number of the following interesting
applications and generalizations.

i) We studied interaction vertices for bosonic fields. It would be
interesting to extend the methods developed in this paper to the case
of fermionic fields and apply these results to interaction vertices
of the closed superstring field theory.

ii) The light-cone gauge formulation of free fields in $AdS$ space
was developed in \cite{Metsaev:1999ui}-\cite{Metsaev:2003cu}. It
would be interesting to extend the methods in this paper to study
cubic interaction vertices for fields propagating in $AdS$ space.
This should be relatively straightforward because the light-cone
gauge formulation of the field dynamics in $AdS$ space provides
certain simplifications.

iii) Another interesting application is related to certain massless
(nonsupersymmetric) triplets in $d=11$, the dimension of M-theory. It
was found in \cite{Pengpan:1998qn} that some irreps of the $so(9)$
algebra naturally group together into triplets to be referred to as
Euler triplets which are such that bosonic and fermionic degrees of
freedom match up the same way as in $11d$ supergravity (see also
\cite{Brink:2002zq}-\cite{Brink:1999te}). Later on it was conjectured
that these triplets might be organized in a relativistic theory so
that this theory would presumably be finite. The methods we developed
in this paper and those we used for studying $11d$ supergravity
\cite{Metsaev:2004wv} admit a straightforward generalization to
higher spin Euler triplets. We hope to return to these problems in
future publications.

\medskip

{\bf Acknowledgments}. We thank N. Boulanger for informative
communications and A. Semikhatov for useful comments. This work was
supported by the INTAS project 03-51-6346, by the RFBR Grant
No.05-02-17654, RFBR Grant for Leading Scientific Schools, Grant No.
LSS-4401.2006.2 and Russian Science Support Foundation.

\setcounter{section}{0} \setcounter{subsection}{0}
\appendix{Derivation of expressions for $\XX^I$ \rf{cubeq06}.}

In this Appendix we derive formulas \rf{cubeq04}, \rf{cubeq06}. To
this end we note the relation
\be \label{1appen01} p_a^I  =
-\frac{\beta_a\check{\beta}_a}{3\hat{\beta}}\,\Po^I
+\frac{\beta_a}{3}\, \Pt^I \,,\qquad\quad \Pt^I \equiv \sum_{a=1}^3
\frac{p_a^I}{\beta_a}\,. \ee
Making use of \rf{1appen01} it is easy to derive the following
helpful relations:
\beq
\label{1appen03} && \sum_{a=1}^3 p_a^I \check{p}_a^J ={\Po }^I \Pt^J
-{\Po }^J \Pt^I \,,\qquad  \check{p}_a^I  \equiv p_{a+1}^I -
p_{a+2}^I\,,
\\
\label{1appen04} && \sum_{a=1}^3 \check{\beta}_a p_a^- =- \Pt^I {\Po
}^I - \sum_{a=1}^3 \frac{\check{\beta}_a}{2\beta_a} \mas_a^2 \,,
\\
\label{1appen05} && \sum_{a=1}^3 \frac{1}{\beta_a}M^{\sma IJ}p_a^J
=-\frac{1}{3\hat{\beta}}\sum_{a=1}^3\check{\beta}_a M^{\sma IJ}{\Po
}^J +\frac{1}{3}{\bf M}^{IJ} \Pt^J\,. \eeq
In Eq.\rf{cubver3} the operator $\Jbf^{-I\dagger}$ \rf{cubeq05} is
realized as differential operator with respect to the momenta
$p_a^I$, $\beta_a$, which acts on the vertex
$p_\smp3^-(p_a,\beta_a;\alpha)= p_\smp3^-(\Po,\beta_a;\alpha)$. To
derive formulas \rf{cubeq04}, \rf{cubeq06} we should realize the
operator $\Jbf^{-I\dagger}$ as differential operator with respect to
the momenta $\Po^I$, $\beta_a$, which acts on vertex
$p_\smp3^-(\Po,\beta_a;\alpha)$ \rf{p2v}. To this end we consider an
action of different pieces of the operator $\Jbf^{-I\dagger}$
\rf{cubeq05} on the vertex $p_\smp3^-(p_a,\beta_a;\alpha)=
p_\smp3^-(\Po,\beta_a;\alpha)$ and obtain the relations
\beq
\label{1appen06} \sum_{a=1}^3 p_a^- \partial_{p_a^I} \,
p_\smp3^-(p_b,\beta_b;\alpha)
& = & \frac{1}{3}\sum_{a=1}^3 \check{\beta}_a p_a^-\partial_{{\Po
}^I}\, p_\smp3^-(\Po,\beta_b;\alpha)
\nonumber\\
& = & \Bigl( - \frac{1}{3}\Pt^J {\Po }^J - \sum_{a=1}^3
\frac{\check{\beta}_a}{6\beta_a} \mas_a^2\Bigr) \partial_{{\Po }^I}\,
p_\smp3^-(\Po,\beta_b;\alpha) \,,
\\
\label{1appen07}
\sum_{a=1}^3 p_a^I \partial_{\beta_a}\, p_\smp3^-(p_b,\beta_b;\alpha)
&=&\sum_{a=1}^3 \Bigl(-\frac{1}{3\hat{\beta}}{\Po
}^I\check{\beta}_a\beta_a\partial_{\beta_a} + \frac{1}{3} \Pt^I
\beta_a\partial_{\beta_a}\Bigr)p_\smp3^-(p_b,\beta_b;\alpha)
\nonumber\\
&=& \frac{1}{3}({\Po }^J \Pt^I  - {\Po }^I \Pt^J)
\partial_{{\Po }^J} \,
p_\smp3^-(\Po,\beta_b;\alpha)
\nonumber\\
&+& \sum_{a=1}^3 \Bigl(-\frac{1}{3\hat{\beta}}{\Po }^I
\check{\beta}_a\beta_a\partial_{\beta_a} +\frac{1}{3}\Pt^I
\beta_a\partial_{\beta_a}\Bigr) \, p_\smp3^-(\Po,\beta_b;\alpha)\,.
\eeq
In the 2nd line of \rf{1appen06} we use formula \rf{1appen04}. In the
first line of \rf{1appen07} we use formula \rf{1appen01}, while in
the 2nd and 3rd lines of \rf{1appen07} we use the relations
\be  \sum_{a=1}^3 \beta_a\partial_{\beta_a}{\Po }^I = \Po^I \,,
\qquad \sum_{a=1}^3\check{\beta}_a\beta_a\partial_{\beta_a}{\Po }^I
=\hat{\beta} \Pt^I\,.  \ee
Collecting the expressions \rf{1appen05}-\rf{1appen07} in
$\Jbf^{-I\dagger}$ \rf{cubeq05} and taking into account
Eqs.\rf{kinsod}, \rf{honcon04} we note that all terms proportional to
the momentum $\Pt^I$ are cancelled. The remaining terms then lead to
the desired representation for $\Jbf^{-I\dagger}|p_\smp3^-\rangle$
given in \rf{cubeq04}, \rf{cubeq06}.

\appendix{ Field redefinitions in light-cone approach}

In this appendix we discuss field redefinitions in the framework of
Hamiltonian light-cone approach. Let $\phi$ be generic field. We make
a field redefinition $\phi\rightarrow \phiwt$:
\be\label{appfr01} \phiwt  = \phi + \sum_{n=2}^\infty
\phiwt_\smpn[\phi] \,, \ee
where $\phiwt_\smpn$ stands for $n$ - point contribution (having $n$
powers of the generic field $\phi$) to $\phiwt$. The $\phiwt_\smpn$
is restricted to be local (with respect to transverse directions)
functional in $\phi$. In the Hamiltonian light-cone approach we are
allowed to make the field redefinitions \rf{appfr01} that satisfy the
following two basic requirements:
{\bf i}) The field $\phiwt$ should satisfy the light-cone canonical
commutator \rf{bascomrel}. Field redefinitions that respect the
light-cone canonical commutator \rf{bascomrel} will be referred to as
light-cone canonical transformations;
{\bf ii}) The field redefinitions \rf{appfr01} should preserve
structure of the kinematical generators \rf{fierep}, i.e. the field
$\phiwt$  should satisfy the equation $\Gwt^{kin}=G^{kin}[\phi]$,
where $\Gwt\equiv G[\phiwt]$.

The field redefinitions \rf{appfr01} that satisfy these requirements
can be introduced by using a standard procedure. Let $F[\phi]$ be
generating functional of the light-cone canonical transformations.
This functional has an expansion
\be\label{appfr03} F[\phi] = \sum_{n=3}^\infty F_\smpn[\phi] \,, \ee
where $F_\smpn[\phi]$ stands for $n$ - point contribution to
$F[\phi]$ (see \rf{F1} for explicit expression). We then introduce
one parametric flow
\be \partial_s \phi_s = [\phi_s,F_s]\,, \qquad \quad F_s \equiv
F[\phi_s] \,,\ee
and note that the generic field $\phi$ and the canonically
transformed field $\phiwt$ \rf{appfr01} are given by
\be\label{appfr03N1} \phi \equiv \phi_{s=0}\,,\qquad  \phiwt \equiv
\phi_{s=1}\,. \ee
It is easy to check that the field $\phiwt$ \rf{appfr03N1} satisfies
the light-cone canonical commutator \rf{bascomrel}.

We now focus on the light-cone canonical transformations that
maintain the kinematical generators. Any generator $G = G[\phi]$
being canonically transformed takes the form
\be\label{appBcantravar} \Gwt = G_{s=1}\,, \qquad G_s\equiv
G[\phi_s]\,,\ee
where $G_s$ satisfies the differential equation and the initial
condition
\be \label{appBB6} \partial_s G_s = [G_s,F_s]\,,\qquad G_{s=0} = G
\,.\ee
Making use of \rf{appBB6} and the Taylor series expansion for $\Gwt$
\be \Gwt = \sum_{n=1}^\infty \frac{1}{n!} \partial_s^n G_s|_{s=1}\,,
\ee
we obtain the following expansion for the canonically transformed
generator $\Gwt$ \rf{appBcantravar}:
\be \Gwt  \,\, = \,\,   \sum_{n=1}^\infty \frac{1}{n!} [\ldots [
G,F],\ldots F]\,\, = \,\, G + [G,F] + \frac{1}{2}[[G,F],F] + \ldots
\,\,\,.\ee
Then using the expansion for $F[\phi]$  \rf{appfr03} and expansions
for the generic generator $G$ and the canonically transformed
generator $\Gwt$,
\be G = \sum_{n=2}^\infty G_\smpn\,, \qquad  \Gwt = \sum_{n=2}^\infty
\Gwt_\smpn\,, \ee
we obtain the following relations for the leading terms:
\beq \Gwt_\smpt &=&  G_\smpt\,,
\\
\label{appc10}\Gwt_\smp3 &=&  G_\smp3 + [G_\smpt,F_\smp3] \,,
\\
\Gwt_\smpf &=&  G_\smpf + [G_\smpt,F_\smpf] +  [G_\smp3,F_\smp3] +
\frac{1}{2} [[G_\smpt,F_\smp3],F_\smp3] \,.    \eeq
Now we are ready to find the light-cone canonical transformations
that maintain the kinematical generators. Since the generic
kinematical generators $G^{kin} = G_\smpt^{kin}$ are quadratic in the
field $\phi$, all that is required is to find generating function
$F[\phi]$ that satisfies the equation:
\be \Gwt_\smpt^{kin}  = G_\smpt^{kin}\,,\qquad \Gwt_\smpn = 0\,,
\quad \hbox{ for all } \ \ n \geq 3\,.  \ee
{}It is clear that suitable generating functional $F=F[\phi]$ should
satisfy the equations:
\be \label{appB15} [G_\smpt^{kin},F]= 0\,.\ee
{}By using the representation for the $n$ - point contribution to
$F[\phi]$ (see expansion \rf{appfr03})
\beq \label{F1} && F_\smpn  = \int d\Gamma_n  \langle \Phi_\smpn|
f_\smpn \rangle \,, \qquad n\geq 3 \,,\eeq
and procedure of Section 2 we find from \rf{appB15} that the
densities $f_\smpn$ are functions of $\Po_{ab}^I$, $\beta_a$,
$\alpha$,
\beq \label{fdensity01} && f_\smpn = f_\smpn({\Po }_{ab},\beta_a;
\alpha)\,, \eeq
(where $\alpha$ stands for spin D.o.F) and satisfy the equations
\beq
\label{fappen03} && \Bigl(\sum_{\{ a b \}}
\Po_{ab}^I\partial_{\Po_{ab}^J} - \Po_{ab}^J\partial_{\Po_{ab}^I} +
\sum_{a=1}^n M^{\sma IJ} \Bigr) |f_\smpn\rangle =0\,,
\\
\label{fappen02} && \Bigl(\sum_{\{ a b \}}
\Po_{ab}^I\partial_{\Po_{ab}^I} + \sum_{a=1}^n
\beta_a\partial_{\beta_a}\Bigr) |f_\smpn\rangle = |f_\smpn\rangle \,.
\eeq
We finish with the discussion of dependence of the cubic Hamiltonian
$P_\smp3^-$ on the field redefinitions. From \rf{appc10}, we see that
field redefinitions at cubic order are entirely governed by density
$f_\smp3$. The density $f_\smp3 = f_\smp3(\Po,\beta_a;\alpha)$ (see
\rf{fdensity01} and \rf{po122331}) satisfies equations (see
\rf{fappen03}, \rf{fappen02} for $n=3$)
\be
{\bf J}^{IJ}|f_\smp3\rangle =0\,,\qquad\quad
(\Po^I\partial_{\Po^I} +
\sum_{a=1}^3\beta_a\partial_{\beta_a})|f_\smp3\rangle  =
|f_\smp3\rangle\,,\ee
where the angular momentum $\Jbf^{IJ}$ is given in \rf{JIJp3}.
{}Adopting relation \rf{appc10} for the case of $G=P^-$ and making
use of formulas \rf{pm1}, \rf{F1} for $n=3$, it is seen that the
cubic interaction vertex $p_\smp3^-$, being subjected to the field
redefinitions, takes the form
\be\label{appintverred} \widetilde{p}\,{}_\smp3^- = p_\smp3^- -
\Pbf^- f_\smp3\,, \ee
where $\Pbf^-$ is given in \rf{cubver15}. {}From \rf{appintverred}
and expression for $\Pbf^-$ \rf{cubver15}, it is clear that all
$(\Po^I\Po^I)^q$- terms, $q=1,2,\ldots$, in the density $
\widetilde{p}{}_\smp3^-$ are scheme dependent and can therefore be
removed (or created) by appropriate choice of $f_\smp3$.

\appendix{ Derivation of the locality equations \rf{harver01}}

We start with the operator $\XX^I$ given in
\rf{cubeq06}-\rf{harver04} and note that $\XX^I$ \rf{cubeq06} can be
decomposed into non-harmonic and harmonic parts
\beq
&& \label{jIhardec}  \XX^I = \XX_{non-harm}^I + \XX_{harm}^I\,,
\\
\label{jnon-harmexp}&& \XX_{non-harm}^I   \equiv 2\hat\beta\Pbf^-
\frac{1}{2\kh+N} X^{IJ}
\partial_{\Po^J}\,,
\\
\label{j-harmexp} && \XX_{harm}^I   \equiv    X^{IJ} \PP^J +  X^I +
\Bigl(X\delta^{IJ} + \frac{\hat\beta}{2\kh + N}\sum_{a=1}^3
\frac{\mas_a^2}{\beta_a}X^{IJ}\Bigr)
\partial_{\Po^J}\,, \eeq
where $\Pbf^-$ is given in \rf{cubver15}. We note that the action of
$\XX_{harm}^I$ \rf{j-harmexp} on harmonic polynomial in $\Po^I$ gives
a harmonic polynomial. This is easily seen from the relations
\be  \Delta \PP^I = \Bigl( \Po^I - \Po^J\Po^J \frac{1}{2\kh + N +
4}\partial_{\Po^I}\Bigr) \Delta\,, \qquad \Delta\partial_{\Po^I} =
\partial_{\Po^I} \Delta\,,\qquad  \Delta \equiv
\partial_{\Po^I}\partial_{\Po^I}\,.\ee
Using \rf{jIhardec}-\rf{j-harmexp}, we obtain
\be\label{appbb05} \XX^I |p_\smp3^-\rangle = 2\hat\beta\Pbf^-
\frac{1}{2\kh+N} X^{IJ}\partial_{\Po^J} |p_\smp3^-\rangle +
\XX_{harm}^I |p_\smp3^-\rangle\,. \ee
Since the vertex $|p_\smp3^-\rangle$ is chosen to be a harmonic
polynomial in $\Po^I$, the expression $\XX_{harm}^I
|p_\smp3^-\rangle$ in \rf{appbb05} is also a harmonic polynomial in
$\Po^I$. But harmonic polynomial in $\Po^I$ cannot be represented as
$\Pbf^- V$, where $V$ is a polynomial in $\Po^I$. This implies that
in order to respect the light-cone locality condition
\rf{basequ0003}, we should impose the equations $\XX_{harm}^I
|p_\smp3^-\rangle = 0$, which are nothing but the locality equations
\rf{harver01}. The locality equations and relations \rf{appbb05},
\rf{cubver13} imply the representation for $|j_\smp3^{-I}\rangle$
given in \rf{clorepj3}.

\appendix{Derivation of cubic interaction vertices }

{\bf Derivation of vertex} \rf{0002}. Equations to be solved are
obtainable from Eqs.\rf{loc1}, \rf{Gandef} by setting
$\mas_1=\mas_2=\mas_3=0$ in \rf{Gandef}. Since the ket-vectors of
massless fields \rf{intver16n2} are independent of the scalar
oscillators $\alpha_n$, the scalar oscillators $\alpha_n^\sma$ in
\rf{varrep8} do not contribute to the Hamiltonian \rf{pm1} and
therefore we ignore dependence on $\alpha_n^\sma$-terms in
\rf{varrep8}%
\footnote{ Derivatives with respect to the scalar oscillators
$\alpha_n^\sma$  vanish in \rf{loc1}, \rf{Gandef} when
$\mas_a\rightarrow 0$, i.e. ignoring dependence on $\alpha_n^\sma$ in
the cubic interaction vertex is a self-consistent procedure in
solving Eq.\rf{loc1}.}.
Thus, we start with the vertex and equations
\beq
\label{appd03N1} && p_\smp3^- = p_\smp3^-(B_n^\sma;
\alpha_{mn}^\smaaplusone,\alpha_{mn}^\smaa)\,,
\\
&& \label{appd03} \sum_{m=1}^\nnu  ( B_m^\smaplusone
\partial_{\alpha_{nm}^\smaaplusone } - B_m^\smaplustwo
\partial_{\alpha_{mn}^\smaplustwoa } )p_\smp3^-=0\,.  \eeq
To analyze Eqs.\rf{appd03} it is convenient to introduce new
variables defined by
\be \label{appd06} x_{mn}^\smaaplusone \equiv
\frac{\alpha_{mn}^\smaaplusone}{B_m^\sma B_n^\smaplusone}\,. \ee
In terms of these new variables, vertex $p_\smp3^-$ \rf{appd03N1} and
Eqs.\rf{appd03} take the form
\beq
\label{appd07} && p_\smp3^- = p_\smp3^- (B_n^\sma;
x_{mn}^\smonetwo,x_{mn}^\smtwothree,x_{mn}^\smthreeone;\,
\alpha_{mn}^\smaa)\,,
\\
\label{5}&& \sum_{m = 1}^\nnu  \bigl( \partial_{x_{nm}^\smonetwo }
 -
\partial_{x_{mn}^\smthreeone}\bigr) p_\smp3^- =0 \,,
\\
\label{6}&& \sum_{m=1}^\nnu  \bigl( \partial_{x_{nm}^\smtwothree } -
\partial_{x_{mn}^\smonetwo }\bigr) p_\smp3^- =0\,,
\\
\label{7}&& \sum_{m=1}^\nnu  \bigl( \partial_{x_{nm}^\smthreeone} -
\partial_{x_{mn}^\smtwothree } \bigr) p_\smp3^- = 0 \,. \eeq
We solve Eqs.\rf{5}-\rf{7} in turn.

{\bf i)} To find general solution of Eq.\rf{5} we consider the
appropriate characteristic equations
\be \label{appd08} dx_{n1}^\smonetwo =\ldots=dx_{n\,\nnu }^\smonetwo
= - dx_{1n}^\smthreeone =\ldots=-dx_{\nnu\, n}^\smthreeone\,. \ee
Integrals of these equations are given by
\be \label{var1} \tilde{y}_{nn'}^\smonetwo\equiv x_{nn'}^\smonetwo -
x_{n1}^\smonetwo\,, \quad n' = 2,\ldots, \nnu\,;
\qquad\quad
\tilde{y}_{mn}^\smthreeone \equiv x_{mn}^\smthreeone +
x_{n1}^\smonetwo\,, \ee
and this implies that the general solution of Eqs.\rf{5} takes the
form
\be\label{appd10} p_\smp3^- = p_\smp3^-(B_n^\sma;\,
\tilde{y}_{mn'}^\smonetwo\,, x_{mn}^\smtwothree\,,
\tilde{y}_{mn}^\smthreeone;\, \alpha_{mn}^\smaa)\,. \ee

{\bf ii)} Now we are going to find restrictions imposed by Eqs.\rf{7}
on the general solution \rf{appd10}. To this end we rewrite
Eqs.\rf{7} in terms of vertex $p_\smp3^-$  \rf{appd10}:
\be \label{appd12} \sum_{m = 1}^\nnu  \bigl(
\partial_{\tilde{y}_{nm}^\smthreeone} -
\partial_{ x_{mn}^\smtwothree } \bigr) p_\smp3^- =0 \,. \ee
As before, we solve Eqs.\rf{appd12} by using the method of
characteristic equations and we find
\be \label{appd14} p_\smp3^- = p_\smp3^-(B_n^\sma;\, \tilde{y}_{m
n'}^\smonetwo,\, \tilde{y}_{n' n}^\smtwothree,\, y_{mn}^\smthreeone\,
;\, \alpha_{mn}^\smaa)\,, \ee
where new variables $\tilde{y}_{n' n}^\smtwothree$ and
$y_{mn}^\smthreeone$ are defined by
\be \label{var7}
\tilde{y}_{n' n}^\smtwothree \equiv x_{n' n}^\smtwothree - x_{1
n}^\smtwothree\,,\qquad
y_{mn}^\smthreeone \equiv \tilde{y}_{mn}^\smthreeone  +
x_{1m}^\smtwothree\,.
\ee

{\bf iii)} To solve the remaining  Eqs.\rf{6} we should rewrite these
equations in terms of vertex $p_\smp3^-$  \rf{appd14}. To this end by
using the relations (see \rf{var1}, \rf{var7})
\be \label{var8}
\tilde{y}_{mn'}^\smonetwo \equiv x_{mn'}^\smonetwo -
x_{m1}^\smonetwo\,,\qquad
\tilde{y}_{n' n}^\smtwothree = x_{n' n}^\smtwothree - x_{1
n}^\smtwothree\,, \qquad
y_{mn}^\smthreeone = x_{mn}^\smthreeone + x_{n 1}^\smonetwo
 + x_{1m}^\smtwothree \,,
\ee
we prove that Eqs.\rf{6} amount to the following equations for
vertex $p_\smp3^-$  \rf{appd14}:
\beq \label{appd18} && \sum_{m=1}^\nnu   \bigl(
\partial_{\tilde{y}_{m n'}^\smonetwo} -
\partial_{\tilde{y}_{n' m}^\smtwothree}\bigr)  p_\smp3^- = 0\,,
\qquad n' =2,\ldots, \nnu\,. \eeq
Using the method of characteristic equations we obtain the solution
of Eqs.\rf{appd18}
\be \label{appd21}  p_\smp3^- = p_\smp3^- (B_n^\sma;
y_{m'n'}^\smonetwo,\, y_{m' n}^\smtwothree,\, y_{mn}^\smthreeone\,;\,
\alpha_{mn}^\smaa)\,, \ee
where new variables $y_{m'n'}^\smonetwo$ and $y_{m' n}^\smtwothree$
are defined by
\be \label{appd20} y_{m'n'}^\smonetwo = \tilde{y}_{m'n'}^\smonetwo -
\tilde{y}_{1n'}^\smonetwo\,, \qquad
y_{m' n}^\smtwothree = \tilde{y}_{m' n}^\smtwothree + \tilde{y}_{1
m'}^\smonetwo\,,\qquad m',n'=2,\ldots,\nnu\,. \ee

Thus, the solution of Eqs.\rf{5}-\rf{7} is given by formula
\rf{appd21}. Representation of the variables $y_{m'n'}^\smonetwo$,
$y_{m' n}^\smtwothree$, $y_{mn}^\smthreeone$ \rf{appd21} in terms of
the generic variables $x_{mn}^\smab$ \rf{appd07} is given by
$$
y_{m'n'}^\smonetwo = x_{m'n'}^\smonetwo - x_{m' 1}^\smonetwo - x_{1
n'}^\smonetwo + x_{11}^\smonetwo\,,\qquad \ \ \
y_{m' n}^\smtwothree =  x_{m' n}^\smtwothree - x_{1n}^\smtwothree +
x_{1m'}^\smonetwo - x_{11}^\smonetwo \,,
$$
\be
\label{appd22}  y_{mn}^\smthreeone = x_{mn}^\smthreeone +
x_{n1}^\smonetwo + x_{1m}^\smtwothree \,.
\ee
To express $p_\smp3^-$ \rf{appd21}  in terms of more convenient new
independent variables we note the relations
\beq
\label{appd27}&&
y_{m'n'}^\smonetwo =x_{m'n' 1}-x_{m' 11} -x_{1 n' 1}+x_{111}\,,
\\
\label{appd26} && y_{m' 1}^\smtwothree =x_{1 m' 1}-x_{111}\,, \quad
y_{m'n'}^\smtwothree = x_{1m'n'} - x_{11n'}\,,
\\
\label{appd25} && y_{11}^\smthreeone=x_{111}\,, \quad
y_{1m'}^\smthreeone = x_{m'11}\,, \quad
y_{m'1}^\smthreeone = x_{11m'}\,, \quad
y_{m'n'}^\smthreeone = x_{n' 1 m'}\,,
\eeq
where the new independent variables
\be \label{appd28} x_{111}\,,\quad  x_{m' 11}\,, \quad x_{1 m' 1}\,,
\quad x_{11m'}\,,  \quad x_{1m'n'} \quad x_{m'1n'} \,,\quad x_{m'n'
1}\,, \ee
are defined by the relations
\be
\label{appd29} x_{mnq}\equiv x_{mn}^\smonetwo + x_{nq}^\smtwothree +
x_{qm}^\smthreeone\,. \ee
Note that a number of the independent variables \rf{appd28} is equal
to $3\nnu^2 - 3\nnu + 1$, while the variables $x_{mnq}$ \rf{appd29}
constitute overflow basis of variables. We prefer to exploit overflow
basis of the variables \rf{appd29} because this basis is defined
symmetrically with respect to all sort of oscillators. The variables
\rf{appd29}, being non-polynomial with respect to oscillators, admit
the representation
\be \label{appd30} x_{mnq} = \frac{Z_{mnq}}{B_m^\smone B_n^\smtwo
B_q^\smthree}\,,\ee
where the cubic forms $Z_{mnq}$ \rf{0003} are polynomial in
oscillators. In view of \rf{appd30}, we can use the variables
$B_n^\sma$ and $Z_{mnq}$ in place of $B_n^\sma$ and $x_{mnq}$. This
gives vertex in \rf{0002}.

{\bf Derivation of vertex} \rf{00m2}. Equations to be solved are
obtainable from Eqs.\rf{loc1}, \rf{Gandef} by plugging values of
$\mas_a$ \rf{00m1} in Eq.\rf{Gandef}. Since the ket-vectors of
massless fields \rf{intver16n2} are independent of the scalar
oscillators $\alpha_n$, the scalar oscillators $\alpha_n^\smone$,
$\alpha_n^\smtwo$ in \rf{varrep8} do not contribute to the
Hamiltonian \rf{pm1} and we can therefore ignore dependence on
$\alpha_n^\smone$-, $\alpha_n^\smtwo$-terms in \rf{varrep8}%
\footnote{ Derivatives with respect to the scalar oscillators
$\alpha_n^\smone$, $\alpha_n^\smtwo$ vanish in \rf{loc1}, \rf{Gandef}
when $\mas_1 = \mas_2 = 0$, i.e. ignoring dependence on
$\alpha_n^\smone$, $\alpha_n^\smtwo$ in the cubic interaction vertex
is a self-consistent procedure in solving Eq.\rf{loc1}.}. Thus, we
start with the vertex and equations
\beq
\label{app00mN1} && p_\smp3^- = p_\smp3^-(B_n^\sma,
\alpha_n^\smthree\,, \alpha_{mn}^\smaaplusone,
\alpha_{mn}^\smoneone\,, \alpha_{mn}^\smtwotwo\,,
Q_{mn}^\smthreethree )\,,
\\
\label{app00m1}&& \Bigl\{\frac{1}{2}\mas_3^2\partial_{B_n^\smone} +
\sum_{m=1}^\nu  B_m^\smtwo \partial_{\alpha_{nm}^\smonetwo} -
(B_m^\smthree - \frac{1}{2}\mas_3 \alpha_m^\smthree )
\partial_{\alpha_{mn}^{\smthreeone}}\Bigr\}p_\smp3^-=0\,,
\\
\label{app00m2}&& \Bigl\{
-\frac{1}{2}\mas_3^2\partial_{B_n^\smtwo}+\sum_{m=1}^\nu
(B_m^\smthree +\frac{1}{2}\mas_3
\alpha_m^\smthree)\partial_{\alpha_{nm}^\smtwothree} - B_m^\smone
\partial_{\alpha_{mn}^\smonetwo}\Bigr\} p_\smp3^-=0\,,
\\
\label{app00m3}&& \Bigl\{
\mas_3\partial_{\alpha_n^\smthree}+\sum_{m=1}^\nu B_m^\smone
\partial_{\alpha_{nm}^\smthreeone} - B_m^\smtwo
\partial_{\alpha_{mn}^\smtwothree } \Bigr\}p_\smp3^-=0\,. \eeq
Introducing in place of $\alpha_{mn}^\smtwothree$,
$\alpha_{mn}^\smthreeone$ new variables
$\widetilde{Q}_{mn}^\smtwothree$, $\widetilde{Q}_{mn}^\smthreeone$
defined by
\be \label{app00m35N1}
\widetilde{Q}_{mn}^\smtwothree = \alpha_{mn}^\smtwothree
+\frac{\alpha_n^\smthree}{\mas_3} B_m^\smtwo\,,\qquad
\widetilde{Q}_{mn}^\smthreeone = \alpha_{mn}^\smthreeone -
\frac{\alpha_m^\smthree}{\mas_3} B_n^\smone\,,\ee
we recast vertex \rf{app00mN1} and Eqs.\rf{app00m1}-\rf{app00m3} into
the form
\beq
\label{app00m34}
&& p_\smp3^- = p_\smp3^-(B_n^\sma\,, \alpha_n^\smthree,
\alpha_{mn}^\smonetwo\,, \widetilde{Q}_{mn}^\smtwothree,
\widetilde{Q}_{mn}^\smthreeone, \alpha_{mn}^\smoneone\,,
\alpha_{mn}^\smtwotwo\,, Q_{mn}^\smthreethree)\,,
\\[3pt]
\label{app00m4}&& \Bigl\{\frac{1}{2}\mas_3^2\partial_{B_n^\smone} +
\sum_{m=1}^\nnu B_m^\smtwo
\partial_{\alpha_{nm}^\smonetwo}- B_m^\smthree
\partial_{\widetilde{Q}_{mn}^\smthreeone}\Bigr\}p_\smp3^-=0\,,
\\
\label{app00m5}&& \Bigl\{
-\frac{1}{2}\mas_3^2\partial_{B_n^2}+\sum_{m=1}^\nnu B_m^\smthree
\partial_{\widetilde{Q}_{nm}^\smtwothree } - B_m^\smone
\partial_{\alpha_{mn}^\smonetwo}\Bigr\}p_\smp3^-=0\,,
\\
\label{app00m5N1}&& \mas_3\partial_{\alpha_n^\smthree} p_\smp3^-=0\,.
\eeq
Equations \rf{app00m5N1} tell us that vertex $p_\smp3^-$
\rf{app00m34} does not depend on $\alpha_n^\smthree$. Keeping this in
mind and introducing in place of $\alpha_{mn}^\smonetwo$,
$\widetilde{Q}_{mn}^\smtwothree$, $\widetilde{Q}_{mn}^\smthreeone$
new variables $Q_{mn}^\smaaplusone$, $a=1,2,3$, defined by
$$
Q_{mn}^\smonetwo = \alpha_{mn}^\smonetwo - \frac{2}{\mas_3^2}
B_m^\smone B_n^\smtwo\,,
$$
\be  \label{app00m37}
Q_{mn}^\smtwothree = \widetilde{Q}_{mn}^\smtwothree +
\frac{2}{\mas_3^2} B_m^\smtwo B_n^\smthree\,,\qquad
Q_{mn}^\smthreeone = \widetilde{Q}_{mn}^\smthreeone +
\frac{2}{\mas_3^2} B_m^\smthree B_n^\smone\,,\ee
we recast vertex $p_\smp3^-$ \rf{app00m34} and the remaining
Eqs.\rf{app00m4}, \rf{app00m5} into the form
\beq
\label{app00m35}
&& p_\smp3^- = p_\smp3^-(B_n^\sma\,, Q_{mn}^\smonetwo\,,
Q_{mn}^\smtwothree, Q_{mn}^\smthreeone, \alpha_{mn}^\smoneone\,,
\alpha_{mn}^\smtwotwo\,, Q_{mn}^\smthreethree)\,,
\\[3pt]
\label{app00m4NN1}&&  \mas_3^2\partial_{B_n^\smone} p_\smp3^-=0\,,
\qquad \mas_3^2\partial_{B_n^\smtwo} p_\smp3^-=0\,.
\eeq
Equations \rf{app00m4NN1} tell us that vertex $p_\smp3^-$
\rf{app00m35} does not depend on $B_n^\smone$ and $B_n^\smtwo$.
Keeping this in mind and inserting $\widetilde{Q}_{mn}^\smtwothree$,
$\widetilde{Q}_{mn}^\smthreeone$ \rf{app00m35N1} in expressions for
$Q_{mn}^\smtwothree$, $Q_{mn}^\smthreeone$ \rf{app00m37} we obtain
vertex \rf{00m2}-\rf{00m5}.

{\bf Derivation of vertex} \rf{mm09ex01}. Equations to be solved are
obtainable from Eqs.\rf{loc1}, \rf{Gandef} by plugging values of
$\mas_a$ \rf{mm08} in Eq.\rf{Gandef}. Since the ket-vector of
massless field \rf{intver16n2} is independent of the scalar
oscillators $\alpha_n$, the scalar oscillators in $\alpha_n^\smthree$
\rf{varrep8} do not contribute to the Hamiltonian \rf{pm1} and we can
therefore ignore dependence on $\alpha_n^\smthree$-terms in
\rf{varrep8}%
\footnote{ Derivatives with respect to the scalar oscillators
$\alpha_n^\smthree$ vanish in \rf{loc1}, \rf{Gandef} when $\mas_3 =
0$, i.e. ignoring dependence on $\alpha_n^\smthree$ in the cubic
interaction vertex is a self-consistent procedure in solving
Eq.\rf{loc1}.}. Thus, we start with the vertex and equations
\beq
&& \label{appmm073} p_\smp3^- = p_\smp3^-(B_n^\sma,
\alpha_n^\smone\,, \alpha_n^\smtwo\,, \alpha_{mn}^\smaaplusone,
Q_{mn}^\smoneone\,, Q_{mn}^\smtwotwo\,, \alpha_{mn}^\smthreethree
)\,,
\\
\label{appmm014}&& \Bigl\{ -\frac{1}{2} \mas_2^2
\partial_{B_n^\smone} + \mas_1 \partial_{\alpha_n^\smone}
+\sum_{m=1}^\nnu (B_m^\smtwo + \frac{1}{2}\mas_2\alpha_m^\smtwo)
\partial_{\alpha_{nm}^\smonetwo }-B_m^\smthree
\partial_{\alpha_{mn}^\smthreeone}\Bigr\}p_\smp3^-=0\,,
\\
\label{appmm015}&& \Bigl\{ \frac{1}{2}\mas_1^2\partial_{B_n^\smtwo} +
\mas_2
\partial_{\alpha_n^\smtwo}
+ \sum_{m=1}^\nnu B_m^\smthree\partial_{\alpha_{nm}^\smtwothree
}-(B_m^\smone -\frac{1}{2} \mas_1\alpha_m^\smone)
\partial_{\alpha_{mn}^\smonetwo }\Bigr\}p_\smp3^-=0\,,
\\
\label{appmm016}&& \Bigl\{ -\frac{\mas_1^2 -
\mas_2^2}{2}\partial_{B_n^\smthree} +\sum_{m=1}^\nnu (B_m^\smone +
\frac{1}{2}\mas_1\alpha_m^\smone) \partial_{\alpha_{nm}^\smthreeone}
- (B_m^\smtwo - \frac{1}{2}\mas_2\alpha_m^\smtwo)
\partial_{\alpha_{mn}^\smtwothree }\Bigr\}p_\smp3^-=0\,. \hspace{1.5cm}
\eeq
Introducing in place of $B_m^\smone$, $B_m^\smtwo$ the variables
$L_m^\smone$, $L_m^\smtwo$ defined in \rf{mm012} we recast vertex
$p_\smp3^-$ \rf{appmm073} and Eqs.\rf{appmm014}-\rf{appmm016}  into
the form
\beq
\label{appmm074} && {}\hspace{-0.4cm} p_\smp3^- =
p_\smp3^-(L_n^\smone, L_n^\smtwo, B_n^\smthree, \alpha_n^\smone\,
\alpha_n^\smtwo\,, \alpha_{mn}^\smaaplusone, Q_{mn}^\smoneone\,,
Q_{mn}^\smtwotwo\,, \alpha_{mn}^\smthreethree )\,,
\\[3pt]
\label{appmm017}&& {}\hspace{-0.5cm}\Bigl\{ \mas_1
\partial_{\alpha_n^\smone}
+\sum_{m=1}^\nnu (L_m^\smtwo+\frac{\mas_1^2 +
\mas_2^2}{2\mas_2}\alpha_m^\smtwo)
\partial_{\alpha_{nm}^\smonetwo }-B_m^\smthree
\partial_{\alpha_{mn}^\smthreeone}\Bigr\}p_\smp3^-=0\,,
\\
\label{appmm018}&&{}\hspace{-0.5cm} \Bigl\{\mas_2
\partial_{\alpha_n^\smtwo}
+\sum_{m=1}^\nnu B_m^\smthree\partial_{\alpha_{nm}^\smtwothree
}-(L_m^\smone -\frac{\mas_1^2 + \mas_2^2}{2\mas_1}\alpha_m^\smone)
\partial_{\alpha_{mn}^\smonetwo }\Bigr\}p_\smp3^-=0\,,
\\
\label{appmm019}&& {}\hspace{-0.5cm}\Bigl\{\! -\frac{\mas_1^2\! -
\mas_2^2}{2}\partial_{B_n^\smthree}\! + \!\! \sum_{m=1}^\nnu
(L_m^\smone + \! \frac{\mas_1^2\! -
\mas_2^2}{2\mas_1}\alpha_m^\smone)
\partial_{\alpha_{nm}^\smthreeone}\! - \! (L_m^\smtwo\!
+ \!\frac{\mas_1^2\! - \mas_2^2}{2\mas_2} \alpha_m^\smtwo)
\partial_{\alpha_{mn}^\smtwothree }\!\Bigr\}p_\smp3^-=0. \hspace{1.2cm}
\eeq
Introducing in place of $\alpha_{mn}^\smaaplusone$, $a=1,2,3$ new
variables $Q_{mn}^\smonetwo$, $\widetilde{Q}_{mn}^\smtwothree$,
$\widetilde{Q}_{mn}^\smthreeone$ defined by
$$
Q_{mn}^\smonetwo  = \alpha_{mn}^\smonetwo
+\frac{\alpha_n^\smtwo}{\mas_2} L_m^\smone
-\frac{\alpha_m^\smone}{\mas_1} L_n^\smtwo -\frac{\mas_1^2 +
\mas_2^2}{2\mas_1\mas_2}\alpha_m^\smone \alpha_n^\smtwo\,,
$$
\be
\label{appmm021}
\widetilde{Q}_{mn}^\smtwothree = \alpha_{mn}^\smtwothree
-\frac{\alpha_m^\smtwo}{\mas_2} B_n^\smthree\,, \qquad
\widetilde{Q}_{mn}^\smthreeone = \alpha_{mn}^\smthreeone
+\frac{\alpha_n^\smone}{\mas_1} B_m^\smthree\,, \ee
we recast vertex $p_\smp3^-$ \rf{appmm074} and
Eqs.\rf{appmm017}-\rf{appmm019} into the form
\beq
\label{appmm075} && p_\smp3^- = p_\smp3^-(L_n^\smone, L_n^\smtwo,
B_n^\smthree, \alpha_n^\smone\, \alpha_n^\smtwo\,, Q_{mn}^\smonetwo,
\widetilde{Q}_{mn}^\smtwothree, \widetilde{Q}_{mn}^\smthreeone,
 Q_{mn}^\smoneone\,, Q_{mn}^\smtwotwo\,, \alpha_{mn}^\smthreethree)\,,
\\[3pt]
\label{appmm076} &&
\mas_1\partial_{\alpha_n^\smone} p_\smp3^- = 0\,,\qquad
\mas_2\partial_{\alpha_n^\smtwo} p_\smp3^- = 0\,,
\\
&& \label{appmm023}
\Bigl\{ - \frac{\mas_1^2 - \mas_2^2}{2}\partial_{B_n^\smthree}
+\sum_{m=1}^\nnu L_m^\smone
\partial_{\widetilde{Q}_{nm}^\smthreeone}- L_m^\smtwo
\partial_{\widetilde{Q}_{mn}^\smtwothree }\Bigr\}p_\smp3^-=0\,.
\eeq
Equations \rf{appmm076} tell us that vertex $p_\smp3^-$ \rf{appmm075}
does not depend on $\alpha_n^\smone$, $\alpha_n^\smtwo$. Keeping this
in mind and introducing in place of $\widetilde{Q}_{mn}^\smtwothree$,
$\widetilde{Q}_{mn}^\smthreeone$ the new variables
$Q_{mn}^\smtwothree$, $Q_{mn}^\smthreeone$ defined by
\be \label{appmm078N1}
Q_{mn}^\smtwothree  = \tilde{Q}_{mn}^\smtwothree -\frac{2}{\mas_1^2 -
\mas_2^2}B_n^\smthree L_m^\smtwo\,,\qquad
Q_{mn}^\smthreeone = \tilde{Q}_{mn}^\smthreeone + \frac{2}{\mas_1^2 -
\mas_2^2}B_m^\smthree L_n^\smone\,, \ee
we recast vertex $p_\smp3^-$ \rf{appmm075} and the remaining
Eqs.\rf{appmm023} into the form
\beq
\label{appmm078}
&& p_\smp3^- = p_\smp3^-(L_n^\smone, L_n^\smtwo, B_n^\smthree,
Q_{mn}^\smonetwo, Q_{mn}^\smtwothree, Q_{mn}^\smthreeone,
Q_{mn}^\smoneone\,, Q_{mn}^\smtwotwo\,, \alpha_{mn}^\smthreethree)\,,
\\[3pt]
\label{appmm079} && (\mas_1^2 - \mas_2^2)\partial_{B_n^\smthree}
p_\smp3^-=0\,. \eeq
Equations \rf{appmm079} tell us that vertex $p_\smp3^-$ \rf{appmm078}
does not depend on $B_n^\smthree$. Keeping this in mind and inserting
$\widetilde{Q}_{mn}^\smtwothree$, $\widetilde{Q}_{mn}^\smthreeone$
\rf{appmm021} in expressions for $Q_{mn}^\smtwothree$,
$Q_{mn}^\smthreeone$ \rf{appmm078N1} we obtain vertex in
\rf{mm09ex01}-\rf{mm011}.

{\bf Derivation of vertex} \rf{intvereqmas01}. Vertex and appropriate
equations to be solved are obtainable from
\rf{appmm073}-\rf{appmm016} by setting $\mas_1=\mas_2=\mas$.
Repeating procedure, we used to solve equations
\rf{appmm014}-\rf{appmm016}, we obtain vertex \rf{appmm075} and
Eqs.\rf{appmm076}, while Eqs.\rf{appmm023} take the form
\be \label{appmm080} \sum_{m=1}^\nnu  \bigl( L_m^\smone
\partial_{\widetilde{Q}_{nm}^\smthreeone}- L_m^\smtwo
\partial_{\widetilde{Q}_{mn}^\smtwothree } \bigr) p_\smp3^-=0\,.
\ee
The solution of Eqs.\rf{appmm076}, \rf{appmm080} is given by
\beq
\label{appmm081} && p_\smp3^- = p_\smp3^-(L_n^\smone, L_n^\smtwo,
B_n^\smthree, Q_{mn}^\smonetwo,\,, Q_{mn}^\smoneone\,,
Q_{mn}^\smtwotwo\,,
\alpha_{mn}^\smthreethree\,;\,\widetilde{Z}_{mnq})\,,
\\
\label{appmm082} && \widetilde{Z}_{mnq} \equiv L_m^\smone
\widetilde{Q}_{nq}^\smtwothree + L_n^\smtwo
\widetilde{Q}_{qm}^\smthreeone\,.
\eeq
In place of the form $\widetilde{Z}_{mnq}$ \rf{appmm082}, we prefer
to use the form $Z_{mnq}$ defined by
\be\label{ZZtild} Z_{mnq} = \widetilde{Z}_{mnq} + Q_{mn}^\smonetwo
B_q^\smthree\,. \ee
Inserting $\widetilde{Z}_{mnq}$ \rf{appmm082} and
$\widetilde{Q}_{mn}^\smtwothree$, $\widetilde{Q}_{mn}^\smthreeone$
\rf{appmm021} in \rf{ZZtild} we obtain vertex
\rf{intvereqmas01}-\rf{eqmas00005}.

{\bf Derivation of vertex} \rf{mmm1}. Equations to be solved are
given in \rf{loc1}, \rf{Gandef} in which we keep $\mas_a\ne 0$,
$a=1,2,3$. If we introduce variables $L_n^\sma$ \rf{mmm3} in place of
$B_n^\sma$, then vertex \rf{varrep8} and Eqs.\rf{loc1} take the form
\beq
\label{appmmm54}
&& p_\smp3^- = p_\smp3^-(L_n^\sma, \alpha_n^\sma\,,
\alpha_{mn}^\smaaplusone, Q_{mn}^\smaa )\,,
\\[3pt]
\label{appmmm55}
&& \Bigl\{ \mas_a
\partial_{\alpha_n^\sma}+\sum_{m=1}^\nnu
(L_m^\smaplusone + \frac{\mas_{a+1}^2 + \mas_a^2 -
\mas_{a+2}^2}{2\mas_{a+1}}\alpha_m^\smaplusone )
\partial_{\alpha_{nm}^\smaaplusone }\bigr.
\nonumber\\
&&\bigl. \hspace{2cm} -(L_m^\smaplustwo - \frac{\mas_{a+2}^2 +
\mas_a^2 - \mas_{a+1}^2}{2\mas_{a+2}} \alpha_m^\smaplustwo)
\partial_{\alpha_{mn}^\smaplustwoa}\Bigr\}p_\smp3^- =0\,.\eeq
We see that $\partial_{B_n^\sma}$-terms in Eqs.\rf{loc1} are
cancelled in Eqs.\rf{appmmm55}. In fact, it is desire to cancel the
$\partial_{B_n^\sma}$-terms that motivates usage of the variables
$L_n^\sma$. Now, in place of variables $\alpha_{mn}^\smaaplusone$ in
\rf{appmmm54}, we introduce new variables $Q_{mn}^\smaaplusone$
defined by
\be \label{appmmm56} Q_{mn}^\smaaplusone \equiv
\alpha_{mn}^\smaaplusone +
\frac{\alpha_n^\smaplusone}{\mas_{a+1}}L_m^\sma -
\frac{\alpha_m^\sma}{\mas_a}L_n^\smaplusone + \frac{\mas_{a+2}^2 -
\mas_a^2 - \mas_{a+1}^2}{2\mas_a \mas_{a+1}}\alpha_m^\sma
\alpha_n^\smaplusone\,. \ee
In terms of new variables, vertex $p_\smp3^-$ \rf{appmmm54} and
Eqs.\rf{appmmm55} take the form
\beq
&& \label{appmmm57} p_\smp3^- = p_\smp3^-(L_n^\sma, \alpha_n^\sma\,,
Q_{mn}^\smaaplusone, Q_{mn}^\smaa )\,,
\\[3pt]
&& \label{appmmm58} \mas_a\partial_{\alpha_n^\sma} p_\smp3^- =
0\,,\qquad a=1,2,3\,.
\eeq
Because of $\mas_a\ne 0$ Eqs.\rf{appmmm58} tell us that vertex
$p_\smp3^-$ \rf{appmmm57} does not depend on $\alpha_n^\sma$,
$a=1,2,3$. Inserting $L_n^\sma$ \rf{mmm3} in \rf{appmmm56} we obtain
vertex $p_\smp3^-$ given in \rf{mmm1}-\rf{mmm2}.

\appendix{Derivation of relation \rf{d422} and Eq.\rf{d427}.}

To derive the representation \rf{d422} we use $RL$, $Ri$ and $ij$
parts of Eqs.\rf{d43} given in \rf{d417}. Acting with the angular
momentum $\Jbf^{IJ}$ \rf{JIJp3} on vertex $p_\smp3^-$ \rf{p3int}, we
find the relations
\beq && \Jbf^{RL} p_\smp3^- = ({\Po }^L)^k E_q\Bigl(\Mbf^{RL} +
2\rho\partial_\rho - k \Bigr)\widetilde{V}\,,
\\
&& \Jbf^{Ri} p_\smp3^- =q^i \Jbf^{RL} p_\smp3^- +({\Po }^L)^k E_q
\Bigl(\Mbf^{Ri}  - \rho \Mbf^{Li} + q^j
\Mbf^{ij}\Bigr)\widetilde{V}\,, \ \ \
\\
&& \Jbf^{ij} p_\smp3^- =({\Po }^L)^k E_q \Mbf^{ij}\widetilde{V}\,.
\eeq
From these relations, it is easily seen that Eqs.\rf{d417} amount to
the following equations:
\beq \label{tv1} && (\Mbf^{RL} + 2\rho\partial_\rho
-k)\widetilde{V}=0\,,
\\
\label{tv2} && (\Mbf^{Ri} -\rho \Mbf^{Li})\widetilde{V}=0\,,
\\
\label{tv3}&& \Mbf^{ij} \widetilde{V}=0\,. \eeq
{}From relations \rf{newvar} and the fact that the vertex $p_\smp3^-$
is degree $k$ homogeneous polynomial in ${\Po }^I$, it follows that
$\widetilde{V}$ should be degree $k$ polynomial in the momentum
$\rho$, i.e. we can use the expansion
\be\label{tvexp} \widetilde{V}(\rho,\beta_a\,;\, \alpha)
=\sum_{n=0}^k \rho^n \widetilde{V}_n(\beta_a\,;\, \alpha)\,. \ee
Plugging this expansion in Eq.\rf{tv2} we get the following
equations:
\beq &&\label{vnvn-1} \Mbf^{Ri} \widetilde{V}_n = \Mbf^{Li}
\widetilde{V}_{n-1}\,,\qquad n=1,\ldots, k\,,
\\
\label{vnvn-2} && \Mbf^{Ri} \widetilde{V}_0=0\,,
\\
\label{vnvn-3} && \Mbf^{Li} \widetilde{V}_k=0\,, \eeq
while Eqs.\rf{tv1}, \rf{tv3} lead to the respective equations for
$\widetilde{V}_0$ given in \rf{d424}, \rf{d426}. In view of
Eq.\rf{vnvn-2}, we therefore conclude that $\widetilde{V}_0$ should
satisfy Eqs.\rf{d424}-\rf{d426}. Now we focus on Eqs.\rf{vnvn-1},
which tell us that $\widetilde{V}_n$ can be expressed in terms of
$\widetilde{V}_0$. Making use of Eqs.\rf{vnvn-1}, \rf{tv3} one can
make sure that $\widetilde{V}_n$ can be presented in the form
\be\label{vnv0} \widetilde{V}_n =f_n (\Mbf^{Lj} \Mbf^{Lj} )^n
\widetilde{V}_0\,, \qquad f_0 =1\,.\ee
Now, making use of Eqs.\rf{d424}-\rf{d426} and commutators
\be [\Mbf^{Ri},\,(\Mbf^{Lj} \Mbf^{Lj})^n] =( \Mbf^{Lj}
\Mbf^{Lj})^{n-1} 2n\bigl( \Mbf^{Lj} \Mbf^{ji} - \Mbf^{Li} \Mbf^{RL}
-(\frac{N'}{2}-n) \Mbf^{Li}\bigr) \ee
we get
\be\label{for11} \Mbf^{Ri} ( \Mbf^{Lj} \Mbf^{Lj})^n\widetilde{V}_0
=-n(N'+2k-2n)( \Mbf^{Lj} \Mbf^{Lj})^{n-1}
\Mbf^{Li}\widetilde{V}_0\,, \ee
where $N' \equiv d-4$. Making use of \rf{for11}, \rf{vnv0} in
Eqs.\rf{vnvn-1} gives the equations for $f_n\,,$
\be \frac{f_{n-1}}{f_n}=-n(N'+2k-2n)\,. \ee
The solution of these equations with $f_0=1$ is easily found to be
\be \label{fnsol}f_n=(-)^n\frac{\Gamma(\frac{N'}{2}+k-n)}{2^n
n!\Gamma(\frac{N'}{2}+k)}\,, \ee
where $\Gamma$ is the Euler gamma function. Collecting all the steps
of the derivation given in relations \rf{tvexp}, \rf{vnv0},
\rf{fnsol} we arrive at the solution given in \rf{d422}, \rf{d423}.
Note also that taking into account the solution for $\widetilde{V}_k$
and Eq.\rf{vnvn-3} we get the additional equation to be imposed on
the vertex $\widetilde{V}_0$:
\be\label{app0010} \Mbf^{Li} (\Mbf^{Lj}\Mbf^{Lj})^k \widetilde{V}_0 =
0\,.\ee
Our experience led us to a conclusion that Eq.\rf{app0010} is
satisfied automatically provided we satisfied all the equations
above-discussed. Therefore, we have not displayed Eq.\rf{app0010} in
Section \ref{sod-4sec}.

Having derived vertex \rf{p3v0}, we are ready to demonstrate that
this vertex satisfies the harmonic equation \rf{d44N1}. To this end
we rewrite operator $\partial_{\Po^I}\partial_{\Po^I}$ in terms of
the momenta $\Po^L$, $q^i$, $\rho$ \rf{newvar}
\be \label{Dalamnew} \partial_{\Po^I}\partial_{\Po^I} =
(\Po^L)^{-2}\left( (2 \Po^L\partial_{\Po^L}+ N-4  -
2\rho\partial_\rho)\partial_\rho +
\partial_{q^i}\partial_{q^i}\right)\,.\ee
Acing with $\partial_{\Po^I}\partial_{\Po^I}$ \rf{Dalamnew} on vertex
\rf{p3v0} and using solution for $f_n$ \rf{fnsol}, we obtain
\be \partial_{\Po^I}\partial_{\Po^I}|p_\smp3^-\rangle  =
(\Po^L)^{k-2} E_q  (-\rho)^k f_k (\Mbf^{Lj}\Mbf^{Lj})^{k+1}
\widetilde{V}_0\,.\ee
From this formula and \rf{app0010}, we see that the harmonic equation
\rf{d44N1} is satisfied indeed.

We now demonstrate that the locality equation \rf{Llocequ01} leads to
Eq.\rf{d427}. Making use of notation adopted in \rf{j-harmexp} we
rewrite the locality equation \rf{Llocequ01} as
\be\label{applocequ}   \XX_{harm}^L | p_\smp3^-\rangle = 0\,, \qquad
\XX_{harm}^L \equiv X^{LR}\PP^L + X^{Li}\PP^i\,. \ee
The operator $\XX_{harm}^L$, being differential operator in the
momentum $\Po^I$, can be rewritten in terms of the momenta $\Po^L$,
$q^i$, $\rho$ \rf{newvar}. To this end we use chain rules
\beq
\label{appPPL}&& \PP^L =\Po^L (1 - \frac{2}{2\kh + N -
2}\rho\partial_\rho)\,,
\\
\label{appPPi} && \PP^i =\Po^L \Bigl(q^i (1 - \frac{2}{2\kh +
N-2}\rho\partial_\rho) - \frac{2}{2\kh
+N-2}\rho\partial_{q^i}\Bigr)\,, \eeq
where the operator $\kh$ \rf{khdef} takes the form
$\kh = \Po^L\partial_{\Po^L}$.
The differential operators $\PP^L$, $\PP^i$ in \rf{applocequ} are
acting on the vertex that is degree $k$ monomial in $\Po^L$
\rf{p3v001}. Therefore in expressions for $\PP^L$, $\PP^i$
\rf{appPPL}, \rf{appPPi} we can use the identification $\kh = k$. By
using this identification and expressions for $X^{IJ}$ \rf{harver02}
we recast the operator $\XX_{harm}^L$ into the form
\be  \XX_{harm}^L   =   \sum_{a=1}^3 \check{\beta}_a( \beta_a
\partial_{\beta_a} + M^{\sma RL} - q^i M^{\sma Li})\PP^L
+ \frac{2\Po^L \rho}{2k+N-2} \check{\beta}_a
 M^{\sma Li} \partial_{q^i}\,.  \ee
Acting with this operator on vertex $|p_\smp3^-\rangle$ \rf{p3v0}, we
obtain sequence of the relations
\beq
\label{app009} \XX_{harm}^L |p_\smp3^-\rangle &  = &
 (\Po^L)^k \widehat{E}_q \sum_{a=1}^3 \Bigl(\check{\beta}_a(
\beta_a
\partial_{\beta_a} + M^{\sma RL})\PP^L
- \frac{2\Po^L \rho  \check{\beta}_a  M^{\sma Li}}{2k+N-2}
 \Mbf^{Li}\Bigr)\widehat{E}_\rho  \widetilde{V}_0
\nonumber\\
&   = &  (\Po^L)^k \widehat{E}_q \widehat{E}_\rho \PP^L \sum_{a=1}^3
\check{\beta}_a( \beta_a
\partial_{\beta_a} + M^{\sma RL})\widetilde{V}_0
\nonumber\\
& - &\frac{2(\Po^L \rho)^{k+1}\widehat{E}_q}{2k+N-2}f_k
\sum_{a=1}^3\check{\beta}_a
 M^{\sma Li} \Mbf^{Li} (\Mbf^{Lj}\Mbf^{Lj})^k  \widetilde{V}_0\,,   \eeq
where $f_k$ is given in \rf{fnsol}. From the 2nd line of relations
\rf{app009} we see that Eq.\rf{Llocequ01} (or \rf{applocequ}) gives
the desired Eq.\rf{d427}. Expression in the 3rd line in \rf{app009}
is equal to zero due to Eq.\rf{app0010}.


\end{document}